\documentclass[preprint,12pt]{elsarticle}
\usepackage{hyperref}

\usepackage{amssymb}
\usepackage{amsmath}
\usepackage{xcolor}

\usepackage{lineno}

\usepackage{tikz}
\usepackage{multirow}
\usepackage{float}
\usepackage[compat=1.1.0]{tikz-feynhand}
\newcommand{\ep}{\ensuremath{\varepsilon}}
\newcommand{\nickel}[1]{\texttt{#1}}
\newcommand{\Z}{\ensuremath{\mathcal{Z}}}
\newcommand{\Q}{\ensuremath{\mathfrak{n}}}

\newcommand{\KRP}{\ensuremath{\mathcal{KR}^\prime}}
\newcommand{\KRS}{\ensuremath{\mathcal{K\overline{R}}^*}}
\newcommand{\MS}{\ensuremath{\overline{\text{MS}}}}
\newcommand{\auxdia}[2]{\texttt{#1-#2}}

\journal{Nuclear Physics B}
\begin{document}

\begin{frontmatter}



\title{On the six-loop scaling dimensions \\ of the $(\phi^2)^n$ operators in $d=3$}


\author[JINR]{A.V.~Bednyakov}
\ead{bednya@jinr.ru}
\author[JINR,SPbU]{M.V.~Kompaniets}
\ead{m.kompaniets@spbu.ru}
\author[JINR,SPbU]{A.V.~Trenogin}
\ead{av_trenogin@theor.jinr.ru}

\affiliation[JINR]{organization={Joint Institute for Nuclear Research},
            addressline={Joliot-Curie, 6}, 
            city={Dubna},
            postcode={141980}, 
            state={Moscow region},
            country={Russia}}
\affiliation[SPbU]{organization={Saint Petersburg State University},
            addressline={7/9 Universitetskaya nab.}, 
            city={St. Petersburg},
            postcode={199034}, 
            country={Russia}}

\begin{abstract}
	We consider a class of singlet operators $(\phi^2)^n$ in the three-dimensional $O(N)$ model with $\lambda^2 \phi^6$ interaction. Recently \cite{Antipin:2025ekk}, the corresponding anomalous dimensions $\gamma_{2n}$ were computed by semiclassical methods and the all-loop result for the leading-$n$ corrections in the small $\lambda$ limit was found.
	In this paper, we obtain the six-loop expressions not only for the leading-$n$ contribution but also for the subleading one. While the leading correction confirms the predictions of recent semiclassical calculation, the subleading one is a new result and will serve as a future welcome check for all-loop expressions.
	As an important by-product of our calculation, we provide a full dependence on $n$ of the four-loop $\gamma_{2n}$ in the $O(N)$ case.
\end{abstract}




\end{frontmatter}


\section{Introduction}

In high-energy physics, quantum field theory (QFT) is the main tool used to describe various phenomena observed in experiment. 
A major role in QFTs is played by the renormalization group (RG), which, among other things, encodes the scale dependence of the Lagrangian parameters. 
Renormalizable QFT models, such as the Standard Model (SM), have a finite set of parameters giving rise to a high predictive power across a wide range of energy scales. 
Nevertheless, it is believed that renormalizable QFTs in $d$ space-time dimensions are nothing else but (low-energy) effective field theories (EFT), in which we neglect an infinite tower of local composite operators $\{O_i(x)\}$.
A common approach is to order the latter by their canonical dimensions $d_{O_i}>d$. Such operators enter into the Lagrangian with couplings (Wilson coefficients) $c_i$ of negative mass dimension $d_{c_i} = d - d_{O_i}<0$, and one can easily convince oneself that their contributions are suppressed in the infra-red (IR) region. Nevertheless, one can consistently include corrections due to these operators by   
truncating the infinite set and keeping only those $O_i$ that have $d_{O_i} \leq d_{\text{max}}$ for some fixed $d_{\text{max}}$ (see, e.g., Refs.~\cite{Isidori:2023pyp,Henriksson:2025vyi} and references therein).

For phenomenological applications, it is important to know how the Wilson coefficients $c_i$ change with renormalization scale $\mu$. The corresponding RG functions are directly related to the operator anomalous dimensions (ADs) and can be calculated 
in perturbation theory (PT) at weak coupling $\lambda$. 

One can also apply QFT to study second-order phase transitions (see, e.g., the review~\cite{Pelissetto:2000ek} and the textbook~\cite{Vasiliev:2006}). Anomalous dimensions evaluated at a infra-red fixed point (FP) $\lambda^*$ 
become observable quantities corresponding to various critical exponents $\Delta$. One can routinely use PT to compute the latter as a series in $\ep$ \cite{Wilson:1971dc}, which is a deviation of physical dimension $d=d_0 - 2\ep$ from the logarithmic one $d_0$, in which the coupling $\lambda$ is dimensionless. 

At criticality, physical systems usually exhibit universal behavior, which can also be described by a conformal field theory (CFT), see, e.g., the recent \cite{Henriksson:2025vyi} and references therein.  
The application of the CFT methods to critical phenomena can go beyond the usual PT and allows one to study the strong-coupling regime of the model.
Due to this, comparison between perturbative and non-perturbative results plays an important role in understanding QFT models.      
Establishing a connection between the quantities computed in PT and non-perturbatively 
works both ways. On the one hand, non-perturbative results can extend 
the applicability of expressions computed at a fixed order of PT (e.g., by certain re-summation of PT series). 
On the other hand, such a comparison can provide a convenient (QFT) interpretation of quantities computed within CFTs. 

Among recent examples, let us mention a vast literature on the so-called large-charge expansion (see, e.g,. Ref.~\cite{Alvarez-Gaume:2016vff,Gaume:2020bmp,Antipin:2020abu}), in which one introduces a 't Hooft-like coupling $\lambda Q$, rewrites the scaling dimension of the lowest-lying operator with charge $Q$ as 
\begin{align}
	\Delta_Q = \sum\limits_{j=-1}^\infty \frac{\Delta_j(\lambda^* Q)}{Q^j}
	\label{eq:large_Q_exp}
\end{align}
and derives all-loop results for the leading-$Q$ ($\Delta_{-1}$) and subleading-$Q$ ($\Delta_0$) terms via a semiclassical calculation.
It is worth mentioning that a direct comparison with diagrammatic computations (see, e.g, Ref.~\cite{Badel:2019khk,Badel:2019oxl,Jack:2020wvs,Jin:2022nqq,Bednyakov:2022guj,Huang:2024hsn}) beyond the one-loop order demonstrates that the anomalous dimensions obtained by the above-mentioned 
non-perturbative method precisely match standard PT calculations even away from the FP.

Recently, the semiclassical method was extended to account for the spinless neutral operators of the form $(\phi^2)^n$ \cite{Antipin:2025ekk}, and a similar expansion 
\eqref{eq:large_Q_exp} for the scaling dimension $\Delta_{2n}$ was computed in the leading-$n$ \cite{Antipin:2025ekk} approximation for $d_0=3,4,6$. 
The authors of Ref.~\cite{Antipin:2025ekk} compared their leading-$n$ prediction with available lowest-order results and found perfect agreement. 
The subleading-$n$ terms for Ising CFT at $d_0=4$ were derived in \cite{Antipin:2025ilv} together with a bunch of new results for the operators with spin.

In this paper, we restrict ourselves to $d_0 = 3$ and consider both leading-$n$ and subleading-$n$ contributions to the anomalous dimensions $\gamma_{2n}$ of spinless $(\phi^2)^n$ operators\footnote{Strictly speaking, an eigenoperator with a certain scaling dimension, which in the lowest order coincides with $(\phi^2)^n$.} in the $O(N)$ model with $\lambda^2 \phi^6$ interaction. To do this, we use the following Euclidean Lagrangain:
\begin{align}
	\mathcal{L} = \frac{Z_1}{2} (\partial_\nu \phi \cdot \partial_\nu \phi) + \frac{Z_3}{6} \lambda^2\mu^{4\ep}(\phi^2)^3, \qquad [\phi] = \frac{1}{2} - \ep,
	\label{eq:Lag}
\end{align}
where we suppress $O(N)$ indices of the scalar field $\phi = \{\phi^a, a=1,\ldots,N\}$ in the fundamental representation of the group. 
The same model with an additional mass term can be applied in statistical physics to systems with tricritical points,  as demonstrated in Ref. \cite{Vasiliev:2006}. These points have been experimentally observed in various systems (see Refs. \cite{Vasiliev:2006,b:PTandCP9} and the references therein). These systems show different asymptotic behaviors when approaching the tricritical point, depending on the trajectory taken in the physical parameter space. These behaviors arise naturally from the analysis of the $\phi^4+\phi^6$ model\footnote{It should be noted that the critical and tricritical points also differ in the behavior 
when a system approaches them. When approaching the critical point, the coupling of the $\phi^4$ interaction tends to a nonzero constant, while in the vicinity of the tricritical point it tends to zero. As a conslquence, various situations become possible when approaching the tricritical point.} \cite{Vasiliev:2006,Riedel:1972_1,Riedel:1972_2}. For a trajectory for which the $\phi^4$ interaction near the tricritical point is more significant than the $\phi^6$ one, the latter becomes irrelevant, and  the corresponding \emph{modified}  critical behavior can be studied by means of the $\phi^4$ model\footnote{The corresponding \emph{modified} critical exponents are not the same as in the usual critical situation, in which the $\phi^4$ coupling remains a nonzero constant at the fixed point. 
Nevertheless, the modified exponents can be easily recomputed from the critical ones \cite{Vasiliev:2006}.}. On the other hand, if the $\phi^6$ interaction dominates, this behavior becomes tricritical and can be described by the $\phi^6$ model with the corresponding tricritical exponents \cite{a:STEPHEN1973, LewisAdams:1978, Hager:1999, Hager:2002, Adzhemyan:2026}. Additionally, in Ref. \cite{Vasiliev:2006}, a situation was explored when both interactions are equally significant (so-called combined tricriticality\cite{Vasiliev:2006}). However, this possibility has not been well studied yet and calls for an additional investigation (see Ref.~\cite{Adzhemyan:2026}). Let us also mention that the phenomenon was also studied using various simulations in different models, e.g., in the Blum-Capel or Blum-Emery-Griffith models (see Refs. \cite{Henriksson:2025kws,a:Moueddene_2024_d2,a:Moueddene_2024_d3}).

The renormalization constants $Z_1$ and $Z_3$ entering into \eqref{eq:Lag} are known up to six loops \cite{LewisAdams:1978,Hager:1999,Hager:2002,Adzhemyan:2026} and are introduced to render all the Green functions involving the $\phi$ fields finite. In what follows we routinely use dimensional regularization \cite{tHooft:1972tcz}, i.e., we work in $d = 3-2\ep$, together with the (modified) minimal subtraction scheme \MS~\cite{Bardeen:1978yd}.  
In the latter scheme, the renormalization constants are functions of the dimensionless coupling $\lambda$ that depends on the renormalization scale $\mu$.

In perturbation theory (i.e., small $\lambda$ limit) the anomalous dimension can be represented as\footnote{We use $\Q$ (the number of fields in the operator) as an argument of the polynomials $P_{2l}$ for convenience enabling easy comparison with Ref.~\cite{Antipin:2025ekk}.}
\begin{align}
	\gamma_{2n} = \Q \sum\limits_{l=1}^\infty \lambda^{2l}  P_{2l}(\Q), \qquad P_{m}(x) = \sum\limits_{k=0}^{m} C_{m,k}(N)\cdot x^{m-k}, \qquad \Q\equiv 2n,
	\label{eq:gamma_n_PT}
\end{align}
 where the sum over $l$ corresponds to loop expansion, and $P_{m}(x)$ is a degree-$m$ polynomial with the coefficients $C_{m,k}$ depending on $N$. 
 One can see that at each order of PT all $C_{2l,k}$ with $k=0,\ldots,2l$ can be determined by considering the anomalous dimensions $\gamma_{2k}$ for fixed $k=1,\ldots,2l+1$.
Due to the $O(N)$ symmetry\footnote{In the case of $N=1$, we can use half-integer $n$ as a constraint on $\gamma_{2n}$.}, this is not an easy task at high PT orders. For example, at six loops one needs to consider the family of all operators from $\phi^2$ up to $(\phi^2)^7$ to reconstruct the full dependence on $n$.

The model \eqref{eq:Lag} exhibits a fixed point \cite{LewisAdams:1978,Hager:1999,Hager:2002} 
\begin{align}
	\frac{\lambda^{*2}}{8\pi^2} = \frac{\ep}{3 N+22} + \mathcal{O}(\ep^2)
	\label{eq:lambda_fp_leading}
\end{align}
resulting in a Conformal Field Theory (CFT). Substituting \eqref{eq:lambda_fp_leading} into \eqref{eq:gamma_n_PT}, we obtain  ($\Q\equiv 2n$)
\begin{align}\label{eq:phi_n_fp}
	\gamma_{2n}(\lambda^*) \equiv \gamma_{2n}^* = \Q \sum\limits_{l=1}^\infty (2\ep)^{l}  P^*_{2l}(\Q), \quad P^*_{m}(x) = \sum\limits_{k=0}^{m} \frac{D_{m,k}(N)}{(3N+22)^{m-1}}\cdot x^{m-k} \end{align}
that defines a scaling dimension of the corresponding operator 
\begin{align}
	\Delta_{2n} = n (1 - 2 \ep) + \gamma^*_{2n}
	\label{eq:Delta_2n}
\end{align}
and represents an important ingredient of the CFT data. The all-loop result of Ref.~\cite{Antipin:2025ekk} expanded in the weak coupling limit allows one to compute all the leading-$n$ coefficients, i.e., all $C_{2l,0}$ (and $D_{2l,0}$). 

The main aim of the paper is to use  a diagrammatic approach to obtain the expressions for both leading $C_{2l,0}$ and subleading $C_{2l,1}$ up to six loops. As a consequence, our results not only confirm the findings of Ref.~\cite{Antipin:2025ekk} but also provide a useful input and a cross check for subsequent semi-classical studies \cite{Antipin:2025rsr}.

The paper is organized as follows. In Section~\ref{sec:ren_ops}, 
we discuss
the renormalization of the $(\phi^2)^n$ operators along with the subtlety associated with their mixing with operators involving $2(n-2)$ fields. To compute the required contributions, we introduce a set of auxiliary operators together with auxiliary diagrams, which we describe in Sec.~\ref{sec:comp_tech}. Our results for the six-loop leading and subleading-$n$ corrections to $\gamma_{2n}$ and for the complete dependence on $n$ of the four-loop $\gamma^{(4)}_{2n}$ can be found in Sec.~\ref{sec:results}.  We conclude in Sec.~\ref{sec:conclusions}. In Appendices, one can find additional information on operators with $2(n-2)$ fields (\ref{app:On2_primary}), our treatment of quadratically divergent diagrams (\ref{app:KRp_quadratic}), and all required counterterms for six-loop diagrams (\ref{app:KRP_tables}).

\section{Renormalization of $(\phi^2)^n$: leading-$n$ and subleading-$n$ orders\label{sec:ren_ops}} 
We study the local $(\phi^2)^n$ operator which, following Ref.~\cite{Cao:2021cdt}, can be represented graphically as  
\tikzset{node distance = 2mm}
\begin{align}
	O^{(n)}_{n} = (\phi^2)^n \rightarrow
	\underbrace{
		\begin{tikzpicture}[baseline=-2pt]
			\begin{feynhand}
				\vertex (o)  {};
				\vertex [dot] (a1) [above = of o.center] {};
				\vertex [dot] (a2) [below = of o.center] {};
			\end{feynhand}
		\end{tikzpicture}
		~
		\begin{tikzpicture}[baseline=-2pt]
			\begin{feynhand}
				\vertex (o)  {};
				\vertex [dot] (a1) [above = of o.center] {};
				\vertex [dot] (a2) [below = of o.center] {};
			\end{feynhand}
		\end{tikzpicture}
\cdots
		\begin{tikzpicture}[baseline=-2pt]
			\begin{feynhand}
				\vertex (o)  {};
				\vertex [dot] (a1) [above = of o.center] {};
				\vertex [dot] (a2) [below = of o.center] {};
			\end{feynhand}
		\end{tikzpicture}
		~
		\begin{tikzpicture}[baseline=-2pt]
			\begin{feynhand}
				\vertex (o)  {};
				\vertex [dot] (a1) [above = of o.center] {};
				\vertex [dot] (a2) [below = of o.center] {};
			\end{feynhand}
		\end{tikzpicture}
}_n.
	\label{eq:On_op_def}
\end{align}
Here and in what follows we use $O^{(n)}_{m,k}$ to denote the $k$-th\footnote{We omit $k$ if there is a single operator for fixed $n$ and $m$.} operator with canonical dimension $n$ (in $d=3$) constructed from $2m$ fields.
The picture 
shows the structure of the operator by representing each field $\phi = \{\phi^a, a=1,\ldots,N\}$  by a dot. The fields/dots are arranged into $n$ columns corresponding to $O(N)$ contractions $\phi^a \phi^a = \phi^2$. %
The corresponding Feynman rule looks like
\begin{align}
O^{(n)}_{n} \to (2n)!! \left[ \delta^{a_1a_2} \cdots \delta^{a_{2n-1}a_{2n}} + \text{perms}\right], 
	\label{eq:feynman_rule_On}
\end{align}
	where $a_1,\ldots,a_{2n}$ are $O(N)$ indices of the external legs, and there are $(2n-1)!! = (2n)!/(2n)!!$ terms in the brackets.

To compute the anomalous dimension (matrix), we consider 
one-particle-irreducible (1PI) Green functions with an operator insertion, which can be written as
\begin{align}
	\Gamma^{a_1,\ldots,a_{k}}_{k}[O](Q,q_1,\ldots,q_{k}) & = \int d^D x  e^{-i Q x} \left[\prod\limits_{i=1}^{k} d^D x_{i} e^{-i q_i x_i} \right] 
	\times
	\nonumber\\
	& \times\langle  O(x) \phi^{a_1}(x_1) \ldots \phi^{a_{k}}(x_{k}) \rangle_{\text{1PI}}.
	\label{eq:Op_insertions}
\end{align}
Here we restrict ourselves to the singlet composite operators $O(x)$, with $\phi^{a_i}$ being fields (order parameter) in fundamental representation of the $O(N)$ group. To be as general as possible, we introduce a nonzero momentum $Q$ flowing into the operator vertex. 
For brevity, we use the condensed notation $$\Gamma_k[O]=\sum\limits_l \Gamma^{(2l)}_k[O]$$ for \eqref{eq:Op_insertions} and its $2l$-loop contribution, thus  omitting the external $O(N)$ indices together with the momentum dependence of the Green functions.

When the operator \eqref{eq:On_op_def} is inserted into the Green functions \eqref{eq:Op_insertions} with a different number of $\phi$'s, new ultra-violet (UV) divergences are generated, thus requiring the operator to be renormalized in addition to the fields and couplings of the model.
A common situation is the mixing between a family of the operators that results in the following relation between renormalized and bare operators:     
\begin{align}
	[O_i]_R = Z_{ij} (O_j)_{\text{bare}}, \qquad Z_{ij} = \sum\limits_{l=1}^\infty Z_{ij}^{(2l)}, \qquad Z_{ij}^{(2l)} = \sum\limits_{k=1}^l \frac{Z^{(2l,k)}_{ij}}{\ep^k}.
	\label{eq:ops_renorm}
\end{align}
Here the bare operators are constructed from the bare fields $\phi_{\text{bare}} = Z_1^{1/2} \phi$, and the sum over $l$ corresponds to loop expansion. The anomalous-dimension matrix (in the minimal scheme)
can be computed from 
\begin{align}
	\gamma_{ij} & \equiv - \frac{d Z_{ik}}{d \ln \mu} Z^{-1}_{kj} 
	= \sum\limits_{l=1}^{\infty} \gamma^{(2l)}_{ij}, \qquad \gamma^{(2l)}_{ij} =  2 (2 l) Z^{(2l,1)}_{ij}.
	\label{eq:gamma_ij_defs}
\end{align}
In what follows we routinely use the last equation that relates the $2l$-loop contribution $\gamma_{ij}^{(2l)}$ to the anomalous dimension with the coefficient of the first pole in $\ep$ of the renormalization constant $Z_{ij}^{(2l,1)}$.

A standard approach \cite{Vasiliev:2006} to calculate anomalous dimensions from \eqref{eq:ops_renorm} and \eqref{eq:gamma_ij_defs} requires one to use a closed system of composite operators $O_i$ with canonical dimensions equal to (or less than\footnote{e.g., if we consider massive theory or allow multiple insertions of certain operators}) the dimension of our interest.  
For example, in $d=3$ one can replace four $\phi$ fields by two derivatives without affecting the canonical dimension of the operator. 

In general, the system of operators can be quite large and a convenient choice can significantly simplify the computations (see, e.g., recent Ref.~\cite{Henriksson:2025vyi} and references therein).
For example, 
one can set the operator momentum $Q=0$ in \eqref{eq:Op_insertions}, i.e., instead of ``unintegrated'' operators we consider a set of ``integrated'' ones\footnote{This precisely corresponds to the procedure when one introduces (constant) couplings $c_i$ for a set of operators $O_i$ and add the terms $c_i O_i$ to the interaction Lagrangian.  The anomalous dimension matrix $\gamma_{ij}$ of $O_i$ in this case can be extracted from the linear terms of the $c_i$ beta functions.}. For the latter case, we do not need to deal with operators that are total derivatives, since they are 
proportional to $Q$ in momentum space. Such operators correspond to descendants of primaries in CFT and do not add new information to the CFT data with their anomalous dimensions related to that of primaries.

If we fix $n$ large enough, we have to consider insertions of $O^{(n)}_{n}$ not only into $2n$-point functions, which correspond to logarithmically divergent graphs, but also into $2(n-2)$, $2(n-4)$, etc. that lead to quadratically and higher divergent diagrams. As a consequence, the number of operators that can appear in the right-hand side (RHS) of \eqref{eq:ops_renorm} proliferate significantly (one can use Hilbert series to count independent operators, see, e.g., \cite{Cao:2021cdt}). Nevertheless, not all the operators appear as counterterms at lower loops. Due to this, we do not consider a general mixing matrix but use loop-by-loop approach, adding necessary operators when a new divergent contribution appears.

\def\scalefigure{1.2}

\begin{figure}[t]
	\begin{center}
	\begin{tabular}{cccc}
	\begin{tikzpicture}[baseline=-2pt,scale=\scalefigure]
		\def\radius{1cm}
		\begin{feynhand}
			\vertex[crossdot] (o) at (0,0) {};

			\def\NumNodes{3}

			\def\DeltaAngle{25}
			\def\ReferenceAngle{120}

			\foreach \i in {1, ..., \NumNodes} {
        		\pgfmathsetmacro{\angle}{\ReferenceAngle-\DeltaAngle + (\i-1)*(2*\DeltaAngle)/(\NumNodes-1)}
				\vertex  (phi1\i) at (\angle:0.8\radius) {};
				\propag[red] (phi1\i) to (o);
			}

			\node at (-0.4\radius,0.1\radius) {$\vdots$};

			\def\DeltaAngle{25}
			\def\ReferenceAngle{240}

			\foreach \i in {1, ..., \NumNodes} {
        		\pgfmathsetmacro{\angle}{\ReferenceAngle-\DeltaAngle + (\i-1)*(2*\DeltaAngle)/(\NumNodes-1)}
				\vertex  (phi2\i) at (\angle:0.8\radius) {};
				\propag[red] (phi2\i) to (o);
			}

			\vertex[dot] (v1) at (1,0) {}; 
			\propag[plain] (o) to (v1);
			\propag[plain] (o) to [out=45, in=  180-45] (v1);
			\propag[plain] (o) to [out=-45, in=-180+45] (v1);
			\vertex (a1) [above right = 0.7\radius of v1] {}; 
			\vertex (a2) [right = 0.7\radius of v1] {}; 
			\vertex (a3) [below right = 0.7\radius of v1] {}; 
			\propag[plain] (v1) to (a1);
			\propag[plain] (v1) to (a2);
			\propag[plain] (v1) to (a3);
%
		\end{feynhand}
	\end{tikzpicture}
	&
	\begin{tikzpicture}[baseline=-2pt,scale=\scalefigure]
		\def\radius{1cm}
		\begin{feynhand}
			\vertex[crossdot] (o) at (0,0) {};

			\def\NumNodes{3}

			\def\DeltaAngle{25}
			\def\ReferenceAngle{120}

			\foreach \i in {1, ..., \NumNodes} {
        		\pgfmathsetmacro{\angle}{\ReferenceAngle-\DeltaAngle + (\i-1)*(2*\DeltaAngle)/(\NumNodes-1)}
				\vertex  (phi1\i) at (\angle:0.8\radius) {};
				\propag[red] (phi1\i) to (o);
			}

			\node at (-0.4\radius,0.1\radius) {$\vdots$};

			\def\DeltaAngle{25}
			\def\ReferenceAngle{240}

			\foreach \i in {1, ..., \NumNodes} {
        		\pgfmathsetmacro{\angle}{\ReferenceAngle-\DeltaAngle + (\i-1)*(2*\DeltaAngle)/(\NumNodes-1)}
				\vertex  (phi2\i) at (\angle:0.8\radius) {};
				\propag[red] (phi2\i) to (o);
			}

			\vertex[dot] (v1) at (1,0) {}; 

			\propag[plain] (o) to (v1);
			\propag[plain] (o) to [out=30, in=  180-30] (v1);
			\propag[plain] (o) to [out=-30, in=-180+30] (v1);

			\propag[plain] (o) to [out=60, in=  180-60] (v1);
			\propag[plain] (o) to [out=-60, in=-180+60] (v1);
			\vertex (a2) [right = 0.7\radius of v1] {}; 
			\propag[plain] (v1) to (a2);
%
		\end{feynhand}
	\end{tikzpicture}
	&
	\begin{tikzpicture}[baseline=-2pt,scale=\scalefigure]
		\def\radius{1cm}
		\begin{feynhand}
			\vertex[crossdot] (o) at (0,0) {};

			\def\NumNodes{3}

			\def\DeltaAngle{25}
			\def\ReferenceAngle{120}

			\foreach \i in {1, ..., \NumNodes} {
        		\pgfmathsetmacro{\angle}{\ReferenceAngle-\DeltaAngle + (\i-1)*(2*\DeltaAngle)/(\NumNodes-1)}
				\vertex  (phi1\i) at (\angle:0.8\radius) {};
				\propag[red] (phi1\i) to (o);
			}

			\node at (-0.4\radius,0.1\radius) {$\vdots$};

			\def\DeltaAngle{25}
			\def\ReferenceAngle{240}

			\foreach \i in {1, ..., \NumNodes} {
        		\pgfmathsetmacro{\angle}{\ReferenceAngle-\DeltaAngle + (\i-1)*(2*\DeltaAngle)/(\NumNodes-1)}
				\vertex  (phi2\i) at (\angle:0.8\radius) {};
				\propag[red] (phi2\i) to (o);
			}

			\vertex[dot] (v1) at ([shift=({25:0.9\radius})]o) {}; 
			\vertex[dot] (v2) at ([shift=({-25:0.9\radius})]o) {}; 

			\propag[plain] (o) to (v1);
			\propag[plain] (o) to [out=25+20, in=  180+25-20] (v1);
			\propag[plain] (o) to [out=25-20, in=  180+25+20] (v1);

			\propag[plain] (o) to [out=-25+10, in=  180-25-10] (v2);
			\propag[plain] (o) to [out=-25-10, in=  180-25+10] (v2);

			\propag[plain] (o) to [out=-25+30, in=  180-25-30] (v2);
			\propag[plain] (o) to [out=-25-30, in=  180-25+30] (v2);

			\propag[plain] (v1) to (v2);

			\vertex (a1) [above right = 0.3\radius and 0.7\radius of v1] {}; 
			\vertex (a2) [below  right = 0.3\radius and 0.7\radius of v1] {}; 
			\vertex (a3) [below  right = 0.2\radius and 0.7\radius of v2] {}; 
			\propag[plain] (v1) to (a1);
			\propag[plain] (v1) to (a2);
			\propag[plain] (v2) to (a3);
%
		\end{feynhand}
	\end{tikzpicture}
	&
	\begin{tikzpicture}[baseline=-2pt,scale=\scalefigure]
		\def\radius{1cm}
		\begin{feynhand}
			\vertex[crossdot] (o) at (0,0) {};

			\def\NumNodes{3}

			\def\DeltaAngle{25}
			\def\ReferenceAngle{120}

			\foreach \i in {1, ..., \NumNodes} {
        		\pgfmathsetmacro{\angle}{\ReferenceAngle-\DeltaAngle + (\i-1)*(2*\DeltaAngle)/(\NumNodes-1)}
				\vertex  (phi1\i) at (\angle:0.8\radius) {};
				\propag[red] (phi1\i) to (o);
			}

			\node at (-0.4\radius,0.1\radius) {$\vdots$};

			\def\DeltaAngle{25}
			\def\ReferenceAngle{240}

			\foreach \i in {1, ..., \NumNodes} {
        		\pgfmathsetmacro{\angle}{\ReferenceAngle-\DeltaAngle + (\i-1)*(2*\DeltaAngle)/(\NumNodes-1)}
				\vertex  (phi2\i) at (\angle:0.8\radius) {};
				\propag[red] (phi2\i) to (o);
			}

			\vertex[dot] (v1) at ([shift=({25:0.9\radius})]o) {}; 
			\vertex[dot] (v2) at ([shift=({-25:0.9\radius})]o) {}; 

			\propag[plain] (o) to (v1);
			\propag[plain] (o) to [out=25+20, in=  180+25-20] (v1);
			\propag[plain] (o) to [out=25-20, in=  180+25+20] (v1);

			\propag[plain] (o) to (v2);
			\propag[plain] (o) to [out=-25+20, in=  180-25-20] (v2);
			\propag[plain] (o) to [out=-25-20, in=  180-25+20] (v2);

		\propag[plain] (v1) to [out=-90+25, in=90-25] (v2);
		\propag[plain] (v1) to [out=-90-25, in=90+25] (v2);

			\vertex (a1) [above right = 0.2\radius and 0.7\radius of v1] {}; 
			\vertex (a3) [below  right = 0.2\radius and 0.7\radius of v2] {}; 
			\propag[plain] (v1) to (a1);
			\propag[plain] (v2) to (a3);
%
		\end{feynhand}
	\end{tikzpicture}

	\\
	(a) & (b)  & (c) & (d)
\end{tabular}
\end{center}
\caption{Examples of Feynman diagrams giving rise to divergent contribution to $\Gamma_{2n}[O^{(n)}_{n}]$ at two loops $(a)$, and to $\Gamma_{2n-4}[O^{(n)}_{n}]$ at four $(b)$ and six $(c,d)$ loops. The ``spectator'' legs of the operator are marked in red. The corresponding auxiliary graphs $\Gamma_{3}[\tilde O^{(n)}_{3}]$, $\Gamma_{1}[\tilde O^{(n)}_{5}]$, $\Gamma_3[\tilde O^{(n)}_7]$, and $\Gamma_2[\tilde O^{(n)}_{6}]$ for $(a)$, $(b)$ $(c)$, and $(d)$, respectively, are obtained by stripping off the ``spectator'' lines.}
\label{fig:dias_examples}
\end{figure}
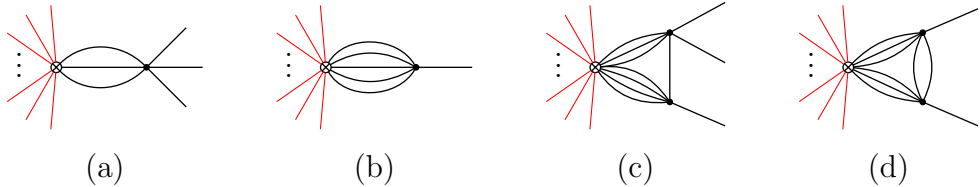

At two and higher loops, the 1PI $2n$-point function with one $O^{(n)}_{n}$ insertion is logarithmically divergent. 
In Fig.~\ref{fig:dias_examples}a, we show a two-loop diagram contributing to $\Gamma_{2n}[O^{(n)}_{n}]$.  One can see that at the lowest PT order only three (``active'' legs) out of $2n$ legs of the operator enter the 1PI part, while $2n-3$  (marked in red) are just ``spectators'', i.e., they are directly connected to the external lines $\phi^{a_i}$.

Having this in mind, we follow \cite{Jack:2020wvs} and consider only the class of UV divergent diagrams that give rise to leading-$n$ and subleading-$n$ terms in the anomalous dimension $\gamma_{2n}$: 
at $2l$ loops we select those graphs that have $2l+1$ and $2l$ ``active'' operator legs. We also ignore self-energy insertions into external legs, since they contribute only linearly in $n$ \cite{Jack:2020wvs}.

To compute the necessary corrections 
to $\gamma_{2n}$, we apply the $\Z$-operation to the selected diagrams. The latter associate a local counterterm with a 1PI Green function and in the minimal scheme is given by
\begin{align}
    \Z  \equiv - \KRP
\end{align}
with $\mathcal{R}^\prime$ being incomplete BPHZ $\mathcal{R}$-operation, and $\mathcal{K}$ extracting the pole part from the result. 
Since $\KRP$ of the graphs with cut vertices (vertex-reducible) factorizes and does not produce first poles in $\ep$, we safely ignore these diagrams.

There is a number of differences between our computation and that of Ref.~\cite{Jack:2020wvs}. First of all, we have a larger number of logarithmically divergent graphs (see~\ref{app:KRP_tables}), since the fixed-charge operators correspond to traceless symmetric products of $\phi^a$ so the diagrams leading to contractions between the fixed-$Q$ operator $O(N)$ indices vanish. 
Moreover, charged operators do not mix under renormalization with operators involving derivatives. 

In our case, at four and higher loops we have a nonzero contribution to the mixing between operators originating from the quadratically divergent graphs $\Gamma_{2n-4}[O^{(n)}_{n}]$ (see Fig.~\ref{fig:dias_examples}b-d). As a consequence, we have to consider all possible operators involving two derivatives, which can be written as   
\begin{align}
	O^{(n)}_{n-2,1} & = 
	(\phi^2)^{n-3} (\phi \partial^2 \phi) 
			&&
	\rightarrow 
			&
		\begin{tikzpicture}[baseline=-2pt]
			\begin{feynhand}
				\vertex (o)  {};
				\vertex [dot] (a1) [above = of o.center] {};
				\vertex [dot] (a2) [below = of o.center] {};
				\vertex (o1)  [above = of a1.center] {};
				\propag [plain] (a1.center) to[in=-180, out=180] (o1.center) to [in=0, out=0] (a1.center);
			\end{feynhand}
		\end{tikzpicture}
		\,
	\underbrace{
		\begin{tikzpicture}[baseline=-2pt]
			\begin{feynhand}
				\vertex (o)  {};
				\vertex [dot] (a1) [above = of o.center] {};
				\vertex [dot] (a2) [below = of o.center] {};
			\end{feynhand}
		\end{tikzpicture}
\cdots
		\begin{tikzpicture}[baseline=-2pt]
			\begin{feynhand}
				\vertex (o)  {};
				\vertex [dot] (a1) [above = of o.center] {};
				\vertex [dot] (a2) [below = of o.center] {};
			\end{feynhand}
		\end{tikzpicture}
		~
		\begin{tikzpicture}[baseline=-2pt]
			\begin{feynhand}
				\vertex (o)  {};
				\vertex [dot] (a1) [above = of o.center] {};
				\vertex [dot] (a2) [below = of o.center] {};
			\end{feynhand}
		\end{tikzpicture}
	}_{n-3},
	\label{eq:On1}\\
	O^{(n)}_{n-2,2} & = 
	(\phi^2)^{n-3} (\partial_\mu \phi \partial_\mu \phi)
			&&
	\rightarrow 
			&
		\begin{tikzpicture}[baseline=-2pt]
			\begin{feynhand}
				\vertex (o)  {};
				\vertex [dot] (a1) [above = of o.center] {};
				\vertex [dot] (a2) [below = of o.center] {};
				\propag [plain] (a1.center) to (a2.center);
			\end{feynhand}
		\end{tikzpicture}
		\,
	\underbrace{
		\begin{tikzpicture}[baseline=-2pt]
			\begin{feynhand}
				\vertex (o)  {};
				\vertex [dot] (a1) [above = of o.center] {};
				\vertex [dot] (a2) [below = of o.center] {};
			\end{feynhand}
		\end{tikzpicture}
\cdots
		\begin{tikzpicture}[baseline=-2pt]
			\begin{feynhand}
				\vertex (o)  {};
				\vertex [dot] (a1) [above = of o.center] {};
				\vertex [dot] (a2) [below = of o.center] {};
			\end{feynhand}
		\end{tikzpicture}
		~
		\begin{tikzpicture}[baseline=-2pt]
			\begin{feynhand}
				\vertex (o)  {};
				\vertex [dot] (a1) [above = of o.center] {};
				\vertex [dot] (a2) [below = of o.center] {};
			\end{feynhand}
		\end{tikzpicture}
	}_{n-3}
	, \label{eq:On2} \\
	O^{(n)}_{n-2,3} & = 
	(\phi^2)^{n-4} (\phi \partial_\mu \phi)^2
			&&
\rightarrow 
			&
		\begin{tikzpicture}[baseline=-2pt]
			\begin{feynhand}
				\vertex (o)  {};
				\vertex [dot] (a1) [above = of o.center] {};
				\vertex [dot] (a2) [below = of o.center] {};
				\vertex [dot] (b1) [right = 8pt of a1] {};
				\vertex [dot] (b2) [right = 8pt of a2] {};
				\propag [plain] (a1.center) to (b1.center);
			\end{feynhand}
		\end{tikzpicture}
		~
	\underbrace{
		\begin{tikzpicture}[baseline=-2pt]
			\begin{feynhand}
				\vertex (o)  {};
				\vertex [dot] (a1) [above = of o.center] {};
				\vertex [dot] (a2) [below = of o.center] {};
			\end{feynhand}
		\end{tikzpicture}
\cdots
		\begin{tikzpicture}[baseline=-2pt]
			\begin{feynhand}
				\vertex (o)  {};
				\vertex [dot] (a1) [above = of o.center] {};
				\vertex [dot] (a2) [below = of o.center] {};
			\end{feynhand}
		\end{tikzpicture}
		~
		\begin{tikzpicture}[baseline=-2pt]
			\begin{feynhand}
				\vertex (o)  {};
				\vertex [dot] (a1) [above = of o.center] {};
				\vertex [dot] (a2) [below = of o.center] {};
			\end{feynhand}
		\end{tikzpicture}
	}_{n-4}.
	\label{eq:On3} 
\end{align}
	Here the lines represent Lorentz contractions of two derivatives $\partial_\mu$ acting on the corresponding fields. 
	As it was mentioned earlier, we do not include the total-derivative operator $\partial^2 (\phi^2)^{n-2}$ in this list.  

 The application of $\Z$ to the relevant $2l$-loop divergent 1PI Green function with operator insertions can naturally be written 
 as 
\begin{align}
	\Z\Gamma^{(2l)}_{2n}[O_n^{(n)}] &= \tilde Z^{(2l)}_{n,n} \cdot O^{(n)}_{n}  + \mathcal{O}(n^{2l-1}),  \\
	\Z\Gamma^{(2l)}_{2(n-2)}[O_n^{(n)}] & = \sum_i \tilde Z^{(2l)}_{n,n-2,i} \cdot O^{(n)}_{n-2,i} + \mathcal{O}(n^{2l-1}),
	\label{eq:Z_ops_offshell}
\end{align}
where 
the renormalization constants $\tilde Z^{(2l)}_{n,n}$ and $\tilde Z^{(2l)}_{n,n-2,i}$ contain poles in $\varepsilon$ and are linear combinations of $n^{2l+1}$ (leading) and $n^{2l}$ (subleading) terms.

While the set $\{O^{(n)}_{n}, O^{(n)}_{n-2,i}\}$ is directly related to diagram computations, a more convenient basis of operators consists of conformal primaries (``physical'' operators) and the so-called EOM-operators, which are redundant due to the equations of motion (EOM). 
The former are annihilated by the generator of special conformal transformation $K_\mu$, while the latter can be dropped from the calculation (see, e.g., \cite{Vasiliev:2006} and the recent \cite{Cao:2021cdt} for a more elaborate treatment).

To construct the (necessary part) of the physical basis, we notice that $O^{(n)}_{n}$ is itself a conformal primary. To find a primary among the  $O^{(n)}_{n-2,i}$ operators, we first use the integration by parts\footnote{This can also be checked from the requirement $\partial^2 [(\phi^2)^{n-2}] = 0$ or by considering the corresponding Feynman rules (see~\ref{app:On2_primary}) and assuming $Q=q_1 + \ldots + q_{2n-4} = 0$.}
\begin{align}
	O^{(n)}_{n-2,1} + O^{(n)}_{n-2,2} + 2 (n-3) O^{(n)}_{n-2,3} = 0
	\label{eq:On1On2On3_IBP}
\end{align}
reducing the number of independent operators from three to two. Then we construct a conformal primary combination $\bar O^{(n)}_{n-2}$ of operators $O^{(n)}_{n-2,k}$ annihilated by the generator $K_\mu$. Following \cite{RoosmaleNepveu:2024zlz}, we consider the ansatz 
\begin{align}
	\bar O^{(n)}_{n-2}  = a O^{(n)}_{n-2,2} + b O^{(n)}_{n-2,3}
	\label{eq:conformal_primare_anzats}
\end{align}
and obtain a constraint on the coefficients $a$ and $b$ from the requirement  (see~\ref{app:On2_primary} for more details)
\begin{align}
K_\mu \bar O^{(n)}_{n-2} = 0 \Rightarrow a + b = 0.
\label{eq:conformal_constraint}
\end{align}
As a consequence, we confirm \cite{RoosmaleNepveu:2024zlz} that 
\begin{align}
	\bar O^{(n)}_{n-2}  = b (O^{(n)}_{n-2,3} - O^{(n)}_{n-2,2}) = b (\phi^2)^{n-4} \left[ (\phi \partial_\mu \phi)^2 - \phi^2 (\partial_\mu \phi \partial_\mu \phi)\right], 
	\label{eq:conformal_primary}
\end{align}
where $b$ is an arbitrary constant which, as we will see, is irrelevant for our computations, so we set $b = 1$. Given \eqref{eq:On1On2On3_IBP} and \eqref{eq:conformal_primary}, one can express any operator $O^{(n)}_{n-2,k}$ in terms of the 
primary $\bar O^{(n)}_{n-2}$ and $O^{(n)}_{n-2,1}$: 
\begin{align}
O^{(n)}_{n-2,2} & =  -\frac{1}{2n-5} \left[2 (n-3) \bar O^{(n)}_{n-2} + O^{(n)}_{n-2,1}\right] \nonumber\\
		& = -\bar O^{(n)}_{n-2} + \frac{\bar O^{(n)}_{n-2} - O^{(n)}_{n-2,1}}{2 n} + \mathcal{O}(n^{-2}),  \\    
O^{(n)}_{n-2,3} & =  \frac{1}{2n-5} \left[\bar O^{(n)}_{n-2} - O^{(n)}_{n-2,1}\right] = \frac{\bar O^{(n)}_{n-2} - O^{(n)}_{n-2,1}}{2n} + \mathcal{O}(n^{-2}),
\end{align}
where we neglect the subsubleading-$n$ terms. The operator $O^{(n)}_{n-2,1}$ is redundant due to EOMs (or can be removed by field redefinitions), i.e.,
\begin{align}
	O^{(n)}_{n-2,1} = (\phi^2)^{n-3} (\phi \partial^2 \phi) & = (\phi^2)^{n-3} \phi^a  \underbrace{\left[ \partial^2 \phi^a - \lambda^2 \phi^a (\phi^2)^2\right]}_{\text{EOM}} + \lambda^2 (\phi^2)^n 
\end{align}
giving rise to an additional contribution to the anomalous dimension of the $O^{(n)}_{n}$ operator. Equivalently, we can define the class of EOM operators $\mathcal{E}^{(n)}$
\begin{align}
	\mathcal{E}^{(n)} & = O^{(n)}_{n-2,1} - \lambda^2 O^{(n)}_{n} \quad \Rightarrow \quad \mathcal{E}^{(n)} |_{n-2} = O^{(n)}_{n-2,1}, \quad \mathcal{E}^{(n)} |_{n} = -\lambda^2 O^{(n)}_{n},  
\end{align}
so that the tree-level insertion of $\mathcal{E}^{(n)}$ into the $n$ point function denoted as $\mathcal{E}^{(n)}|_n$ is equivalent to the insertion of $-\lambda^2 O^{(n)}_{n}$, while the insertion of $\mathcal{E}^{(n)}|_{n-2}$ is equivalent to $O^{(n)}_{n-2,1}$. 

Explicit computations (see below) show that up to six loops in $\phi^6$ theory  it is sufficient to consider only these operators to obtain the leading and subleading-$n$ corrections to $\gamma_{2n}$. Moreover, following the reasoning of Ref.~\cite{Bern:2019wie,Cao:2021cdt}, one can convince oneself that we can ignore the mixing with $\bar O^{(n)}_{n-2}$. 
At higher loop orders, one needs to take into account operators with a higher number of derivatives along with nontrivial mixing with conformal primaries.

Rewriting RHS of \eqref{eq:Z_ops_offshell} in terms of $O^{(n)}_n$, $\bar O^{(n)}_{n-2}$, and the EOM-operators $\mathcal{E}^{(n)}$, we obtain 
\begin{align}
	\Z\Gamma^{(2l)}_{2n}[O_n^{(n)}] &= 
	Z^{(2l)}_{n,n\phantom{-2}} \cdot O^{(n)}_{n} + Z^{(2l)}_{n,\mathcal{E}} \cdot \mathcal{E}^{(n)}|_{n\phantom{-2}}+ \mathcal{O}(n^{2l-1}),  \\
	\Z\Gamma^{(2l)}_{2n-4}[O_n^{(n)}] & = 
	Z^{(2l)}_{n,n-2} \cdot \bar O^{(n)}_{n-2} + Z^{(2l)}_{n,\mathcal{E}} \cdot \mathcal{E}^{(n)}|_{n-2}+ \mathcal{O}(n^{2l-1}).
	\label{eq:Z_ops_phys}
\end{align}
By comparing \eqref{eq:Z_ops_offshell} and \eqref{eq:Z_ops_phys}
\begin{align}
	Z^{(2l)}_{n,n-2} \bar O^{(n)}_{n-2} & + Z^{(2l)}_{n,\mathcal{E}} \mathcal{E}^{(n)}|_{n-2} = \sum_i \tilde Z^{(2l)}_{n,n-2,i} \cdot O^{(n)}_{n-2,i} 
	\\ &=
	\tilde Z^{(2l)}_{n,n-2,1} \mathcal{E}^{(n)}|_{n-2}  
	+
	\frac{\tilde Z^{(2l)}_{n,n-2,2}}{2n-5} \left[-(2 n - 6) \bar O^{(n)}_{n-2} - \mathcal{E}^{(n)}|_{n-2} \right] \nonumber \\
	& + 
	\frac{\tilde Z^{(2l)}_{n,n-2,3}}{2n -5} \left[ \bar O^{(n)}_{n-2} - \mathcal{E}^{(n)}|_{n-2} \right] \nonumber \\
	& = \frac{\bar O^{(n)}_{n-2}}{2n-5} \left[ - (2n-6) \tilde Z^{(2l)}_{n,n-2,2}  + \tilde Z^{(2l)}_{n,n-2,3}\right]  \nonumber \\
	& + \frac{\mathcal{E}^{(n)}|_{n-2}}{2n-5} 
	\left[ (2n-5) \tilde  Z^{(2l)}_{n,n-2,1} - \tilde Z^{(2l)}_{n,n-2,2} -  \tilde Z^{(2l)}_{n,n-2,3}
	\right],
\end{align}
 we find the relations between the renormalization constants in the ``physical'' $(Z_{n,k})$ and the initial bases $(\tilde Z_{n,k})$:
\begin{align}
	Z^{(2l)}_{n,n-2} & = \frac{1}{2n-5} \left[ -(2n-6) \tilde Z^{(2l)}_{n,n-2,2}  + \tilde Z^{(2l)}_{n,n-2,3}\right] \nonumber \\
			 & =  - \tilde Z^{(2l)}_{n,n-2,2} + \frac{\tilde Z^{(2l)}_{n,n-2,2} + \tilde Z^{(2l)}_{n,n-2,3}}{2n} + \mathcal{O}(n^{2l-1}), \\
	Z^{(2l)}_{n,\mathcal{E}} & = \frac{1}{2n-5} \left[ (2n-5) \tilde Z^{(2l)}_{n,n-2,1}  - \tilde Z^{(2l)}_{n,n-2,2} - \tilde Z^{(2l)}_{n,n-2,3}\right] \nonumber \\
			   & =  \tilde Z^{(2l)}_{n,n-2,1} - \frac{\tilde Z^{(2l)}_{n,n-2,2} +  \tilde Z^{(2l)}_{n,n-2,3}}{2n} + \mathcal{O}(n^{2l-1}).
\end{align}
The ``physical'' renormalization constant $Z_{n,n}$ of the conformal primary $O^{(n)}_n$ can be found from 
\begin{align}
	Z_{n,n} - \lambda^2 Z_{n,\mathcal{E}} = \tilde Z_{n,n},
\end{align}
so the $2l$-loop contribution in the leading-$n$ and subleading-$n$ orders is given by 
\begin{align}
	Z^{(2l)}_{n,n} = \tilde Z^{(2l)}_{n,n} + \lambda^2 \left[ \tilde Z^{(2l)}_{n,n-2,1} - \frac{\tilde Z^{(2l)}_{n,n-2,2} +  \tilde Z^{(2l)}_{n,n-2,3}}{2n} \right] + \mathcal{O}(n^{2l-1}).
	\label{eq:Znn_phys}
\end{align}
This equation turns out to be sufficient to extract relevant corrections to $\gamma_{2n}$ up to six loops.

\section{Computational technique\label{sec:comp_tech}} 

To account for the fact that we can choose any $k$ ``active'' lines (in our two-loop example in Fig.~\ref{fig:dias_examples}a $k=3$) out of $2n$ operator legs, we use the following trick. We introduce a set of auxiliary operators $\tilde O^{(n)}_{k}$ involving dummy fields $\hat \phi$ with $k$ external lines that have the Feynman rules
\begin{align}
	\tilde O^{(n)}_{k} \to \frac{1}{(\hat \phi^2)^n} \left[\prod\limits_{i=1}^{k} \frac{\partial}{\partial \hat \phi^{a_i}}\right]  \left(\hat\phi^2 \right)^n.
	\label{eq:Onk_feynman_rule}
\end{align}
For example,
$\tilde O^{(n)}_{3}$ corresponds to
\begin{align}
	\frac{(2n-4) (2n-2) (2n)}{(\hat\phi^2)^3} \hat \phi^{a} \hat\phi^b \hat \phi^c
	+ \frac{(2n-2)(2n)}{(\hat\phi^2)^2} \left[ \hat\phi^a \delta^{bc} + \hat\phi^b \delta^{ac} + \hat\phi^c \delta^{ab}\right]. 
	\label{eq:Onk3_feynman_rule}
\end{align}
Here the first term gives rise to the leading-$n$ contribution and accounts for the choice, when all three ``active'' legs with $O(N)$ indices $a$, $b$ and $c$ belong to different contracted pairs in the $(\phi^2)^n$ operator\footnote{Equivalently, the belong to different Kronecker deltas in the Feynman rule \eqref{eq:feynman_rule_On} and, when inserted into  the $(2n)$-point 1PI function, are contracted with some of the ``spectator'' legs (represented by $\hat \phi$).}. The subsequent terms in \eqref{eq:Onk3_feynman_rule} contribute at the subleading-$n$ order and correspond to the situation when two ``active'' legs, e.g., $b$ and $c$ entering into $\delta^{bc}$ belong to a single contracted pair $\phi^2$ in the initial operator, while the third leg with index $a$ is contracted with a spectator one. By considering these auxiliary operators, we can ignore the ``spectator'' legs, e.g., replace the diagram shown in Fig.~\ref{fig:dias_examples}a by an equivalent diagram corresponding to $\Gamma_{3}[\tilde O^{(n)}_3]$, which we for brevity denote as  \auxdia{3}{3}.

Obviously, for two-, four-, and six-loop contributions to $\gamma_{2n}$ we need to consider the logarithmically divergent diagrams corresponding to \auxdia{3}{3},  \auxdia{5}{5} and \auxdia{4}{4},  \auxdia{7}{7} and \auxdia{6}{6} auxiliary graphs, respectively \cite{Jack:2020wvs}.

To generate relevant auxiliary diagrams, we use independently the recent version 4.0 of \texttt{qgraf} \cite{Nogueira:1991ex} and the python library \texttt{GraphState} \cite{Batkovich:2014bla}. 
We encode the corresponding expressions as products of three factors, which schematically can be represented as 
$$
	[ \text{ symmetry factor } ]  \times [ \text{ } O(N) ~ \text{factor } ] \times [ \text{ loop integral } ].
$$
Here the $O(N)$ factor originates from the Feynman rules, and the loop integral is represented by a graph corresponding to the product of propagator denominators. 
It is worth mentioning that we do not explicitly consider loop integration  but use pre-computed tables for the results of $\KRP$ (see~\ref{app:KRP_tables}), which 
directly associate a counterterm with the corresponding graph given in the Nickel notation\footnote{For example,   {\nickel{e\,$\cdots$e111|eee|}} and {\nickel{eee111||}} for $\Gamma^{(2)}_{2n}[O^{(n)}_n]$ and auxiliary $\Gamma^{(2)}_3[\tilde O^{(n)}_{3}]$ in Fig.~\ref{fig:dias_examples}a, respectively.}. 



To deal with $O(N)$ indices, we use \texttt{FORM}~\cite{Vermaseren:2000nd,Tentyukov:2007mu,Kuipers:2012rf,Ruijl:2017dtg}.  While the initial expression for $\Gamma_k[\tilde O^{(n)}_k]$ is a tensor involving $k$ open $O(N)$ indices $a_{1},\ldots,a_{k}$, 
we contract the latter with $\hat \phi^{a_1}\ldots \hat \phi^{a_k}/k!$ 
to account for permutations of $k$ external legs and 
cancel the powers of $(\hat\phi^2)$ appearing in the denominators due the Feynman rule \eqref{eq:Onk_feynman_rule}.

As it was mentioned earlier, we should also compute divergent $\Gamma_{2n-4}[O^{(n)}_n]$. For this purpose,  we consider the single \auxdia{5}{1} auxiliary graph at four loops (corresponding to Fig.\ref{fig:dias_examples}b), and the six-loop auxiliary diagrams\footnote{Of course, \auxdia{5}{1} also contributes at six loops but at the subsubleading order, so we ignore it.} \auxdia{7}{3} 
and \auxdia{6}{2} (examples are given in Fig.~\ref{fig:dias_examples}c-d). 
Obviously, we \emph{are not allowed} to nullify the momentum $Q$ flowing into the auxiliary operator\footnote{For example, in this case, we completely miss the contribution \auxdia{5}{1} in Fig.~\ref{fig:dias_examples}b.} because it is related to the total momentum of the spectator legs. We summarize the types of the auxiliary graphs considered in this paper in Table~\ref{tab:aux_dias_types}. 

Contrary to the case of logarithmically divergent diagrams, for which we can use the infra-red rearrangement trick \cite{Vladimirov:1979zm} to compute $\KRP$ (i.e., nullify all but 2 external momenta), for quadratically divergent diagrams the corresponding $\KRP$ operation depends on the external momenta of the graph. As a consequence, we should keep this dependence to account for the mixing with the $O^{(n)}_{n-2,i}$  operators involving $2(n-2)$ fields (see~\ref{app:KRp_quadratic} for our treatment of such diagrams).  
To distinguish contributions associated with the operators $O^{(n)}_{n-2,i}$, we first contract $\Gamma_{k}[\tilde O^{(n)}_{k+4}]$ 
with the product of \emph{different} auxiliary fields $\hat \phi_{1}^{a_1}\ldots\hat \phi_{k}^{a_{k}}/k!$, then make the identification
\begin{align}
	(q^2_i) & \to  - (\hat \phi^2)^2 O^{(n)}_{n-2,1},  & 
	\\
	(q_i q_j) (\hat \phi_i \cdot \hat \phi_j) & \to - (\hat \phi^2)^3 O^{(n)}_{n-2,2}, \\
	(q_i q_j) (\hat \phi_i \cdot \phi) (\hat \phi_j \cdot \phi)& \to - (\hat \phi^2)^4 O^{(n)}_{n-2,3}, & \phi & = \{ \hat \phi, \hat \phi_k \}, \quad k \neq i \neq j,
	\label{eq:quadratic_ops_id}
\end{align}
and finally replace all the remaining $\hat \phi_i$ by $\hat \phi$ to remove completely the dependence on dummy fields $\hat \phi$.  
In the above expressions the momentum $q_i$ and the field $\hat \phi_i$ are associated with the $i$-th external non-operator leg.   
\begin{table}[t]
	\begin{center}
	\begin{tabular}{c|c|c|c}
		& 2 loops & 4 loops & 6 loops \\
		\hline
		leading-$n$ & \auxdia{3}{3} & \auxdia{5}{5}, \textcolor{red}{\auxdia{5}{1}}\phantom{, \auxdia{4}{4}} & \auxdia{7}{7}, \textcolor{red}{\auxdia{7}{3}}\phantom{, \auxdia{6}{6}, \auxdia{6}{2}} \\ 
		subleading-$n$ & \auxdia{3}{3} & \auxdia{5}{5}, \textcolor{red}{\auxdia{5}{1}}, \auxdia{4}{4} & \auxdia{7}{7}, \textcolor{red}{\auxdia{7}{3}}, \auxdia{6}{6}, \textcolor{red}{\auxdia{6}{2}} 	
	\end{tabular}
	\caption{Types of auxiliary diagrams needed to compute $\gamma_{2n}$ at the leading and subleading order in $n$ in perturbative theory up to six loop. The notation \auxdia{k}{m} corresponds 
		to the insertion of the operator $\tilde O^{(n)}_{k}$ into the $m$-point 1PI Green function $\Gamma_m[\tilde O^{(n)}_{k}]$. In the case of quadratically divergent diagrams (marked in red), we have to keep the momentum $Q$ flowing into the auxiliary operator vertex nonzero (e.g., for diagrams in Fig.~\ref{fig:dias_examples}b-d).  
	}
	\label{tab:aux_dias_types}
\end{center}
\end{table}

\section{Results\label{sec:results}}
Let us summarize the results of our computations. In the initial operator basis, the coefficients of the first pole in $\ep$ of $\tilde Z^{(2l)}_{n,n}$ are given by
\begin{align}
	2 \cdot 2 \cdot \tilde Z^{(2,1)}_{n,n}  & =  \phantom{-}\frac{\lambda^2}{64\pi^2} \Bigl[\frac{40}{3} \Q^3 -40 \Q^2 + 8 [N\!-\!1] \Q^2 \Bigr] 	
	+ \mathcal{O}(\Q),
	\\
	2 \cdot 4 \cdot \tilde Z^{(4,1)}_{n,n} & =  -\frac{\lambda^4}{(64\pi^2)^2} \Bigl[ 1400 \Q^5 - 900(12 -\pi ^2) \Q^4 \nonumber
	\\
					       & \hspace{3.7cm} + 4 \left(380+9 \pi ^2\right) [N\!-\!1] \Q^4 \Bigr] 
	+ \mathcal{O}(\Q^3), \\
	2 \cdot 6 \cdot  \tilde Z^{(6,1)}_{n,n}&  =  \phantom{-}\frac{\lambda^6}{(64\pi^2)^3} \Bigl[ 
    \frac{2516800}{9}\Q^7 -\frac{400}{9} \left(72812-2490 \pi ^2-405 \pi^4\right) \Q^6 \nonumber \\
	& \hspace{1.05cm} + \frac{16}{9} \left(246460+4794 \pi ^2+81 \pi ^4\right)[N\!-\!1] \Q^6 \Bigr]
	+ \mathcal{O}(\Q^5),
\end{align}
with \auxdia{3}{3} , \auxdia{4}{4} and \auxdia{5}{5}, \auxdia{7}{7} and \auxdia{6}{6} contributing at two loops, at four loops, and at six loops, respectively. Necessary contributions to the nondiagonal elements 
$\tilde Z^{(2l)}_{n,n-2,i}$ can be found from the corresponding divergent auxiliary diagrams.  The mixing of $O^{(n)}_{n}$ into $O^{(n)}_{n-2,1}$ becomes nonzero at the four-loop level (\auxdia{5}{1})
\begin{align}
	2 \cdot 4 \cdot \tilde Z^{(4,1)}_{n,n-2,1} & = \phantom{+}\frac{\lambda^2}{(64\pi^2)^2} \Bigl[\frac{8}{3} \Q^5 - \frac{80}{3} \Q^4 + \frac{16}{3} [N\!-\!1] \Q^4 \Bigr]  + \mathcal{O}(\Q^3),\\
	2 \cdot 6 \cdot \tilde Z^{(6,1)}_{n,n-2,1} & =  -\frac{\lambda^4}{(64\pi^2)^3} \Bigl[\frac{320}{9} \Q^7 +\frac{64}{9}\left(245 + 23 [N\!-\!1]\right) \Q^6 \Bigr] + \mathcal{O}(\Q^5),
\end{align}
with \auxdia{7}{3} and \auxdia{6}{2} contributing at six loops. The mixing of $O^{(n)}_{n}$ into $O^{(n)}_{n-2,2-3}$ starts at six loops\footnote{Actually, according to \eqref{eq:Znn_phys}, we need only leading-$n$ terms in $\tilde Z^{(2l)}_{n,n-2,2-3}$.} 
\begin{align} 
	2 \cdot 6 \cdot \tilde Z^{(6,1)}_{n,n-2,2} & = \phantom{+}\frac{\lambda^4}{(64\pi^2)^3 }\left[\frac{256}{3} \Q^7 - 2304 \Q^6 + 256 [N\!-\!1] \Q^6 \right]  + \mathcal{O}(\Q^5),\\
	2 \cdot 6 \cdot \tilde Z^{(6,1)}_{n,n-2,3} & = -\frac{\lambda^4}{(64\pi^2)^3 } \left[512 \Q^7 - \frac{27392}{3} \Q^6 + 1280 [N\!-\!1] \Q^6 
	\right]\!\! + \mathcal{O}(\Q^5), \!\!
\end{align}
and receives contributions from \auxdia{7}{3} and \auxdia{6}{2}.
Combining these expressions by means of \eqref{eq:Znn_phys}, we compute the required terms in the renormalization constant for $O^{(n)}_n$ for the physical basis: 
\begin{align}
	2 \cdot 2 \cdot Z^{(2,1)}_{n,n}  & =  \phantom{-}\frac{\lambda^2}{64\pi^2} \Bigl[\frac{40}{3} \Q^3 - 40 \Q^2 + 8 [N\!-\!1] \Q^2 \Bigr] 	
	+ \mathcal{O}(\Q),
	\\
	2 \cdot 4 \cdot Z^{(4,1)}_{n,n} & =  - \frac{\lambda^4}{(64\pi^2)^2} \Bigl[ \frac{4192}{3} \Q^5
    - \left(\frac{32320}{3} -900 \pi ^2\right) \Q^4
\nonumber\\
					& \hspace{2.15cm}
    + \left(\frac{4544}{3}+36 \pi ^2\right) [N\!-\!1] \Q^4
	\Bigr]
	+ \mathcal{O}(\Q^3), \\
	2 \cdot 6 \cdot Z^{(6,1)}_{n,n}&  = \phantom{-}  \frac{\lambda^6}{(64\pi^2)^3} \Bigl[ \frac{2516480 }{9} \Q^7 
	- \frac{80}{9} \left(364208-12450 \pi ^2-2025 \pi ^4\right) \Q^6 
		\nonumber \\
	& \hspace{1.0cm} 
		+ \frac{16}{9} \left(246368+4794 \pi ^2+81 \pi ^4\right)[N\!-\!1] \Q^6 
\Bigr]
	+ \mathcal{O}(\Q^5).
\end{align}
This allows us to find the leading $C_{2l,0}$ coefficients in \eqref{eq:gamma_n_PT} for $l=1,2,3$:
\begin{align}
	C_{2,0} & = \frac{5}{24 \pi^2}, &&  
		&C_{4,0} & = -\frac{131}{384 \pi^4}, &&  
		&C_{6,0} & = \frac{4915}{4608 \pi^6}, &&  
		\label{eq:C_leading}
\end{align}
which perfectly match all-order\footnote{We thank the authors of Ref.~\cite{Antipin:2025ekk} for sharing with us their prediction for $C_{6,0}$.} predictions given in Ref.~\cite{Antipin:2025ekk}. In addition, we get the new result for the subleading $C_{2l,1}$ up to six loops:
\begin{align}
	C_{2,1} & = -\frac{5}{8 \pi^2} + \frac{[N\!-\!1]}{8 \pi^2}, \\
	C_{4,1} & = \frac{5 \left(1616 - 135 \pi ^2\right)}{3072 \pi ^4} - \frac{\left(1136+27 \pi ^2\right)}{3072 \pi ^4}[N\!-\!1],\\
C_{6,1} & = -\frac{5 \left(364208 - 12450 \pi ^2 - 2025 \pi^4\right)}{147456 \pi ^6} \nonumber \\
        & \hspace{0.49cm} + \frac{\left(246368+4794 \pi ^2+81 \pi ^4\right)}{147456 \pi ^6} [N\!-\!1].
		\label{eq:C_subleading}
\end{align}

It can be shown \cite{Vasiliev:2006} that the anomalous dimensions of operators for small values of $n\leq 6$ are related to the anomalous dimensions of parameters and the beta function of the $\phi^4+\phi^6$ theory\footnote{relations: $\gamma_2 = -\gamma_\tau = \gamma_{m^2}$, $\gamma_4 = -\gamma_\lambda = \gamma_u$, $\gamma_6 = 4\ep + \partial_u \beta(u) = 2\epsilon + \partial_{\bar{w}_R} \beta(\bar{w}_R)$, $\epsilon = 2\ep$. The first equality in the notation of Ref.~\cite{Vasiliev:2006}; the second -- \cite{Hager:1999, Hager:2002}, where the necessary four-loop results are presented.}. Using the computed leading and subleading coefficients together with the four-loop results for $\gamma_2, \gamma_4$ and $\gamma_6$ known from the literature
\begin{align}
	\gamma_2^{(2)} & + \gamma_2^{(4)} = \Bigl(\frac{5}{4 \pi ^4} + \frac{(N+7)}{12 \pi ^4} [N\!-\!1] \Bigr) \lambda^4, \hspace{2cm} \gamma_2^{(2)} = 0,\\
	\gamma_4^{(2)} & + \gamma_4^{(4)} = 
	\frac{5}{\pi ^2} \lambda^2 
	- \frac{5 \left(1736+135 \pi ^2\right)}{128 \pi ^4} \lambda^4 \nonumber \\
		       &+ \Bigl( \frac{\lambda^2}{\pi^2} - \frac{8 (85 N+991)+3 \pi ^2 (N^2 + 23N + 211)}{384 \pi ^4} \lambda^4\Bigr) [N\!-\!1] , \\
	\gamma_6^{(2)} & + \gamma_6^{(4)} = 
\frac{25}{\pi ^2} \lambda^2 - \frac{45 \left(2248+225 \pi ^2\right)}{128 \pi ^4} \lambda^4 \nonumber \\
   & + \Bigl( \frac{3\lambda^2}{\pi^2} - \frac{ 24 (53 N+911) + 3\pi ^2 (N^2+35 N+655)}{128 \pi ^4} \lambda^4 \Bigr) [N\!-\!1]. 
   \label{eq:gamma_2_4_6}
\end{align}
allows us to reconstruct the full dependence on $n$ of $\gamma_{2n}$ \eqref{eq:gamma_n_PT} up to four loops:
\begin{align}
	C_{2,2} &= \frac{5}{12 \pi ^2} - \frac{[N\!-\!1]}{4 \pi ^2}, \label{eq:C22}  \\
    C_{4,2} &= - \frac{5 \left(2948-405 \pi ^2\right)}{1536 \pi ^4} - \frac{28 (5 N\!-\!121) + 9 \pi^2 (N+19)}{1536 \pi ^4} [N\!-\!1], \label{eq:C42}\\
	    C_{4,3} &= \frac{25 \left(1976-297 \pi ^2\right)}{3072 \pi ^4} + \frac{8 (117 N-2153) -3 \pi ^2 (N^2-13 N-725)}{3072 \pi ^4} [N\!-\!1], \label{eq:C43}\\
	    C_{4,4} &= - \frac{4472-675 \pi^2}{512 \pi ^4} -\frac{ 8 (13 N-361) -\pi ^2 (N^2-N-461)}{512 \pi ^4} [N\!-\!1]. \label{eq:C44}
\end{align}
At the fixed point
\begin{align}
	\frac{\lambda^{*2}}{8\pi^2} & = \frac{\ep}{3 N+22} + \frac{15 \left(2248+225 \pi ^2\right)}{8 (3 N+22)^3} \ep^2 \nonumber \\
    &+ \frac{8 (53 N+911)+\pi ^2 (N^2+35 N+655)}{8 (3 N+22)^3} [N-1] \ep^2 + O(\ep^3)
\end{align}
we find the coefficients of \eqref{eq:phi_n_fp} contributing to $\Delta_{2n}$ at two loops:
\begin{align}
	D_{2,0} &= \frac{5}{6}, && 
		& D_{2,1} &= -\frac{5}{2} + \frac{[N\!-\!1]}{2}, &&
		& D_{2,2} &= \frac{5}{3}  - [N\!-\!1]  
\end{align}
and four loops:
\begin{align}
	D_{4,0} =& -\frac{131}{24} \Bigl( 3N + 22 \Bigr), \\
	D_{4,1} =&\hspace{0.51cm} \frac{1}{192} \Bigl( 3 N+22 \Bigr) \Bigl\{ 5 \left(1616-135 \pi ^2\right) - \left(1136+27 \pi ^2\right) [N\!-\!1] \Bigr\}, \\
	D_{4,2} =& -  \frac{25}{24} \left(1999 - 675 \pi ^2\right) -\frac{1}{8}  \left(35 N^2-767 N-5563\right) [N\!-\!1] \nonumber \\
	    & \hspace{3.85cm} -\frac{\pi^2}{48}   \left(11 N^2+268 N-2794\right) [N\!-\!1], \\
    D_{4,3} =& \frac{25}{192} \left(8936 - 11475 \pi ^2\right) 
    + 
    \frac{1}{24} \left(669 N^2-327 N-36347\right) [N\!-\!1] \nonumber\\
	      & \hspace{3.08cm}- \frac{\pi^2}{64} \left(N^3-75 N^2-3351 N-7415\right) [N\!-\!1],  \\
   D_{4,4} =& \frac{75}{32} \left(8+375 \pi ^2\right) 
   -\frac{1}{12} \left(435 N^2+2227 N-8081\right) [N\!-\!1] \nonumber \\
	     & \hspace{2.71cm} +\frac{\pi^2}{96} \left(3 N^3-137 N^2-7585 N-34121\right) [N\!-\!1].
\end{align}
The two-loop result is
\begin{equation}
    \gamma^{*(2)}_{2n} = \Q P^*_2(\Q) (2\ep) = \frac{\Q (\Q-2) (5\Q + 3N - 8)}{6 (3N+22)} (2\ep)
\end{equation}
and agrees with the Ref. \cite{Basu:2015gpa}. The four-loop expression\footnote{Full four-loop result for the $O(N)$ case and the computed $C_{ij}$ and $D_{ij}$ can be found in supplementary material.} for $N = 1$ is

\begin{align}\label{eq:phi_n_4l}
    \gamma^{*(2)}_{2n} + \gamma^{*(4)}_{2n} &= \frac{\Q (\Q-1) (\Q-2)}{30} (2\ep) - \frac{9 \pi^2}{1600} \Q (\Q-5) (\Q-2) (\Q-1)(2\ep)^2\nonumber \\
					    & - \frac{\Q}{15000} \left(131 \Q^4-1010 \Q^3+1999 \Q^2-1117 \Q-18\right) (2\ep)^2.
\end{align}
and matches the result given in Refs. \cite{O_Dwyer_2008, Henriksson:2025kws}. At the same time, it can be verified that \eqref{eq:phi_n_4l} gives the correct four-loop result for $\gamma^*_{1} = \frac{\eta}{2}$ (see Refs. \cite{LewisAdams:1978, Hager:1999}).

Only the $C_{6,0}$ and $C_{6,1}$ coefficients contribute to the six-loop leading and subleading terms in \eqref{eq:phi_n_fp}; thus, $D_{6,0}$ and $D_{6,1}$ can be found:
\begin{align}
    D_{6,0} &= \frac{4915}{72} (3N+22)^2, \\
    D_{6, 1} &= -\frac{1}{2304} (3 N+22)^2 \Bigl\{ 5 \left(364208-12450 \pi ^2-2025 \pi ^4\right) \nonumber \\
	     & \hspace{3.82cm} - \left(246368+4794 \pi ^2+81 \pi ^4\right) [N\!-\!1] \Bigr\}.
\end{align}
\section{Conclusions\label{sec:conclusions}}

We have considered a restricted set of Feynman diagrams contributing to the leading-$n$ and subleading-$n$ corrections to the anomalous dimension $\gamma_{2n}$ of the $(\phi^2)^n$ operator in $d=3-2\ep$ dimensions.
To deal with operator insertions at generic $n$, we split the $2n$ operator legs into  ``spectator'' and ``active'' ones. The former are directly connected to external lines, while the latter correspond to loop propagators and can give rise to UV divergence. 
At $2l$ loops, one can only have at most $2l+1$ ``active'' legs, giving rise to at least $2n-2l-1$ spectators for a sufficiently large $n$. 
The leading-$n$ and subleading-$n$ terms in $2l$ loop anomalous dimensions correspond precisely to diagrams with $2l+1$ and $2l$ active legs, respectively. 
This fact allows us to introduce a set of auxiliary operators \eqref{eq:Onk_feynman_rule} involving only the required number of (active) legs and account for a combinatorial factor related to spectators in the corresponding Feynman rule. 

We took into account the nontrivial mixing of neutral operators, which is absent in the fixed-charge case, and obtained the expressions for the leading $C_{2l,0}$ and subleading $C_{2l,1}$ coefficients at four ($l=2$) and six ($l=3$) loops. The leading coefficients are in full agreement with the ones derived from the all-loop result in Ref.~\cite{Antipin:2025ekk}.
The four- and six-loop subleading-$n$ terms are our new finding. In subsequent studies, it can be compared to subleading semiclassical corrections to $\Delta_{2n}$ \cite{Antipin:2025rsr} . 
The latter can be obtained along the lines of Ref.~\cite{Antipin:2025ilv}. 

In addition, we utilized the four-loop results for $\gamma_{2}$, $\gamma_{4}$ and $\gamma_{6}$, which are known from the literature (see, Eqs.~\eqref{eq:gamma_2_4_6} and Refs.~\cite{Hager:2002,Hager:1999}), 
to find 
the full dependence on $n$ of the four-loop anomalous dimension $\gamma_{2n}$ in the $O(N)$ model for $d=3$. At the fixed point, it constitutes an important addition to the CFT data. 
We plan to use the same method for the $l(N)$ $\phi^4$-model in $d=4$ and cross-check the result of Ref.~\cite{Antipin:2025ilv} for the subleading-$n$ terms in the four-loop approximation. 
At this loop order in $\phi^4$, one has to take into account the operator with four derivatives corresponding to the auxiliary diagram \auxdia{5}{1} but considered in $d=4$. 

Let us also mention that one can try to extend the technique to account for further terms in the large-$n$ expansion: by considering \auxdia{3}{3} auxiliary diagrams at four loops along with the subsubleading terms in \auxdia{5}{5}, \auxdia{5}{1}, and \auxdia{4}{4},  we were able to reproduce the correct expression for $C_{4,2}$ \eqref{eq:C42}.

Finally, let us close by a speculation that while fixed-charged operators may find their high-energy physics application in processes with many charged particles, 
neutral operators considered here may potlntially be related to the multi-particle production of states with zero net charge (see, e.g,. Ref.~\cite{Demidov:2018czx} and references therein). 

\section*{Acknowledgements}
We thank O.~Antipin, J.~Bersini, L.~Bork, S.~Fedoruk, N.~Lebedev, and A.~Pikelner for fruitful discussions. We are grateful to the authors of Ref.~\cite{Antipin:2025rsr} for sharing with us the preliminary results of their paper, confirming our diagrammatic computation. M.V.K. gratefully acknowledge the support of Foundation for the Advancement of Theoretical Physics ``BASIS'' through Grant 25-1-2-48-1.

\appendix
\section{Feynman rules of $O^{(n)}_{n-2,i}$ operators and the conformal primary\label{app:On2_primary} }
In this appendix, we follow Refs.~\cite{Cao:2021cdt,RoosmaleNepveu:2024zlz} and derive the expression for the conformal primary $\bar O^{(n)}_{n-2}$. Given the Feynman rules of $O^{(n)}_{n-2,i}$ in momentum space
\begin{align}
	O^{(n)}_{n-2,1} & \rightarrow - (2(n\!-\!3))!! \left(\sum\limits_{m}  q^2_m\right)  \prod\limits_{i=1}^{n-2} \delta^{a_{2i-1} a_{2i}} \label{eq:On21_fr} \\ 
	O^{(n)}_{n-2,2} & \rightarrow - (2(n\!-\!3))!! \left[\sum\limits_{i=1}^{n-2} 2 (q_{2i-1} q_{2i})\right] \prod\limits_{i=1}^{n-2} \delta^{a_{2i-1} a_{2i}} \label{eq:On22_fr} 
	,\\
		O^{(n)}_{n-2,3} & \rightarrow - (2(n\!-\!4))!!  \left[\sum\limits_{m\neq m'} (q_{m} q_{m'}) - \sum\limits_{i=1}^{n-2} 2 (q_{2i-1} q_{2i})\right] \prod\limits_{i=1}^{n-2} \delta^{a_{2i-1} a_{2i}}, 
\label{eq:On23_fr}
\end{align}
	where 
	$q_m$ and $a_m$ for $m=1,\ldots,2(n-2)$ correspond to the momentum and the $O(N)$ index of the $m$-th external leg, respectively.
	For brevity in Eqs.~\eqref{eq:On21_fr}-\eqref{eq:On23_fr} we omit $(2(n-2)-1)!! = \frac{(2(n-2))!}{(2(n-2))!!}$ nontrivial permutations of the $O(N)$ indices and the corresponding momenta.
    
	We use the following representation of $K_\mu$ (see, e.g,. Ref.~\cite{RoosmaleNepveu:2024zlz}))

\begin{align}
	K_\mu = \sum\limits_{i=1}^{k} \left[ \frac{\partial}{\partial q^\mu_i} + 2 q^\nu_i \frac{\partial}{\partial q^\nu_i} \frac{\partial}{\partial q^\mu_i} 
		- q^i_\mu \frac{\partial}{\partial q^\nu_i}\frac{\partial}{\partial q_\nu^i}
	\right],
	\label{eq:Kmu_generator}
\end{align}
	and apply it to each operator in the ansatz \eqref{eq:conformal_primare_anzats}. Taking into account that both $O^{(n)}_{n-2,2}$ and $O^{(n)}_{n-2,3}$
	depend only on $q_i q_j$, $i\neq j$ but not on $q_i^2$, we need only the first term in $K_\mu$.  As a consequence, applying $K^\mu$ to $O^{(n)}_{n-2,2}$ \eqref{eq:On22_fr}, we get 
\begin{align*}
	 & - 2\cdot(2(n-3))!!  \sum\limits_{j=1}^{n-2} \left[
\frac{\partial}{\partial (q_{2j-1})_
\mu}  + 
\frac{\partial}{\partial (q_{2j})_\mu)}\right]  
 \left[\sum\limits_{i=1}^{n-2} (q_{2i-1} q_{2i})\right] \prod\limits_{i=1}^{n-2} \delta^{a_{2i-1} a_{2i}}  
\end{align*}
giving rise to
\begin{align}
	K^\mu O^{(n)}_{n-2,2} \to  & - 2\cdot(2(n-3))!!  \left[\sum\limits_{m=1}^{2(n-2)} q_{m}^{\mu}\right]
	\prod\limits_{i=1}^{n-2} \delta^{a_{2i-1} a_{2i}}  + \text{perms}.
	\label{eq:KmuOn22}
\end{align}
Similarly,  application of $K^\mu$ to  $O^{(n)}_{n-2,3}$ \eqref{eq:On23_fr} leads to
\begin{align*}
	& - 2\cdot(2(n-4))!!  \sum\limits_{k=1}^{2(n-2)} \frac{\partial}{\partial (q_k)_\mu} 
	\left[\frac{1}{2}\sum\limits_{m\neq m'} (q_{m} q_{m'}) - \sum\limits_{i=1}^{n-2} (q_{2i-1} q_{2i})\right] 
	\prod\limits_{i=1}^{n-2} \delta^{a_{2i-1} a_{2i}}  \\
	= & - 2\cdot(2(n-4))!!  \sum\limits_{k=1}^{2(n-2)} \left[\sum\limits_{m\neq k} q_m^\mu - q_k^\mu\right] 
	\prod\limits_{i=1}^{n-2} \delta^{a_{2i-1} a_{2i}} 
\end{align*}
resulting in 
\begin{align}
	K^\mu O^{(n)}_{n-2,3} \to & 
	- 2\cdot(2(n-4))!!  \sum\limits_{k=1}^{2(n-2)} \left[ 2(n-3) q_k^\mu\right] 
	\prod\limits_{i=1}^{n-2} \delta^{a_{2i-1} a_{2i}}  + \text{perms}.
	\label{eq:KmuOn23}
\end{align}
Summing \eqref{eq:KmuOn22} and \eqref{eq:KmuOn23} with arbitrary coefficients $a$ and $b$, we get
\begin{align}
	0 & = a K^\mu O^{(n)}_{n-2,2} + b K^\mu O^{(n)}_{n-2,3} =  \nonumber \\
	& - (a+b) \cdot2\cdot(2(n-3))!! \left[ \left(\sum\limits_{m=1}^{2(n-2)} q_{m}^{\mu}\right) \prod\limits_{i=1}^{n-2} \delta^{a_{2i-1} a_{2i}}  
	    + \text{perms}\right]
\end{align}
leading to the constraint $a=-b$ \eqref{eq:conformal_constraint}.

\section{Counterterms of graphs}

We have used G-functions \cite{GFunctions} to calculate the required diagrams. All the \auxdia{6}{6} diagrams together with \nickel{e11122|222|e|} (\auxdia{6}{2}) are generated during the six-loop renormalization group analysis in the $\phi^6$ theory. The analysis was performed in Refs. \cite{Hager:1999, Hager:2002}, but these papers do not provide individual counterterms for the diagrams. One can only find the counterterm of the \nickel{e11122|222|e|} diagram in Ref. \cite{Hager:1999}. We use this counterterm in Eq.~\eqref{eq:KR_6_2}. Such diagrams were calculated in another paper \cite{Adzhemyan:2026} (yet to be published). As part of the present study, we computed the counterterms for the \auxdia{7}{7} and \auxdia{7}{3} diagrams. Two logarithmic divergent diagrams \nickel{eeee12|22223|33|eee|} and \nickel{eee112|22233|e3|eee|} (\auxdia{7}{7}) were calculated using the $\mathcal{K}\mathcal{R}^*$ operation instead of the $\KRP$ operation. In what follows, we briefly describe how we calculated non-trivial \auxdia{7}{3} and \auxdia{6}{2} diagrams.

\subsection{$\KRP$ for quadratically-divergent graphs\label{app:KRp_quadratic}}
Here we briefly describe the method\footnote{We do not use here a general approach of Ref.~\cite{Herzog:2017bjx}, which is  based on the expansion on external momenta and can deal with arbitrary tensor integrals.} used to compute the superficial divergences of the quadratically divergent \emph{scalar} integrals corresponding to the graph $\Gamma_k$ involving $k$ external momenta $p_1$, \ldots, $p_k$ with $p_1 + \ldots + p_k = 0$ in the \MS~scheme. For such a diagram the application of the $\KRP$ operation is a quadratic polynomial in external (off-shell) momenta with coefficients being poles in $\ep$. 
Due to this, we can write the ansatz for the result as   
\begin{align}
	\KRP \cdot \Gamma_k(p_1, \ldots, p_k) = \sum\limits_{i<j} p_i p_j A^{ij}(\ep),  
\end{align}
since the scalar products $p_i p_j$ with $i\neq j$ form a basis of Lorentz invariants\footnote{Given the conservation of momenta, we have $p_i^2 = -\sum\limits_{j\neq i} (p_i p_j)$.} (see, e.g,. Ref.~\cite{Cao:2021cdt,RoosmaleNepveu:2024zlz}). This representation allows one to compute $A^{ij}$ by setting all but two external momenta to zero, $p_k = 0$ for $k\neq i \neq j$, together with  $p_i = - p_j = P$. With this trick, we obtain a propogator-type  diagram $\Gamma^{(ij)}_2$, in which the momentum $P$ flows in into the leg $i$ and flows out of the leg $j$. The superficial UV divergence is given by
\begin{align}
	\KRS \cdot \Gamma^{(ij)}_2(P) = -P^2 A^{ij}(\ep)  
\end{align}
with the incomplete $\mathcal{R}^*$ operation \cite{Chetyrkin:1982nn,Chetyrkin:1984xa} instead of $\mathcal{R}^\prime$ to account for possible IR divergences\footnote{In our calculation, we do not encounter IR divergences in this procedure.} that can be generated by setting some of the external momenta to zero. As a consequence, by considering all possible  flows of a single external momentum, one can find all the coefficients $A^{ij}$.

In our problem, we use this approach for nontrivial\footnote{That are not propagator-type, in which all external $\phi$ legs are connected to a single vertex.} \auxdia{7}{3} (Fig.~\ref{fig:dias_examples}c) and \auxdia{6}{2} (Fig.~\ref{fig:dias_examples}d) auxiliary diagrams, which are, when the operator leg is taken into account, 4-point and 3-point Green functions:

\def\scalefigure{0.8}

\begin{align}
	& \KRP\left[\begin{tikzpicture}[baseline=-2pt,scale=\scalefigure]
		\def\radius{1cm}
		\setlength{\feynhandarrowsize}{3pt}
		\begin{feynhand}
			\vertex[crossdot] (o) at (0,0) {};
			\vertex[particle,red] (vo) at (-1.1\radius,0) {$Q$};
			\def\DeltaAngle{25}
			\def\ReferenceAngle{240}
			\propag[red,fer] (vo) to (o);
			\vertex[dot] (v1) at ([shift=({25:0.9\radius})]o) {}; 
			\vertex[dot] (v2) at ([shift=({-25:0.9\radius})]o) {}; 
			\propag[plain] (o) to (v1);
			\propag[plain] (o) to [out=25+20, in=  180+25-20] (v1);
			\propag[plain] (o) to [out=25-20, in=  180+25+20] (v1);
			\propag[plain] (o) to [out=-25+10, in=  180-25-10] (v2);
			\propag[plain] (o) to [out=-25-10, in=  180-25+10] (v2);
			\propag[plain] (o) to [out=-25+30, in=  180-25-30] (v2);
			\propag[plain] (o) to [out=-25-30, in=  180-25+30] (v2);
			\propag[plain] (v1) to (v2);
			\vertex (a1) [above right = 0.3\radius and 0.7\radius of v1] {$q_1$}; 
			\vertex (a2) [below  right = 0.3\radius and 0.7\radius of v1] {$q_2$}; 
			\vertex (a3) [below  right = 0.2\radius and 0.7\radius of v2] {$q_3$}; 
			\propag[plain,fer] (v1) to (a1);
			\propag[plain,fer] (v1) to (a2);
			\propag[plain,fer] (v2) to (a3);
		\end{feynhand}
	\end{tikzpicture}
\right]
	 = \frac{(Q, q_1+q_2)}{P^2}  \cdot \KRP\left[ 
\begin{tikzpicture}[baseline=-2pt,scale=\scalefigure]
		\def\radius{1cm}
		\setlength{\feynhandarrowsize}{3pt}
		\begin{feynhand}
			\vertex[crossdot] (o) at (0,0) {};
			\vertex[particle] (vo) at (-1.1\radius,0) {$P$};
			\def\DeltaAngle{25}
			\def\ReferenceAngle{240}
			\propag[fer] (vo) to (o);
			\vertex[dot] (v1) at ([shift=({25:0.9\radius})]o) {}; 
			\vertex[dot] (v2) at ([shift=({-25:0.9\radius})]o) {}; 
			\propag[plain] (o) to (v1);
			\propag[plain] (o) to [out=25+20, in=  180+25-20] (v1);
			\propag[plain] (o) to [out=25-20, in=  180+25+20] (v1);
			\propag[plain] (o) to [out=-25+10, in=  180-25-10] (v2);
			\propag[plain] (o) to [out=-25-10, in=  180-25+10] (v2);
			\propag[plain] (o) to [out=-25+30, in=  180-25-30] (v2);
			\propag[plain] (o) to [out=-25-30, in=  180-25+30] (v2);
			\propag[plain] (v1) to (v2);
			\vertex (a1) [above right = 0.3\radius and 0.7\radius of v1] {$P$}; 
			\propag[plain,fer] (v1) to (a1);
		\end{feynhand}
	\end{tikzpicture}
\right]  \nonumber\\
	& + \frac{(Q q_3)}{P^2} \cdot \KRP \left[ 
		 \begin{tikzpicture}[baseline=-2pt,scale=\scalefigure]
		\def\radius{1cm}
		\setlength{\feynhandarrowsize}{3pt}
		\begin{feynhand}
			\vertex[crossdot] (o) at (0,0) {};
			\vertex[particle] (vo) at (-1.1\radius,0) {$P$};
			\def\DeltaAngle{25}
			\def\ReferenceAngle{240}
			\propag[fer] (vo) to (o);
			\vertex[dot] (v1) at ([shift=({25:0.9\radius})]o) {}; 
			\vertex[dot] (v2) at ([shift=({-25:0.9\radius})]o) {}; 
			\propag[plain] (o) to (v1);
			\propag[plain] (o) to [out=25+20, in=  180+25-20] (v1);
			\propag[plain] (o) to [out=25-20, in=  180+25+20] (v1);
			\propag[plain] (o) to [out=-25+10, in=  180-25-10] (v2);
			\propag[plain] (o) to [out=-25-10, in=  180-25+10] (v2);
			\propag[plain] (o) to [out=-25+30, in=  180-25-30] (v2);
			\propag[plain] (o) to [out=-25-30, in=  180-25+30] (v2);
			\propag[plain] (v1) to (v2);
			\vertex (a3) [below  right = 0.2\radius and 0.7\radius of v2] {$P$}; 
			\propag[plain,fer] (v2) to (a3);
		\end{feynhand}
	\end{tikzpicture}
	\right] - \frac{(q_3, q_1 + q_2)}{P^2} \cdot  \KRP \left[ 
\begin{tikzpicture}[baseline=-2pt,scale=\scalefigure]
		\def\radius{1cm}
		\setlength{\feynhandarrowsize}{3pt}
		\begin{feynhand}
			\vertex[crossdot] (o) at (0,0) {};
			\def\DeltaAngle{25}
			\def\ReferenceAngle{240}
			\vertex[dot] (v1) at ([shift=({25:0.9\radius})]o) {}; 
			\vertex[dot] (v2) at ([shift=({-25:0.9\radius})]o) {}; 
			\propag[plain] (o) to (v1);
			\propag[plain] (o) to [out=25+20, in=  180+25-20] (v1);
			\propag[plain] (o) to [out=25-20, in=  180+25+20] (v1);
			\propag[plain] (o) to [out=-25+10, in=  180-25-10] (v2);
			\propag[plain] (o) to [out=-25-10, in=  180-25+10] (v2);
			\propag[plain] (o) to [out=-25+30, in=  180-25-30] (v2);
			\propag[plain] (o) to [out=-25-30, in=  180-25+30] (v2);
			\propag[plain] (v1) to (v2);
			\vertex (a1) [above right = 0.3\radius and 0.7\radius of v1] {$P$}; 
			\vertex (a3) [below  right = 0.2\radius and 0.7\radius of v2] {$P$}; 
			\propag[plain,fer] (v1) to (a1);
			\propag[plain,fer] (a3) to (v2);
		\end{feynhand}
	\end{tikzpicture}
	\right] \nonumber \\
	& = \frac{1}{(64\pi^2)^3} \left[\frac{4}{9 \ep} (Q, q_1 + q_2)  + \left[\frac{1}{9 \ep^2} - \frac{28}{27 \ep}\right](Q q_3) 
		- \left[\frac{1}{9 \ep^2} + \frac{8}{27 \ep}\right] (q_3, q_1+q_2)\right] \nonumber\\
	 & = \frac{1}{(64\pi^2)^3} \left[\frac{4}{9 \ep} (q_1+q_2)^2 + \left[\frac{1}{9 \ep^2} - \frac{28}{27 \ep}\right] q_3^2 - \frac{24}{27\ep} (q_3,q_1+q_2)\right],
	\label{eq:KR_7_3}
\end{align}
\begin{align}
	  &\KRP \left[\begin{tikzpicture}[baseline=-2pt,scale=\scalefigure]
		\def\radius{1cm}
		\setlength{\feynhandarrowsize}{3pt}
		\begin{feynhand}
			\vertex[crossdot] (o) at (0,0) {};
			\vertex[particle,red] (vo) at (-1.1\radius,0) {$Q$};
			\def\DeltaAngle{25}
			\def\ReferenceAngle{240}
			\propag[red,fer] (vo) to (o);
			\vertex[dot] (v1) at ([shift=({25:0.9\radius})]o) {}; 
			\vertex[dot] (v2) at ([shift=({-25:0.9\radius})]o) {}; 
			\propag[plain] (o) to (v1);
			\propag[plain] (o) to [out=25+20, in=  180+25-20] (v1);
			\propag[plain] (o) to [out=25-20, in=  180+25+20] (v1);
			\propag[plain] (o) to (v2);
			\propag[plain] (o) to [out=-25+20, in=  180-25-20] (v2);
			\propag[plain] (o) to [out=-25-20, in=  180-25+20] (v2);
		\propag[plain] (v1) to [out=-90+25, in=90-25] (v2);
		\propag[plain] (v1) to [out=-90-25, in=90+25] (v2);
			\vertex (a1) [above right = 0.2\radius and 0.6\radius of v1] {$q_1$}; 
			\vertex (a3) [below  right = 0.2\radius and 0.6\radius of v2] {$q_2$}; 
			\propag[plain,fer] (v1) to (a1);
			\propag[plain,fer] (v2) to (a3);
		\end{feynhand}
	\end{tikzpicture}
\right]  = \nonumber \\
	  & = \frac{(Q, q_1+q_2)}{P^2} \cdot \KRP \left[
\begin{tikzpicture}[baseline=-2pt,scale=\scalefigure]
		\def\radius{1cm}
		\setlength{\feynhandarrowsize}{3pt}
		\begin{feynhand}
			\vertex[crossdot] (o) at (0,0) {};
			\vertex[particle] (vo) at (-1.1\radius,0) {$P$};
			\def\DeltaAngle{25}
			\def\ReferenceAngle{240}
			\propag[fer] (vo) to (o);
			\vertex[dot] (v1) at ([shift=({25:0.9\radius})]o) {}; 
			\vertex[dot] (v2) at ([shift=({-25:0.9\radius})]o) {}; 
			\propag[plain] (o) to (v1);
			\propag[plain] (o) to [out=25+20, in=  180+25-20] (v1);
			\propag[plain] (o) to [out=25-20, in=  180+25+20] (v1);
			\propag[plain] (o) to (v2);
			\propag[plain] (o) to [out=-25+20, in=  180-25-20] (v2);
			\propag[plain] (o) to [out=-25-20, in=  180-25+20] (v2);
		\propag[plain] (v1) to [out=-90+25, in=90-25] (v2);
		\propag[plain] (v1) to [out=-90-25, in=90+25] (v2);
			\vertex (a1) [above right = 0.2\radius and 0.6\radius of v1] {$P$}; 
			\propag[plain,fer] (v1) to (a1);
		\end{feynhand}
	\end{tikzpicture}
	\right] 
	- \frac{q_1 q_2}{P^2} \cdot \KRP\left[ 
\begin{tikzpicture}[baseline=-2pt,scale=\scalefigure]
		\def\radius{1cm}
		\setlength{\feynhandarrowsize}{3pt}
		\begin{feynhand}
			\vertex[crossdot] (o) at (0,0) {};
			\def\DeltaAngle{25}
			\def\ReferenceAngle{240}
			\vertex[dot] (v1) at ([shift=({25:0.9\radius})]o) {}; 
			\vertex[dot] (v2) at ([shift=({-25:0.9\radius})]o) {}; 
			\propag[plain] (o) to (v1);
			\propag[plain] (o) to [out=25+20, in=  180+25-20] (v1);
			\propag[plain] (o) to [out=25-20, in=  180+25+20] (v1);
			\propag[plain] (o) to (v2);
			\propag[plain] (o) to [out=-25+20, in=  180-25-20] (v2);
			\propag[plain] (o) to [out=-25-20, in=  180-25+20] (v2);
		\propag[plain] (v1) to [out=-90+25, in=90-25] (v2);
		\propag[plain] (v1) to [out=-90-25, in=90+25] (v2);
			\vertex (a1) [above right = 0.2\radius and 0.6\radius of v1] {$P$}; 
			\vertex (a3) [below  right = 0.2\radius and 0.6\radius of v2] {$P$}; 
			\propag[plain,fer] (v1) to (a1);
			\propag[plain,fer] (a3) to (v2);
		\end{feynhand}
	\end{tikzpicture}
	\right] \nonumber\\
	& = 
\frac{1}{(64\pi^2)^3} \left[
	\left(\frac{1}{9\ep^2} - \frac{28}{27\ep}\right) Q^2 
-  \left(\frac{2}{9\ep^2} - \frac{8}{27\ep}\right) (q_1 q_2)\right] 
	\nonumber \\
	& = \frac{1}{(64\pi^2)^3} \left[
 \frac{q_1^2 + q_2^2}{9 \ep^2} - \frac{4[7 (q_1^2 + q_2^2) + 12 q_1 q_2]}{27\ep}\right].
	\label{eq:KR_6_2}
\end{align}

\def\scalefigure{0.7}
\begin{table}[H]
	\begin{center}
{\renewcommand{\arraystretch}{1.5}
\begin{tabular}{|l|c|c|}
	\hline
	Nickel index &  Diagram & $(64 \pi^2)^3 \cdot \KRP$ \\
	\hline
 \nickel{eee123|ee222|333|ee|}   & 
 \begin{tikzpicture}[baseline=(v1),scale=\scalefigure]
\node (v1) at (0., 1.20799) {};
\node (v2) at (0.197937, 0.) {};
\node (v3) at (0.953902, 0.855439) {};
\node (v4) at (1.46066, 0.658784) {};
\node (v5) at (1.47846, 0.997963) {};
\node (v6) at (1.27011, 2.2572) {};
\node (v7) at (1.95253, 0.809024) {};
\node (v8) at (1.96501, 0.768623) {};
\node (v9) at (1.98521, 0.801308) {};
\node (v10) at (1.98521, 0.833994) {};
\node (v11) at (2.00127, 1.45482) {};
\node (v12) at (2.00541, 0.768623) {};
\node (v13) at (2.0179, 0.809024) {};
\node (v14) at (2.00127, 2.2) {};
\node (v15) at (2.73189, 2.25949) {};
\node (v16) at (2.50103, 1.00149) {};
\node (v17) at (2.51915, 0.656242) {};
\node (v18) at (3.03497, 0.856421) {};
\node (v19) at (3.79498, 0.00296865) {};
\node (v20) at (3.98923, 1.21228) {};
\draw (v9.center) to (v11.center);
\draw (v9.center) to (v3.center);
\draw (v9.center) to (v18.center);
\draw (v15.center) to (v11.center);
\draw (v11.center) to (v6.center);
\draw (v11.center) to (v14.center);
\draw (v11.center) to (v3.center);
\draw (v11.center) to (v18.center);
\draw (v2.center) to (v3.center);
\draw (v3.center) to (v1.center);
\draw (v19.center) to (v18.center);
\draw (v18.center) to (v20.center);
\draw (v9.center) .. controls (v4.center)  .. (v3.center);
\draw (v9.center) .. controls (v5.center)  .. (v3.center);
\draw (v9.center) .. controls (v16.center)  .. (v18.center);
\draw (v9.center) .. controls (v17.center)  .. (v18.center);
\filldraw[black] (v15) circle (0.0369369);
\filldraw[black] (v11) circle (0.0369369);
\filldraw[black] (v6) circle (0.0369369);
\filldraw[black] (v14) circle (0.0369369);
\filldraw[black] (v2) circle (0.0369369);
\filldraw[black] (v3) circle (0.0369369);
\filldraw[black] (v1) circle (0.0369369);
\filldraw[black] (v19) circle (0.0369369);
\filldraw[black] (v18) circle (0.0369369);
\filldraw[black] (v20) circle (0.0369369);
\filldraw[red] (v10.center) -- (v7.center) -- (v8.center) -- (v12.center) -- (v13.center) -- (v10.center) -- cycle;
\end{tikzpicture}

				  & $\frac{1}{3 \ep ^3}-\frac{4}{3 \ep ^2}-\frac{8}{3 \ep }$ \\
 \nickel{eee112|22333|ee3|ee|}   & 
 \begin{tikzpicture}[baseline=(v4),scale=\scalefigure]
\node (v1) at (0., 1.04693) {};
\node (v2) at (0.367851, 2.12027) {};
\node (v3) at (0.41019, 0.1) {};
\node (v4) at (1.00935, 1.08731) {};
\node (v5) at (1.49038, 0.799005) {};
\node (v6) at (1.56761, 1.1408) {};
\node (v7) at (2.0216, 0.858882) {};
\node (v8) at (2.03192, 0.825461) {};
\node (v9) at (2.04863, 0.8525) {};
\node (v10) at (2.04863, 0.879538) {};
\node (v11) at (2.04984, 1.13713) {};
\node (v12) at (2.06534, 0.825461) {};
\node (v13) at (2.07567, 0.858882) {};
\node (v14) at (2.21855, 1.08085) {};
\node (v15) at (2.21975, 1.36547) {};
\node (v16) at (1.85606, 2.22183) {};
\node (v17) at (2.66576, 2.2218) {};
\node (v18) at (2.58777, 1.14824) {};
\node (v19) at (2.65806, 0.770492) {};
\node (v20) at (3.1972, 1.06623) {};
\node (v21) at (4.06109, 0.378192) {};
\node (v22) at (4.08275, 1.68364) {};
\draw (v9.center) to (v20.center);
\draw (v1.center) to (v4.center);
\draw (v4.center) to (v2.center);
\draw (v4.center) to (v3.center);
\draw (v4.center) to (v15.center);
\draw (v16.center) to (v15.center);
\draw (v15.center) to (v17.center);
\draw (v15.center) to (v20.center);
\draw (v21.center) to (v20.center);
\draw (v20.center) to (v22.center);
\draw (v9.center) .. controls (v5.center)  .. (v4.center);
\draw (v9.center) .. controls (v6.center)  .. (v4.center);
\draw (v9.center) .. controls (v11.center)  .. (v15.center);
\draw (v9.center) .. controls (v14.center)  .. (v15.center);
\draw (v9.center) .. controls (v18.center)  .. (v20.center);
\draw (v9.center) .. controls (v19.center)  .. (v20.center);
\filldraw[black] (v1) circle (0.0375314);
\filldraw[black] (v4) circle (0.0375314);
\filldraw[black] (v2) circle (0.0375314);
\filldraw[black] (v3) circle (0.0375314);
\filldraw[black] (v16) circle (0.0375314);
\filldraw[black] (v15) circle (0.0375314);
\filldraw[black] (v17) circle (0.0375314);
\filldraw[black] (v21) circle (0.0375314);
\filldraw[black] (v20) circle (0.0375314);
\filldraw[black] (v22) circle (0.0375314);
\filldraw[red] (v10.center) -- (v7.center) -- (v8.center) -- (v12.center) -- (v13.center) -- (v10.center) -- cycle;
\end{tikzpicture}

				  & $\frac{1}{6 \ep ^3}-\frac{2}{\ep ^2}+\frac{32}{3 \ep }$ \\
 {\nickel{eee112|22233|e3|eee|}}   & 
 \begin{tikzpicture}[baseline=(v5),scale=\scalefigure]
\node (v1) at (0., 1.92996) {};
\node (v2) at (0.250417, 0.952577) {};
\node (v3) at (0.528447, 2.73731) {};
\node (v4) at (0.981991, 1.71435) {};
\node (v5) at (1.49742, 1.95188) {};
\node (v6) at (1.53779, 1.59957) {};
\node (v7) at (1.94011, 1.49487) {};
\node (v8) at (2.01898, 1.84518) {};
\node (v9) at (2.03206, 1.80286) {};
\node (v10) at (2.05104, 0.3) {};
\node (v11) at (2.05222, 1.1523) {};
\node (v12) at (2.05322, 1.8371) {};
\node (v13) at (2.05322, 1.87134) {};
\node (v14) at (2.07438, 1.80286) {};
\node (v15) at (2.08746, 1.84518) {};
\node (v16) at (2.16533, 1.49453) {};
\node (v17) at (2.5686, 1.59926) {};
\node (v18) at (2.60916, 1.9516) {};
\node (v19) at (3.12454, 1.71377) {};
\node (v20) at (3.57713, 2.73716) {};
\node (v21) at (3.85427, 0.951204) {};
\node (v22) at (4.10597, 1.92964) {};
\draw (v12.center) to (v11.center);
\draw (v21.center) to (v19.center);
\draw (v19.center) to (v22.center);
\draw (v19.center) to (v20.center);
\draw (v19.center) to (v11.center);
\draw (v1.center) to (v4.center);
\draw (v4.center) to (v2.center);
\draw (v4.center) to (v3.center);
\draw (v4.center) to (v11.center);
\draw (v10.center) to (v11.center);
\draw (v12.center) .. controls (v18.center)  .. (v19.center);
\draw (v12.center) .. controls (v17.center)  .. (v19.center);
\draw (v12.center) .. controls (v6.center)  .. (v4.center);
\draw (v12.center) .. controls (v5.center)  .. (v4.center);
\draw (v12.center) .. controls (v16.center)  .. (v11.center);
\draw (v12.center) .. controls (v7.center)  .. (v11.center);
\filldraw[black] (v21) circle (0.0376781);
\filldraw[black] (v19) circle (0.0376781);
\filldraw[black] (v22) circle (0.0376781);
\filldraw[black] (v20) circle (0.0376781);
\filldraw[black] (v1) circle (0.0376781);
\filldraw[black] (v4) circle (0.0376781);
\filldraw[black] (v2) circle (0.0376781);
\filldraw[black] (v3) circle (0.0376781);
\filldraw[black] (v10) circle (0.0376781);
\filldraw[black] (v11) circle (0.0376781);
\filldraw[red] (v13.center) -- (v8.center) -- (v9.center) -- (v14.center) -- (v15.center) -- (v13.center) -- cycle;
\end{tikzpicture}

						   & $\frac{1}{3 \ep ^3}-\frac{8}{3 \ep ^2}+\frac{8}{3 \ep }$ \\
 \hline
\end{tabular}
}
\end{center}
\caption{
    The first part of the six-loop logarithmically divergent diagrams in Nickel notation contributing to the \auxdia{7}{7} auxiliary graphs and their $\KRP$. We highlight the operator insertion by a red pentagon.
    These three diagrams (c.f., Fig.~2a-c of Ref.~\cite{Jack:2020wvs}) also contribute to the anomalous dimension of the large-charge operators, which are represented by traceless symmetric products of the $\phi^a$ fields.
}
\label{tab:7-7_diagrams_1}
\end{table}

\subsection{Tables of $\KRP$ \label{app:KRP_tables}}

In this appendix, we collect all the relevant six-loop diagrams together with the corresponding results of the $\KRP$ operation 
(Tables~\ref{tab:7-7_diagrams_1}-\ref{tab:6-2_diagrams}). 
We specify the graph by  using either simple Nickel index (for \auxdia{n}{n} diagrams in Tables~\ref{tab:7-7_diagrams_1}-\ref{tab:6-6_diagrams_2}) 
or its extension with colored edges~\cite{Batkovich:2014bla} (for quadratically divergent integrals \auxdia{7}{3} and \auxdia{6}{2} in Tables~\ref{tab:7-3_diagrams} and \ref{tab:6-2_diagrams}, respectively).
In the latter case, we use positive integers $i=\nickel{1,2,3}$ to denote external edges with (outgoing) momentum $q_i$, the label \nickel{-1} specifies the operator insertion (also marked by a red pentagon). The lines marked by \nickel{0} correspond to internal propagators.

It is worth pointing that part of the \auxdia{7}{7} and \auxdia{6}{6}  diagrams  were calculated in Ref. \cite{Jack:2020wvs}. 
Our results are completely consistent with the results of the reference (see Table \ref{tab:7-7_diagrams_1}, \ref{tab:6-6_diagrams_1} in our work and (A19), (A20){\footnote{There is a typo in (A19) and (A20) in the published article: in the left parts should be $(64\pi^2)^3$ instead of $(64\pi^3)^3$. The typo has been corrected in the latest arXiv version.} in Ref.~\cite{Jack:2020wvs}).
	The other diagrams presented in Tables~\ref{tab:7-7_diagrams_2} and \ref{tab:6-6_diagrams_2} give a vanishing contribution to the renormalization of fixed-charge operators and thus were not considered in Ref.~\cite{Jack:2020wvs}.

\begin{table}[H]
	\begin{center}
{\renewcommand{\arraystretch}{1.5}
\begin{tabular}{|l|c|c|}
	\hline
	Nickel index &  Diagram & $(64 \pi^2)^3 \cdot \KRP$ \\
	\hline
	\nickel{eeee12|eee33|33333||}  & 
	\begin{tikzpicture}[baseline=(v4),scale=\scalefigure]
\node (v1) at (0., 1.42886) {};
\node (v2) at (0.125683, 0.372028) {};
\node (v3) at (0.557974, 2.20769) {};
\node (v4) at (0.906277, 0.495219) {};
\node (v5) at (0.956935, 1.0732) {};
\node (v6) at (1.17133, 0.00879174) {};
\node (v7) at (1.18556, -0.0372424) {};
\node (v8) at (1.20857, 0.) {};
\node (v9) at (1.20857, 0.0372424) {};
\node (v10) at (1.23159, -0.0372424) {};
\node (v11) at (1.24582, 0.00879174) {};
\node (v12) at (1.25923, 0.577978) {};
\node (v13) at (1.57192, 0.14725) {};
\node (v14) at (1.57603, 0.0863014) {};
\node (v15) at (1.58428, -0.0362138) {};
\node (v16) at (1.58839, -0.0971624) {};
\node (v17) at (1.95174, 0.0500876) {};
\node (v18) at (2.43065, 1.05768) {};
\node (v19) at (2.52855, 2.28508) {};
\node (v20) at (3.03277, 0.0324619) {};
\node (v21) at (3.22752, 1.85115) {};
\node (v22) at (3.42762, 0.896704) {};
\draw (v8.center) to (v17.center);
\draw (v19.center) to (v18.center);
\draw (v18.center) to (v22.center);
\draw (v18.center) to (v20.center);
\draw (v18.center) to (v21.center);
\draw (v18.center) to (v5.center);
\draw (v18.center) to (v17.center);
\draw (v2.center) to (v5.center);
\draw (v5.center) to (v1.center);
\draw (v5.center) to (v3.center);
\draw (v8.center) .. controls (v4.center)  .. (v5.center);
\draw (v8.center) .. controls (v12.center)  .. (v5.center);
\draw (v8.center) .. controls (v13.center)  .. (v17.center);
\draw (v8.center) .. controls (v14.center)  .. (v17.center);
\draw (v8.center) .. controls (v15.center)  .. (v17.center);
\draw (v8.center) .. controls (v16.center)  .. (v17.center);
\filldraw[black] (v19) circle (0.0332295);
\filldraw[black] (v18) circle (0.0332295);
\filldraw[black] (v22) circle (0.0332295);
\filldraw[black] (v20) circle (0.0332295);
\filldraw[black] (v21) circle (0.0332295);
\filldraw[black] (v2) circle (0.0332295);
\filldraw[black] (v5) circle (0.0332295);
\filldraw[black] (v1) circle (0.0332295);
\filldraw[black] (v3) circle (0.0332295);
\filldraw[black] (v17) circle (0.0332295);
\filldraw[red] (v9.center) -- (v6.center) -- (v7.center) -- (v10.center) -- (v11.center) -- (v9.center) -- cycle;
\end{tikzpicture}
					& $\frac{2}{9 \ep ^2}-\frac{40}{27 \ep }$ \\
	\nickel{eeee12|eee23|33333||} & 
	\begin{tikzpicture}[baseline=(v4),scale=\scalefigure]
\node (v1) at (0., 1.5238) {};
\node (v2) at (0.073507, 0.47163) {};
\node (v3) at (0.601775, 2.26201) {};
\node (v4) at (0.984562, 1.11347) {};
\node (v5) at (1.04996, 0.0990076) {};
\node (v6) at (1.25368, 0.460064) {};
\node (v7) at (1.29721, 0.412334) {};
\node (v8) at (1.3847, 0.31639) {};
\node (v9) at (1.42822, 0.268661) {};
\node (v10) at (1.59256, 0.639014) {};
\node (v11) at (1.6076, 0.590336) {};
\node (v12) at (1.63194, 0.629717) {};
\node (v13) at (1.63194, 0.669098) {};
\node (v14) at (1.65628, 0.590336) {};
\node (v15) at (1.67132, 0.639014) {};
\node (v16) at (2.45157, 1.10783) {};
\node (v17) at (2.68665, 2.29843) {};
\node (v18) at (2.92219, 0.) {};
\node (v19) at (3.35457, 1.73353) {};
\node (v20) at (3.44764, 0.759395) {};
\draw (v12.center) to (v16.center);
\draw (v12.center) to (v4.center);
\draw (v12.center) to (v5.center);
\draw (v17.center) to (v16.center);
\draw (v16.center) to (v20.center);
\draw (v16.center) to (v18.center);
\draw (v16.center) to (v19.center);
\draw (v16.center) to (v4.center);
\draw (v3.center) to (v4.center);
\draw (v4.center) to (v2.center);
\draw (v4.center) to (v1.center);
\draw (v4.center) to (v5.center);
\draw (v12.center) .. controls (v9.center)  .. (v5.center);
\draw (v12.center) .. controls (v8.center)  .. (v5.center);
\draw (v12.center) .. controls (v7.center)  .. (v5.center);
\draw (v12.center) .. controls (v6.center)  .. (v5.center);
\filldraw[black] (v17) circle (0.0333659);
\filldraw[black] (v16) circle (0.0333659);
\filldraw[black] (v20) circle (0.0333659);
\filldraw[black] (v18) circle (0.0333659);
\filldraw[black] (v19) circle (0.0333659);
\filldraw[black] (v3) circle (0.0333659);
\filldraw[black] (v4) circle (0.0333659);
\filldraw[black] (v2) circle (0.0333659);
\filldraw[black] (v1) circle (0.0333659);
\filldraw[black] (v5) circle (0.0333659);
\filldraw[red] (v13.center) -- (v10.center) -- (v11.center) -- (v14.center) -- (v15.center) -- (v13.center) -- cycle;
\end{tikzpicture}

				       & $\frac{4}{9 \ep }$ \\
	{\nickel{eeee12|22223|33|eee|}}   & 
 \begin{tikzpicture}[baseline=(v4),scale=\scalefigure]
\node (v1) at (0., 1.37866) {};
\node (v2) at (0.409802, 2.52972) {};
\node (v3) at (0.411056, 0.229162) {};
\node (v4) at (0.994366, 1.37884) {};
\node (v5) at (1.48425, 1.71922) {};
\node (v6) at (1.59066, 1.362) {};
\node (v7) at (1.97436, 1.37956) {};
\node (v8) at (2.04504, 1.37956) {};
\node (v9) at (2.04826, 1.71) {};
\node (v10) at (2.06059, 1.6701) {};
\node (v11) at (2.08052, 1.05674) {};
\node (v12) at (2.08054, 1.70238) {};
\node (v13) at (2.08054, 1.73466) {};
\node (v14) at (2.1005, 1.6701) {};
\node (v15) at (2.11283, 1.71) {};
\node (v16) at (2.11603, 1.37956) {};
\node (v17) at (2.1867, 1.37956) {};
\node (v18) at (3.20973, 1.3796) {};
\node (v19) at (3.50481, 2.75878) {};
\node (v20) at (3.50588, 0.) {};
\node (v21) at (4.13674, 1.97951) {};
\node (v22) at (4.13817, 0.780056) {};
\draw (v12.center) to (v18.center);
\draw (v19.center) to (v18.center);
\draw (v18.center) to (v20.center);
\draw (v18.center) to (v22.center);
\draw (v18.center) to (v21.center);
\draw (v18.center) to (v11.center);
\draw (v3.center) to (v4.center);
\draw (v4.center) to (v1.center);
\draw (v4.center) to (v2.center);
\draw (v4.center) to (v11.center);
\draw (v12.center) .. controls (v6.center)  .. (v4.center);
\draw (v12.center) .. controls (v5.center)  .. (v4.center);
\draw (v12.center) .. controls (v17.center)  .. (v11.center);
\draw (v12.center) .. controls (v16.center)  .. (v11.center);
\draw (v12.center) .. controls (v8.center)  .. (v11.center);
\draw (v12.center) .. controls (v7.center)  .. (v11.center);
\filldraw[black] (v19) circle (0.0378809);
\filldraw[black] (v18) circle (0.0378809);
\filldraw[black] (v20) circle (0.0378809);
\filldraw[black] (v22) circle (0.0378809);
\filldraw[black] (v21) circle (0.0378809);
\filldraw[black] (v3) circle (0.0378809);
\filldraw[black] (v4) circle (0.0378809);
\filldraw[black] (v1) circle (0.0378809);
\filldraw[black] (v2) circle (0.0378809);
\filldraw[black] (v11) circle (0.0378809);
\filldraw[red] (v13.center) -- (v9.center) -- (v10.center) -- (v14.center) -- (v15.center) -- (v13.center) -- cycle;
\end{tikzpicture}

						   & $\frac{2}{3 \ep ^2}-\frac{4}{3 \ep }$ \\
 \nickel{eeee12|ee333|e3333||}   & 
 \begin{tikzpicture}[baseline=(v4),scale=\scalefigure]
\node (v1) at (0., 1.61171) {};
\node (v2) at (0.233112, 1.08799) {};
\node (v3) at (0.37368, 2.08045) {};
\node (v4) at (0.983604, 1.50564) {};
\node (v5) at (1.11544, 2.17893) {};
\node (v6) at (1.2915, 0.) {};
\node (v7) at (1.3735, 0.699565) {};
\node (v8) at (1.73211, 0.922547) {};
\node (v9) at (1.75286, 0.837213) {};
\node (v10) at (1.77371, 0.751492) {};
\node (v11) at (1.79446, 0.666158) {};
\node (v12) at (2.12441, 0.895907) {};
\node (v13) at (2.13063, 1.22479) {};
\node (v14) at (2.13536, 0.860477) {};
\node (v15) at (2.15307, 0.88914) {};
\node (v16) at (2.15307, 0.917803) {};
\node (v17) at (2.17079, 0.860477) {};
\node (v18) at (2.18174, 0.895907) {};
\node (v19) at (2.3118, 1.50824) {};
\node (v20) at (2.33425, 1.17259) {};
\node (v21) at (2.96831, 2.0166) {};
\node (v22) at (3.2684, 1.41054) {};
\draw (v15.center) to (v19.center);
\draw (v3.center) to (v4.center);
\draw (v4.center) to (v5.center);
\draw (v4.center) to (v1.center);
\draw (v4.center) to (v2.center);
\draw (v4.center) to (v19.center);
\draw (v4.center) to (v7.center);
\draw (v21.center) to (v19.center);
\draw (v19.center) to (v22.center);
\draw (v6.center) to (v7.center);
\draw (v15.center) .. controls (v13.center)  .. (v19.center);
\draw (v15.center) .. controls (v20.center)  .. (v19.center);
\draw (v15.center) .. controls (v11.center)  .. (v7.center);
\draw (v15.center) .. controls (v10.center)  .. (v7.center);
\draw (v15.center) .. controls (v9.center)  .. (v7.center);
\draw (v15.center) .. controls (v8.center)  .. (v7.center);
\filldraw[black] (v3) circle (0.0321339);
\filldraw[black] (v4) circle (0.0321339);
\filldraw[black] (v5) circle (0.0321339);
\filldraw[black] (v1) circle (0.0321339);
\filldraw[black] (v2) circle (0.0321339);
\filldraw[black] (v21) circle (0.0321339);
\filldraw[black] (v19) circle (0.0321339);
\filldraw[black] (v22) circle (0.0321339);
\filldraw[black] (v6) circle (0.0321339);
\filldraw[black] (v7) circle (0.0321339);
\filldraw[red] (v16.center) -- (v12.center) -- (v14.center) -- (v17.center) -- (v18.center) -- (v16.center) -- cycle;
\end{tikzpicture}

				  & $\frac{1}{3 \ep ^2}$ \\
 \nickel{eeee12|ee223|3333|e|}   & 
 \begin{tikzpicture}[baseline=(v5),scale=\scalefigure]
\node (v1) at (0., 0.) {};
\node (v2) at (0.469303, 1.87177) {};
\node (v3) at (0.78828, 0.585823) {};
\node (v4) at (1.163, 2.41928) {};
\node (v5) at (1.19093, 0.837157) {};
\node (v6) at (1.21442, 0.741282) {};
\node (v7) at (1.23802, 0.644973) {};
\node (v8) at (1.26151, 0.549097) {};
\node (v9) at (1.36863, 1.43045) {};
\node (v10) at (1.41279, 1.06685) {};
\node (v11) at (1.62, 1.16404) {};
\node (v12) at (1.62936, 0.808645) {};
\node (v13) at (1.64265, 0.765637) {};
\node (v14) at (1.66416, 0.800431) {};
\node (v15) at (1.66416, 0.835226) {};
\node (v16) at (1.68566, 0.765637) {};
\node (v17) at (1.69895, 0.808645) {};
\node (v18) at (2.65622, 1.05452) {};
\node (v19) at (2.91632, 0.087433) {};
\node (v20) at (3.07117, 1.95351) {};
\node (v21) at (3.56076, 0.571026) {};
\node (v22) at (3.62892, 1.36818) {};
\draw (v14.center) to (v18.center);
\draw (v22.center) to (v18.center);
\draw (v18.center) to (v21.center);
\draw (v18.center) to (v19.center);
\draw (v18.center) to (v20.center);
\draw (v18.center) to (v9.center);
\draw (v4.center) to (v9.center);
\draw (v9.center) to (v2.center);
\draw (v9.center) to (v3.center);
\draw (v1.center) to (v3.center);
\draw (v14.center) .. controls (v10.center)  .. (v9.center);
\draw (v14.center) .. controls (v11.center)  .. (v9.center);
\draw (v14.center) .. controls (v8.center)  .. (v3.center);
\draw (v14.center) .. controls (v7.center)  .. (v3.center);
\draw (v14.center) .. controls (v6.center)  .. (v3.center);
\draw (v14.center) .. controls (v5.center)  .. (v3.center);
\filldraw[black] (v22) circle (0.034586);
\filldraw[black] (v18) circle (0.034586);
\filldraw[black] (v21) circle (0.034586);
\filldraw[black] (v19) circle (0.034586);
\filldraw[black] (v20) circle (0.034586);
\filldraw[black] (v4) circle (0.034586);
\filldraw[black] (v9) circle (0.034586);
\filldraw[black] (v2) circle (0.034586);
\filldraw[black] (v1) circle (0.034586);
\filldraw[black] (v3) circle (0.034586);
\filldraw[red] (v15.center) -- (v12.center) -- (v13.center) -- (v16.center) -- (v17.center) -- (v15.center) -- cycle;
\end{tikzpicture}

				  & $\frac{4}{3 \ep }$ \\
 \nickel{eeee12|e2223|333|ee|}   & 
 \begin{tikzpicture}[baseline=(v5),scale=\scalefigure]
\node (v1) at (0., 1.2652) {};
\node (v2) at (0.106963, 2.17907) {};
\node (v3) at (0.537397, 0.565272) {};
\node (v4) at (0.787971, 2.68582) {};
\node (v5) at (0.979049, 1.57368) {};
\node (v6) at (1.99707, 1.4011) {};
\node (v7) at (2.03845, 1.75241) {};
\node (v8) at (2.05121, 1.71114) {};
\node (v9) at (2.07184, 1.74453) {};
\node (v10) at (2.07184, 1.77792) {};
\node (v11) at (2.09248, 1.71114) {};
\node (v12) at (2.10523, 1.75241) {};
\node (v13) at (2.14074, 1.08032) {};
\node (v14) at (2.21552, 1.42376) {};
\node (v15) at (2.43361, 0.) {};
\node (v16) at (2.57217, 1.54335) {};
\node (v17) at (2.59392, 1.8796) {};
\node (v18) at (3.09425, 1.67841) {};
\node (v19) at (3.81081, 2.4398) {};
\node (v20) at (4.02874, 1.3722) {};
\draw (v9.center) to (v5.center);
\draw (v9.center) to (v18.center);
\draw (v9.center) to (v13.center);
\draw (v3.center) to (v5.center);
\draw (v5.center) to (v2.center);
\draw (v5.center) to (v1.center);
\draw (v5.center) to (v4.center);
\draw (v5.center) to (v13.center);
\draw (v19.center) to (v18.center);
\draw (v18.center) to (v20.center);
\draw (v18.center) to (v13.center);
\draw (v15.center) to (v13.center);
\draw (v9.center) .. controls (v17.center)  .. (v18.center);
\draw (v9.center) .. controls (v16.center)  .. (v18.center);
\draw (v9.center) .. controls (v14.center)  .. (v13.center);
\draw (v9.center) .. controls (v6.center)  .. (v13.center);
\filldraw[black] (v3) circle (0.0371888);
\filldraw[black] (v5) circle (0.0371888);
\filldraw[black] (v2) circle (0.0371888);
\filldraw[black] (v1) circle (0.0371888);
\filldraw[black] (v4) circle (0.0371888);
\filldraw[black] (v19) circle (0.0371888);
\filldraw[black] (v18) circle (0.0371888);
\filldraw[black] (v20) circle (0.0371888);
\filldraw[black] (v15) circle (0.0371888);
\filldraw[black] (v13) circle (0.0371888);
\filldraw[red] (v10.center) -- (v7.center) -- (v8.center) -- (v11.center) -- (v12.center) -- (v10.center) -- cycle;
\end{tikzpicture}

				  & $\frac{1}{3 \ep ^2}-\frac{4}{\ep }$ \\
 \nickel{eee112|23333|eee3|e|}   & 
 \begin{tikzpicture}[baseline=(v5),scale=\scalefigure]
\node (v1) at (0., 0.739611) {};
\node (v2) at (0.0676514, 2.20302) {};
\node (v3) at (0.332284, 0.152289) {};
\node (v4) at (0.649137, 1.5755) {};
\node (v5) at (0.98244, 1.30219) {};
\node (v6) at (0.995403, 0.771658) {};
\node (v7) at (1.01478, 1.3858) {};
\node (v8) at (1.04726, 1.46978) {};
\node (v9) at (1.0796, 1.55338) {};
\node (v10) at (1.12215, 0.) {};
\node (v11) at (1.38001, 1.28783) {};
\node (v12) at (1.39258, 1.24717) {};
\node (v13) at (1.41291, 1.28007) {};
\node (v14) at (1.41291, 1.31296) {};
\node (v15) at (1.43323, 1.24717) {};
\node (v16) at (1.4458, 1.28783) {};
\node (v17) at (1.81543, 1.01387) {};
\node (v18) at (1.89486, 1.30476) {};
\node (v19) at (2.29739, 1.03856) {};
\node (v20) at (2.87007, 1.67456) {};
\node (v21) at (2.97913, 0.480284) {};
\node (v22) at (3.30453, 1.09802) {};
\draw (v13.center) to (v6.center);
\draw (v1.center) to (v6.center);
\draw (v6.center) to (v3.center);
\draw (v6.center) to (v10.center);
\draw (v6.center) to (v19.center);
\draw (v6.center) to (v4.center);
\draw (v20.center) to (v19.center);
\draw (v19.center) to (v22.center);
\draw (v19.center) to (v21.center);
\draw (v2.center) to (v4.center);
\draw (v13.center) .. controls (v18.center)  .. (v19.center);
\draw (v13.center) .. controls (v17.center)  .. (v19.center);
\draw (v13.center) .. controls (v5.center)  .. (v4.center);
\draw (v13.center) .. controls (v7.center)  .. (v4.center);
\draw (v13.center) .. controls (v8.center)  .. (v4.center);
\draw (v13.center) .. controls (v9.center)  .. (v4.center);
\filldraw[black] (v1) circle (0.0323843);
\filldraw[black] (v6) circle (0.0323843);
\filldraw[black] (v3) circle (0.0323843);
\filldraw[black] (v10) circle (0.0323843);
\filldraw[black] (v20) circle (0.0323843);
\filldraw[black] (v19) circle (0.0323843);
\filldraw[black] (v22) circle (0.0323843);
\filldraw[black] (v21) circle (0.0323843);
\filldraw[black] (v2) circle (0.0323843);
\filldraw[black] (v4) circle (0.0323843);
\filldraw[red] (v14.center) -- (v11.center) -- (v12.center) -- (v15.center) -- (v16.center) -- (v14.center) -- cycle;
\end{tikzpicture}

				  & $\frac{2}{3 \ep ^2}-\frac{16}{3 \ep }$ \\
				  \hline
\end{tabular}
}
\end{center}
\caption{
    The second part of the six-loop logarithmically divergent diagrams in Nickel notation contributing to the \auxdia{7}{7} auxiliary graphs and their $\KRP$. We highlight the operator insertion by a red pentagon.
    These graphs vanish in the case of charged operators. 
}
\label{tab:7-7_diagrams_2}
\end{table}

\def\scalefigure{0.6}

\begin{table}[H]
	\begin{center}
{\renewcommand{\arraystretch}{1.5}
\begin{tabular}{|l|c|c|}
	\hline
	Nickel index & Diagram & $(64 \pi^2)^3 \cdot \KRP$ \\
	\hline
	\nickel{ee1112|233|ee33|ee|}  & 
\begin{tikzpicture}[baseline=(v4),scale=\scalefigure]
\node (v1) at (0., 1.00702) {};
\node (v2) at (0.366729, 0.00126923) {};
\node (v3) at (0.979879, 0.85309) {};
\node (v4) at (1.33316, 1.38914) {};
\node (v5) at (1.4533, 0.683937) {};
\node (v6) at (1.46129, 0.997963) {};
\node (v7) at (1.47446, 2.57949) {};
\node (v8) at (1.58238, 1.0748) {};
\node (v9) at (1.89561, 0.838041) {};
\node (v10) at (1.91054, 0.789708) {};
\node (v11) at (1.93471, 0.82881) {};
\node (v12) at (1.93471, 0.867912) {};
\node (v13) at (1.93566, 1.61085) {};
\node (v14) at (1.95887, 0.789708) {};
\node (v15) at (1.97381, 0.838041) {};
\node (v16) at (2.40185, 2.57643) {};
\node (v17) at (2.40812, 0.997305) {};
\node (v18) at (2.41572, 0.683413) {};
\node (v19) at (2.88913, 0.851908) {};
\node (v20) at (3.50328, 0.) {};
\node (v21) at (3.86923, 1.00425) {};
\draw (v11.center) to (v13.center);
\draw (v11.center) to (v19.center);
\draw (v7.center) to (v13.center);
\draw (v13.center) to (v16.center);
\draw (v13.center) to (v19.center);
\draw (v1.center) to (v3.center);
\draw (v3.center) to (v2.center);
\draw (v20.center) to (v19.center);
\draw (v19.center) to (v21.center);
\draw (v11.center) .. controls (v5.center)  .. (v3.center);
\draw (v11.center) .. controls (v6.center)  .. (v3.center);
\draw (v11.center) .. controls (v17.center)  .. (v19.center);
\draw (v11.center) .. controls (v18.center)  .. (v19.center);
\draw (v13.center) .. controls (v8.center)  .. (v3.center);
\draw (v13.center) .. controls (v4.center)  .. (v3.center);
\filldraw[black] (v7) circle (0.0361646);
\filldraw[black] (v13) circle (0.0361646);
\filldraw[black] (v16) circle (0.0361646);
\filldraw[black] (v1) circle (0.0361646);
\filldraw[black] (v3) circle (0.0361646);
\filldraw[black] (v2) circle (0.0361646);
\filldraw[black] (v20) circle (0.0361646);
\filldraw[black] (v19) circle (0.0361646);
\filldraw[black] (v21) circle (0.0361646);
\filldraw[red] (v12.center) -- (v9.center) -- (v10.center) -- (v14.center) -- (v15.center) -- (v12.center) -- cycle;
\end{tikzpicture}

				       & $\frac{\pi ^2 (5-2 \ln2)}{3 \ep }-\frac{\pi ^2}{6 \ep ^2}$ \\
	\nickel{eee123|ee223|333|e|} &
\begin{tikzpicture}[baseline=(v5),scale=\scalefigure]
\node (v1) at (0., 1.50295) {};
\node (v2) at (0.441144, 2.27151) {};
\node (v3) at (1.06485, 1.56221) {};
\node (v4) at (1.26169, 0.) {};
\node (v5) at (1.32698, 1.35697) {};
\node (v6) at (1.39763, 1.55262) {};
\node (v7) at (1.51373, 1.1166) {};
\node (v8) at (1.53318, 0.8442) {};
\node (v9) at (1.63381, 1.3535) {};
\node (v10) at (1.64372, 1.32144) {};
\node (v11) at (1.65975, 1.34738) {};
\node (v12) at (1.65975, 1.37332) {};
\node (v13) at (1.67579, 1.32144) {};
\node (v14) at (1.67921, 1.07498) {};
\node (v15) at (1.6857, 1.3535) {};
\node (v16) at (2.35064, 1.44986) {};
\node (v17) at (2.79427, 2.246) {};
\node (v18) at (3.18515, 0.900172) {};
\node (v19) at (3.40727, 1.65504) {};
\draw (v11.center) to (v16.center);
\draw (v11.center) to (v8.center);
\draw (v17.center) to (v16.center);
\draw (v16.center) to (v18.center);
\draw (v16.center) to (v19.center);
\draw (v16.center) to (v3.center);
\draw (v16.center) to (v8.center);
\draw (v1.center) to (v3.center);
\draw (v3.center) to (v2.center);
\draw (v3.center) to (v8.center);
\draw (v4.center) to (v8.center);
\draw (v11.center) .. controls (v5.center)  .. (v3.center);
\draw (v11.center) .. controls (v6.center)  .. (v3.center);
\draw (v11.center) .. controls (v14.center)  .. (v8.center);
\draw (v11.center) .. controls (v7.center)  .. (v8.center);
\filldraw[black] (v17) circle (0.0330906);
\filldraw[black] (v16) circle (0.0330906);
\filldraw[black] (v18) circle (0.0330906);
\filldraw[black] (v19) circle (0.0330906);
\filldraw[black] (v1) circle (0.0330906);
\filldraw[black] (v3) circle (0.0330906);
\filldraw[black] (v2) circle (0.0330906);
\filldraw[black] (v4) circle (0.0330906);
\filldraw[black] (v8) circle (0.0330906);
\filldraw[red] (v12.center) -- (v9.center) -- (v10.center) -- (v13.center) -- (v15.center) -- (v12.center) -- cycle;
\end{tikzpicture}

				      & $\frac{1}{6 \ep ^3}-\frac{2}{\ep ^2}-\frac{8 \left(\pi ^2-12\right)}{9 \ep }$ \\
	\nickel{eee112|2333|ee33|e|}&
\begin{tikzpicture}[baseline=(v4),scale=\scalefigure]
\node (v1) at (0., 0.90078) {};
\node (v2) at (0.329825, 1.72985) {};
\node (v3) at (0.522585, 0.17061) {};
\node (v4) at (1.01481, 1.02383) {};
\node (v5) at (1.44872, 0.746505) {};
\node (v6) at (1.52863, 1.0582) {};
\node (v7) at (1.92742, 0.78916) {};
\node (v8) at (1.94084, 0.745746) {};
\node (v9) at (1.96254, 0.780868) {};
\node (v10) at (1.96254, 0.815991) {};
\node (v11) at (1.98425, 0.745746) {};
\node (v12) at (1.99767, 0.78916) {};
\node (v13) at (2.23505, 1.42831) {};
\node (v14) at (2.31106, 2.40978) {};
\node (v15) at (2.36434, 0.542481) {};
\node (v16) at (2.39413, 0.911011) {};
\node (v17) at (2.42741, 0.82751) {};
\node (v18) at (2.67012, 1.10642) {};
\node (v19) at (2.82921, 0.589122) {};
\node (v20) at (3.11116, 1.92543) {};
\node (v21) at (3.61468, 0.) {};
\draw (v9.center) to (v13.center);
\draw (v9.center) to (v19.center);
\draw (v3.center) to (v4.center);
\draw (v4.center) to (v2.center);
\draw (v4.center) to (v1.center);
\draw (v4.center) to (v13.center);
\draw (v20.center) to (v13.center);
\draw (v13.center) to (v14.center);
\draw (v21.center) to (v19.center);
\draw (v9.center) .. controls (v5.center)  .. (v4.center);
\draw (v9.center) .. controls (v6.center)  .. (v4.center);
\draw (v9.center) .. controls (v17.center)  .. (v19.center);
\draw (v9.center) .. controls (v15.center)  .. (v19.center);
\draw (v13.center) .. controls (v18.center)  .. (v19.center);
\draw (v13.center) .. controls (v16.center)  .. (v19.center);
\filldraw[black] (v3) circle (0.034491);
\filldraw[black] (v4) circle (0.034491);
\filldraw[black] (v2) circle (0.034491);
\filldraw[black] (v1) circle (0.034491);
\filldraw[black] (v20) circle (0.034491);
\filldraw[black] (v13) circle (0.034491);
\filldraw[black] (v14) circle (0.034491);
\filldraw[black] (v21) circle (0.034491);
\filldraw[black] (v19) circle (0.034491);
\filldraw[red] (v10.center) -- (v7.center) -- (v8.center) -- (v11.center) -- (v12.center) -- (v10.center) -- cycle;
\end{tikzpicture}

				     & $\frac{1}{6 \ep ^3}-\frac{2}{\ep ^2}-\frac{4 \left(\pi ^2-24\right)}{9 \ep }$ \\
	\nickel{eee112|e223|333|ee|} &
\begin{tikzpicture}[baseline=(v5),scale=\scalefigure]
\node (v1) at (0., 1.84448) {};
\node (v2) at (0.294328, 0.851629) {};
\node (v3) at (0.567876, 2.66945) {};
\node (v4) at (1.00075, 1.64691) {};
\node (v5) at (1.45942, 1.2002) {};
\node (v6) at (1.63503, 1.55966) {};
\node (v7) at (1.95797, 1.45234) {};
\node (v8) at (2.01546, 1.81424) {};
\node (v9) at (2.02873, 1.77132) {};
\node (v10) at (2.05019, 1.80605) {};
\node (v11) at (2.05019, 1.84077) {};
\node (v12) at (2.07165, 1.77132) {};
\node (v13) at (2.08491, 1.81424) {};
\node (v14) at (2.0937, 1.11295) {};
\node (v15) at (2.18592, 1.46665) {};
\node (v16) at (2.26642, 0.) {};
\node (v17) at (2.54103, 1.59536) {};
\node (v18) at (2.57027, 1.92784) {};
\node (v19) at (3.06111, 1.71716) {};
\node (v20) at (3.76928, 2.50362) {};
\node (v21) at (4.00417, 1.38578) {};
\draw (v10.center) to (v4.center);
\draw (v10.center) to (v19.center);
\draw (v3.center) to (v4.center);
\draw (v4.center) to (v2.center);
\draw (v4.center) to (v1.center);
\draw (v20.center) to (v19.center);
\draw (v19.center) to (v21.center);
\draw (v19.center) to (v14.center);
\draw (v16.center) to (v14.center);
\draw (v10.center) .. controls (v18.center)  .. (v19.center);
\draw (v10.center) .. controls (v17.center)  .. (v19.center);
\draw (v10.center) .. controls (v15.center)  .. (v14.center);
\draw (v10.center) .. controls (v7.center)  .. (v14.center);
\draw (v4.center) .. controls (v6.center)  .. (v14.center);
\draw (v4.center) .. controls (v5.center)  .. (v14.center);
\filldraw[black] (v3) circle (0.0370323);
\filldraw[black] (v4) circle (0.0370323);
\filldraw[black] (v2) circle (0.0370323);
\filldraw[black] (v1) circle (0.0370323);
\filldraw[black] (v20) circle (0.0370323);
\filldraw[black] (v19) circle (0.0370323);
\filldraw[black] (v21) circle (0.0370323);
\filldraw[black] (v16) circle (0.0370323);
\filldraw[black] (v14) circle (0.0370323);
\filldraw[red] (v11.center) -- (v8.center) -- (v9.center) -- (v12.center) -- (v13.center) -- (v11.center) -- cycle;
\end{tikzpicture}
				      & $\frac{1}{6 \ep ^3}-\frac{2}{\ep ^2}+\frac{32}{3 \ep }$ \\
	\nickel{eee112|2223|33|eee|} &
\begin{tikzpicture}[baseline=(v4),scale=\scalefigure]
\node (v1) at (0., 1.3951) {};
\node (v2) at (0.452938, 2.79079) {};
\node (v3) at (0.453633, 0.) {};
\node (v4) at (1.0115, 1.39548) {};
\node (v5) at (1.48745, 1.02025) {};
\node (v6) at (1.61726, 1.376) {};
\node (v7) at (1.96327, 1.39547) {};
\node (v8) at (2.05374, 1.01009) {};
\node (v9) at (2.06882, 0.961298) {};
\node (v10) at (2.09298, 1.79024) {};
\node (v11) at (2.09321, 1.00077) {};
\node (v12) at (2.09321, 1.04025) {};
\node (v13) at (2.11761, 0.961298) {};
\node (v14) at (2.13268, 1.01009) {};
\node (v15) at (2.22292, 1.39555) {};
\node (v16) at (2.56922, 1.41525) {};
\node (v17) at (2.69883, 1.77113) {};
\node (v18) at (3.17507, 1.39613) {};
\node (v19) at (3.73291, 2.78973) {};
\node (v20) at (3.73354, 0.00101124) {};
\node (v21) at (4.18619, 1.39573) {};
\draw (v11.center) to (v18.center);
\draw (v11.center) to (v10.center);
\draw (v19.center) to (v18.center);
\draw (v18.center) to (v21.center);
\draw (v18.center) to (v20.center);
\draw (v3.center) to (v4.center);
\draw (v4.center) to (v2.center);
\draw (v4.center) to (v1.center);
\draw (v4.center) to (v10.center);
\draw (v11.center) .. controls (v5.center)  .. (v4.center);
\draw (v11.center) .. controls (v6.center)  .. (v4.center);
\draw (v11.center) .. controls (v7.center)  .. (v10.center);
\draw (v11.center) .. controls (v15.center)  .. (v10.center);
\draw (v18.center) .. controls (v16.center)  .. (v10.center);
\draw (v18.center) .. controls (v17.center)  .. (v10.center);
\filldraw[black] (v19) circle (0.0381819);
\filldraw[black] (v18) circle (0.0381819);
\filldraw[black] (v21) circle (0.0381819);
\filldraw[black] (v20) circle (0.0381819);
\filldraw[black] (v3) circle (0.0381819);
\filldraw[black] (v4) circle (0.0381819);
\filldraw[black] (v2) circle (0.0381819);
\filldraw[black] (v1) circle (0.0381819);
\filldraw[black] (v10) circle (0.0381819);
\filldraw[red] (v12.center) -- (v8.center) -- (v9.center) -- (v13.center) -- (v14.center) -- (v12.center) -- cycle;
\end{tikzpicture}
				      & $\frac{1}{3 \ep ^3}-\frac{8}{3 \ep ^2}+\frac{8}{3 \ep }$ \\
	\nickel{eee112|2233|e33|ee|}  & 
\begin{tikzpicture}[baseline=(v4),scale=\scalefigure]
\node (v1) at (0., 1.84448) {};
\node (v2) at (0.294328, 0.851629) {};
\node (v3) at (0.567876, 2.66945) {};
\node (v4) at (1.00075, 1.64691) {};
\node (v5) at (1.4993, 1.89905) {};
\node (v6) at (1.55164, 1.55391) {};
\node (v7) at (1.95797, 1.45234) {};
\node (v8) at (2.01546, 1.81424) {};
\node (v9) at (2.02873, 1.77132) {};
\node (v10) at (2.05019, 1.80605) {};
\node (v11) at (2.05019, 1.84077) {};
\node (v12) at (2.07165, 1.77132) {};
\node (v13) at (2.08491, 1.81424) {};
\node (v14) at (2.0937, 1.11295) {};
\node (v15) at (2.18592, 1.46665) {};
\node (v16) at (2.26642, 0.) {};
\node (v17) at (2.47805, 1.57414) {};
\node (v18) at (2.54103, 1.59536) {};
\node (v19) at (2.57027, 1.92784) {};
\node (v20) at (2.67676, 1.25597) {};
\node (v21) at (3.06111, 1.71716) {};
\node (v22) at (3.76928, 2.50362) {};
\node (v23) at (4.00417, 1.38578) {};
\draw (v3.center) to (v4.center);
\draw (v4.center) to (v2.center);
\draw (v4.center) to (v1.center);
\draw (v4.center) to (v14.center);
\draw (v22.center) to (v21.center);
\draw (v21.center) to (v23.center);
\draw (v16.center) to (v14.center);
\draw (v10.center) .. controls (v6.center)  .. (v4.center);
\draw (v10.center) .. controls (v5.center)  .. (v4.center);
\draw (v10.center) .. controls (v19.center)  .. (v21.center);
\draw (v10.center) .. controls (v18.center)  .. (v21.center);
\draw (v10.center) .. controls (v15.center)  .. (v14.center);
\draw (v10.center) .. controls (v7.center)  .. (v14.center);
\draw (v21.center) .. controls (v20.center)  .. (v14.center);
\draw (v21.center) .. controls (v17.center)  .. (v14.center);
\filldraw[black] (v3) circle (0.0370323);
\filldraw[black] (v4) circle (0.0370323);
\filldraw[black] (v2) circle (0.0370323);
\filldraw[black] (v1) circle (0.0370323);
\filldraw[black] (v22) circle (0.0370323);
\filldraw[black] (v21) circle (0.0370323);
\filldraw[black] (v23) circle (0.0370323);
\filldraw[black] (v16) circle (0.0370323);
\filldraw[black] (v14) circle (0.0370323);
\filldraw[red] (v11.center) -- (v8.center) -- (v9.center) -- (v12.center) -- (v13.center) -- (v11.center) -- cycle;
\end{tikzpicture}

				       & $\frac{\pi ^2 (2 \ln2+3)}{3 \ep }-\frac{\pi ^2}{3 \ep ^2}$ \\
	\nickel{ee1123|2233|ee3|ee|} &
\begin{tikzpicture}[baseline=(v5),scale=\scalefigure]
\node (v1) at (0., 1.48671) {};
\node (v2) at (0.501194, 2.16562) {};
\node (v3) at (1.05882, 1.47388) {};
\node (v4) at (1.13355, 0.) {};
\node (v5) at (1.30476, 1.25389) {};
\node (v6) at (1.38733, 1.44282) {};
\node (v7) at (1.55205, 0.968057) {};
\node (v8) at (1.60787, 1.22884) {};
\node (v9) at (1.61757, 1.19744) {};
\node (v10) at (1.63327, 1.22284) {};
\node (v11) at (1.63327, 1.24824) {};
\node (v12) at (1.63792, 0.714801) {};
\node (v13) at (1.64897, 1.19744) {};
\node (v14) at (1.65868, 1.22884) {};
\node (v15) at (1.71914, 0.969584) {};
\node (v16) at (1.87724, 1.44564) {};
\node (v17) at (1.96186, 1.25734) {};
\node (v18) at (2.15097, 0.0045403) {};
\node (v19) at (2.20583, 1.48014) {};
\node (v20) at (2.7482, 2.17695) {};
\node (v21) at (3.26543, 1.50359) {};
\draw (v18.center) to (v12.center);
\draw (v12.center) to (v4.center);
\draw (v12.center) to (v19.center);
\draw (v12.center) to (v3.center);
\draw (v20.center) to (v19.center);
\draw (v19.center) to (v21.center);
\draw (v19.center) to (v3.center);
\draw (v2.center) to (v3.center);
\draw (v3.center) to (v1.center);
\draw (v10.center) .. controls (v15.center)  .. (v12.center);
\draw (v10.center) .. controls (v7.center)  .. (v12.center);
\draw (v10.center) .. controls (v16.center)  .. (v19.center);
\draw (v10.center) .. controls (v17.center)  .. (v19.center);
\draw (v10.center) .. controls (v5.center)  .. (v3.center);
\draw (v10.center) .. controls (v6.center)  .. (v3.center);
\filldraw[black] (v18) circle (0.0321132);
\filldraw[black] (v12) circle (0.0321132);
\filldraw[black] (v4) circle (0.0321132);
\filldraw[black] (v20) circle (0.0321132);
\filldraw[black] (v19) circle (0.0321132);
\filldraw[black] (v21) circle (0.0321132);
\filldraw[black] (v2) circle (0.0321132);
\filldraw[black] (v3) circle (0.0321132);
\filldraw[black] (v1) circle (0.0321132);
\filldraw[red] (v11.center) -- (v8.center) -- (v9.center) -- (v13.center) -- (v14.center) -- (v11.center) -- cycle;
\end{tikzpicture}
				      & $\frac{\pi ^4}{3 \ep }$ \\
	\nickel{eee112|e333|ee333||}&
\begin{tikzpicture}[baseline=(v4),scale=\scalefigure]
\node (v1) at (0., 1.34627) {};
\node (v2) at (0.268225, 1.91229) {};
\node (v3) at (0.503209, 0.) {};
\node (v4) at (0.812675, 1.01181) {};
\node (v5) at (0.931306, 0.608105) {};
\node (v6) at (0.956827, 1.40712) {};
\node (v7) at (1.07546, 1.00342) {};
\node (v8) at (1.08198, 2.07356) {};
\node (v9) at (1.34252, 0.772554) {};
\node (v10) at (1.35985, 0.496373) {};
\node (v11) at (1.73975, 0.668215) {};
\node (v12) at (1.75171, 0.629503) {};
\node (v13) at (1.77106, 0.660821) {};
\node (v14) at (1.77106, 0.692139) {};
\node (v15) at (1.79042, 0.629503) {};
\node (v16) at (1.80238, 0.668215) {};
\node (v17) at (1.89926, 1.00536) {};
\node (v18) at (2.07874, 0.862008) {};
\node (v19) at (2.20694, 1.20655) {};
\node (v20) at (2.97739, 1.61126) {};
\node (v21) at (3.11035, 0.952854) {};
\draw (v13.center) to (v19.center);
\draw (v13.center) to (v5.center);
\draw (v1.center) to (v6.center);
\draw (v6.center) to (v2.center);
\draw (v6.center) to (v8.center);
\draw (v6.center) to (v19.center);
\draw (v20.center) to (v19.center);
\draw (v19.center) to (v21.center);
\draw (v3.center) to (v5.center);
\draw (v13.center) .. controls (v17.center)  .. (v19.center);
\draw (v13.center) .. controls (v18.center)  .. (v19.center);
\draw (v13.center) .. controls (v10.center)  .. (v5.center);
\draw (v13.center) .. controls (v9.center)  .. (v5.center);
\draw (v6.center) .. controls (v7.center)  .. (v5.center);
\draw (v6.center) .. controls (v4.center)  .. (v5.center);
\filldraw[black] (v1) circle (0.0310256);
\filldraw[black] (v6) circle (0.0310256);
\filldraw[black] (v2) circle (0.0310256);
\filldraw[black] (v8) circle (0.0310256);
\filldraw[black] (v20) circle (0.0310256);
\filldraw[black] (v19) circle (0.0310256);
\filldraw[black] (v21) circle (0.0310256);
\filldraw[black] (v3) circle (0.0310256);
\filldraw[black] (v5) circle (0.0310256);
\filldraw[red] (v14.center) -- (v11.center) -- (v12.center) -- (v15.center) -- (v16.center) -- (v14.center) -- cycle;
\end{tikzpicture}

				     & $\frac{1}{3 \ep ^3}-\frac{4}{3 \ep ^2}+\frac{4 \left(\pi ^2-6\right)}{9 \ep }$ \\
 \hline
\end{tabular}
}
\end{center}
\caption{
    The first part of the six-loop logarithmically divergent diagrams in Nickel notation contributing to the \auxdia{6}{6} auxlliary graphs and their $\KRP$. We highlight the operator insertion by a red pentagon.
	These eight diagrams also contribute to the anomalous dimension of the fixed-charge operators (c.f., Fig.~3a-h of Ref.~\cite{Jack:2020wvs}). 
}
\label{tab:6-6_diagrams_1}
\end{table}
\begin{table}[H]
	\begin{center}
{\renewcommand{\arraystretch}{1.5}
\begin{tabular}{|c|c|c|}
	\hline
	Nickel index & Diagram & $(64 \pi^2)^3 \cdot \KRP$ \\
	\hline
	\nickel{eeee12|ee233|3333||} &
\begin{tikzpicture}[baseline=(v5),scale=\scalefigure]
\node (v1) at (0., 1.52627) {};
\node (v2) at (0.125143, 0.523734) {};
\node (v3) at (0.571883, 2.29912) {};
\node (v4) at (0.869328, 0.) {};
\node (v5) at (1.01123, 1.19062) {};
\node (v6) at (1.8642, 1.71787) {};
\node (v7) at (2.13277, 2.04306) {};
\node (v8) at (2.17956, 1.96887) {};
\node (v9) at (2.22657, 1.89435) {};
\node (v10) at (2.27336, 1.82017) {};
\node (v11) at (2.31171, 1.67852) {};
\node (v12) at (2.40352, 1.16617) {};
\node (v13) at (2.50336, 2.15446) {};
\node (v14) at (2.51809, 2.10678) {};
\node (v15) at (2.54193, 2.14535) {};
\node (v16) at (2.54193, 2.18393) {};
\node (v17) at (2.56577, 2.10678) {};
\node (v18) at (2.58051, 2.15446) {};
\node (v19) at (2.63374, 1.633) {};
\node (v20) at (3.04921, 0.219408) {};
\node (v21) at (3.44868, 1.30399) {};
\draw (v1.center) to (v5.center);
\draw (v5.center) to (v4.center);
\draw (v5.center) to (v3.center);
\draw (v5.center) to (v2.center);
\draw (v5.center) to (v12.center);
\draw (v5.center) to (v6.center);
\draw (v20.center) to (v12.center);
\draw (v12.center) to (v21.center);
\draw (v12.center) to (v6.center);
\draw (v15.center) .. controls (v19.center)  .. (v12.center);
\draw (v15.center) .. controls (v11.center)  .. (v12.center);
\draw (v15.center) .. controls (v10.center)  .. (v6.center);
\draw (v15.center) .. controls (v9.center)  .. (v6.center);
\draw (v15.center) .. controls (v8.center)  .. (v6.center);
\draw (v15.center) .. controls (v7.center)  .. (v6.center);
\filldraw[black] (v1) circle (0.0333729);
\filldraw[black] (v5) circle (0.0333729);
\filldraw[black] (v4) circle (0.0333729);
\filldraw[black] (v3) circle (0.0333729);
\filldraw[black] (v2) circle (0.0333729);
\filldraw[black] (v20) circle (0.0333729);
\filldraw[black] (v12) circle (0.0333729);
\filldraw[black] (v21) circle (0.0333729);
\filldraw[black] (v6) circle (0.0333729);
\filldraw[red] (v16.center) -- (v13.center) -- (v14.center) -- (v17.center) -- (v18.center) -- (v16.center) -- cycle;
\end{tikzpicture}

				      & $0$ \\
	\nickel{eeee12|ee233|3333||} &
\begin{tikzpicture}[baseline=(v5),scale=\scalefigure]
\node (v1) at (0., 1.52362) {};
\node (v2) at (0.124023, 0.521921) {};
\node (v3) at (0.570872, 2.29765) {};
\node (v4) at (0.869113, 0.) {};
\node (v5) at (1.00987, 1.18942) {};
\node (v6) at (1.82378, 1.72516) {};
\node (v7) at (1.83851, 1.67747) {};
\node (v8) at (1.86236, 1.71605) {};
\node (v9) at (1.86236, 1.75463) {};
\node (v10) at (1.8862, 1.67747) {};
\node (v11) at (1.90094, 1.72516) {};
\node (v12) at (2.1317, 2.04084) {};
\node (v13) at (2.17837, 1.96652) {};
\node (v14) at (2.22524, 1.89188) {};
\node (v15) at (2.27191, 1.81756) {};
\node (v16) at (2.31095, 1.67641) {};
\node (v17) at (2.40217, 1.16473) {};
\node (v18) at (2.54125, 2.14235) {};
\node (v19) at (2.63247, 1.63067) {};
\node (v20) at (3.05007, 0.219833) {};
\node (v21) at (3.44647, 1.30264) {};
\draw (v8.center) to (v5.center);
\draw (v8.center) to (v17.center);
\draw (v1.center) to (v5.center);
\draw (v5.center) to (v4.center);
\draw (v5.center) to (v3.center);
\draw (v5.center) to (v2.center);
\draw (v5.center) to (v17.center);
\draw (v21.center) to (v17.center);
\draw (v17.center) to (v20.center);
\draw (v8.center) .. controls (v12.center)  .. (v18.center);
\draw (v8.center) .. controls (v13.center)  .. (v18.center);
\draw (v8.center) .. controls (v14.center)  .. (v18.center);
\draw (v8.center) .. controls (v15.center)  .. (v18.center);
\draw (v17.center) .. controls (v16.center)  .. (v18.center);
\draw (v17.center) .. controls (v19.center)  .. (v18.center);
\filldraw[black] (v1) circle (0.0333579);
\filldraw[black] (v5) circle (0.0333579);
\filldraw[black] (v4) circle (0.0333579);
\filldraw[black] (v3) circle (0.0333579);
\filldraw[black] (v2) circle (0.0333579);
\filldraw[black] (v21) circle (0.0333579);
\filldraw[black] (v17) circle (0.0333579);
\filldraw[black] (v20) circle (0.0333579);
\filldraw[black] (v18) circle (0.0333579);
\filldraw[red] (v9.center) -- (v6.center) -- (v7.center) -- (v10.center) -- (v11.center) -- (v9.center) -- cycle;
\end{tikzpicture}

				      & $0$ \\
	\nickel{eeee12|22233|33|ee|} & 
\begin{tikzpicture}[baseline=(v5),scale=\scalefigure]
\node (v1) at (0., 1.88512) {};
\node (v2) at (0.00165753, 0.724213) {};
\node (v3) at (0.691261, 2.6098) {};
\node (v4) at (0.692049, 0.) {};
\node (v5) at (0.93826, 1.30504) {};
\node (v6) at (1.9547, 1.30546) {};
\node (v7) at (2.02577, 0.999893) {};
\node (v8) at (2.03771, 0.961232) {};
\node (v9) at (2.05704, 0.99251) {};
\node (v10) at (2.05704, 1.02379) {};
\node (v11) at (2.0581, 1.61807) {};
\node (v12) at (2.07637, 0.961232) {};
\node (v13) at (2.08832, 0.999893) {};
\node (v14) at (2.16044, 1.30512) {};
\node (v15) at (2.51593, 1.31657) {};
\node (v16) at (2.51636, 1.29389) {};
\node (v17) at (2.6187, 0.980937) {};
\node (v18) at (2.61932, 1.62918) {};
\node (v19) at (3.07759, 1.305) {};
\node (v20) at (3.91407, 0.61185) {};
\node (v21) at (3.9147, 1.99802) {};
\draw (v9.center) to (v5.center);
\draw (v9.center) to (v11.center);
\draw (v1.center) to (v5.center);
\draw (v5.center) to (v3.center);
\draw (v5.center) to (v4.center);
\draw (v5.center) to (v2.center);
\draw (v5.center) to (v11.center);
\draw (v20.center) to (v19.center);
\draw (v19.center) to (v21.center);
\draw (v9.center) .. controls (v15.center)  .. (v19.center);
\draw (v9.center) .. controls (v17.center)  .. (v19.center);
\draw (v9.center) .. controls (v6.center)  .. (v11.center);
\draw (v9.center) .. controls (v14.center)  .. (v11.center);
\draw (v19.center) .. controls (v16.center)  .. (v11.center);
\draw (v19.center) .. controls (v18.center)  .. (v11.center);
\filldraw[black] (v1) circle (0.0364585);
\filldraw[black] (v5) circle (0.0364585);
\filldraw[black] (v3) circle (0.0364585);
\filldraw[black] (v4) circle (0.0364585);
\filldraw[black] (v2) circle (0.0364585);
\filldraw[black] (v20) circle (0.0364585);
\filldraw[black] (v19) circle (0.0364585);
\filldraw[black] (v21) circle (0.0364585);
\filldraw[black] (v11) circle (0.0364585);
\filldraw[red] (v10.center) -- (v7.center) -- (v8.center) -- (v12.center) -- (v13.center) -- (v10.center) -- cycle;
\end{tikzpicture}

				      & $-\frac{2 \pi ^2}{3 \ep }$ \\
	\nickel{eeee12|e2333|e333||} &
\begin{tikzpicture}[baseline=(v5),scale=\scalefigure]
\node (v1) at (0., 1.23723) {};
\node (v2) at (0.00149711, 0.613326) {};
\node (v3) at (0.655117, 0.210604) {};
\node (v4) at (0.655231, 1.63912) {};
\node (v5) at (0.953793, 0.92509) {};
\node (v6) at (2.08462, 0.628831) {};
\node (v7) at (2.08493, 1.22098) {};
\node (v8) at (2.38113, 0.890438) {};
\node (v9) at (2.38131, 0.959492) {};
\node (v10) at (2.47853, 0.663369) {};
\node (v11) at (2.47866, 1.18646) {};
\node (v12) at (2.67516, 0.) {};
\node (v13) at (2.67629, 1.85003) {};
\node (v14) at (2.74544, 0.931965) {};
\node (v15) at (2.75675, 0.895368) {};
\node (v16) at (2.77505, 0.924976) {};
\node (v17) at (2.77505, 0.954583) {};
\node (v18) at (2.79334, 0.895368) {};
\node (v19) at (2.80465, 0.931965) {};
\draw (v16.center) to (v6.center);
\draw (v16.center) to (v7.center);
\draw (v2.center) to (v5.center);
\draw (v5.center) to (v1.center);
\draw (v5.center) to (v3.center);
\draw (v5.center) to (v4.center);
\draw (v5.center) to (v6.center);
\draw (v5.center) to (v7.center);
\draw (v12.center) to (v6.center);
\draw (v6.center) to (v7.center);
\draw (v13.center) to (v7.center);
\draw (v16.center) .. controls (v10.center)  .. (v6.center);
\draw (v16.center) .. controls (v8.center)  .. (v6.center);
\draw (v16.center) .. controls (v9.center)  .. (v7.center);
\draw (v16.center) .. controls (v11.center)  .. (v7.center);
\filldraw[black] (v2) circle (0.0286043);
\filldraw[black] (v5) circle (0.0286043);
\filldraw[black] (v1) circle (0.0286043);
\filldraw[black] (v3) circle (0.0286043);
\filldraw[black] (v4) circle (0.0286043);
\filldraw[black] (v12) circle (0.0286043);
\filldraw[black] (v6) circle (0.0286043);
\filldraw[black] (v13) circle (0.0286043);
\filldraw[black] (v7) circle (0.0286043);
\filldraw[red] (v17.center) -- (v14.center) -- (v15.center) -- (v18.center) -- (v19.center) -- (v17.center) -- cycle;
\end{tikzpicture}
				      & $\frac{2}{3 \ep ^2}-\frac{8}{3 \ep }$ \\
	\nickel{eeee12|e2233|333|e|} & 
\begin{tikzpicture}[baseline=(v5),scale=\scalefigure]
\node (v1) at (0., 0.510372) {};
\node (v2) at (0.926419, 0.891235) {};
\node (v3) at (1.17989, 2.593) {};
\node (v4) at (1.20383, 1.36971) {};
\node (v5) at (1.39378, 1.04908) {};
\node (v6) at (1.39624, 0.740862) {};
\node (v7) at (1.43368, 1.11165) {};
\node (v8) at (1.67364, 1.21933) {};
\node (v9) at (1.71108, 1.59012) {};
\node (v10) at (1.82819, 0.907066) {};
\node (v11) at (1.84172, 0.863307) {};
\node (v12) at (1.86359, 0.898709) {};
\node (v13) at (1.86359, 0.93411) {};
\node (v14) at (1.88547, 0.863307) {};
\node (v15) at (1.899, 0.907066) {};
\node (v16) at (1.90103, 1.26949) {};
\node (v17) at (2.91316, 1.1598) {};
\node (v18) at (3.10509, 0.) {};
\node (v19) at (3.26216, 2.25889) {};
\node (v20) at (3.81963, 0.568954) {};
\node (v21) at (3.8895, 1.56368) {};
\draw (v12.center) to (v17.center);
\draw (v12.center) to (v2.center);
\draw (v18.center) to (v17.center);
\draw (v17.center) to (v19.center);
\draw (v17.center) to (v20.center);
\draw (v17.center) to (v21.center);
\draw (v17.center) to (v9.center);
\draw (v3.center) to (v9.center);
\draw (v1.center) to (v2.center);
\draw (v12.center) .. controls (v8.center)  .. (v9.center);
\draw (v12.center) .. controls (v16.center)  .. (v9.center);
\draw (v12.center) .. controls (v6.center)  .. (v2.center);
\draw (v12.center) .. controls (v5.center)  .. (v2.center);
\draw (v9.center) .. controls (v7.center)  .. (v2.center);
\draw (v9.center) .. controls (v4.center)  .. (v2.center);
\filldraw[black] (v18) circle (0.0362958);
\filldraw[black] (v17) circle (0.0362958);
\filldraw[black] (v19) circle (0.0362958);
\filldraw[black] (v20) circle (0.0362958);
\filldraw[black] (v21) circle (0.0362958);
\filldraw[black] (v3) circle (0.0362958);
\filldraw[black] (v9) circle (0.0362958);
\filldraw[black] (v1) circle (0.0362958);
\filldraw[black] (v2) circle (0.0362958);
\filldraw[red] (v13.center) -- (v10.center) -- (v11.center) -- (v14.center) -- (v15.center) -- (v13.center) -- cycle;
\end{tikzpicture}
				      & $\frac{1}{3 \ep ^2}-\frac{4}{\ep }$ \\
	\nickel{eee112|eee3|33333||} &
\begin{tikzpicture}[baseline=(v4),scale=\scalefigure]
\node (v1) at (0., 1.44352) {};
\node (v2) at (0.135198, 0.345212) {};
\node (v3) at (0.621651, 2.22002) {};
\node (v4) at (0.957524, 1.0654) {};
\node (v5) at (1.24825, 0.00980982) {};
\node (v6) at (1.2627, -0.0369609) {};
\node (v7) at (1.28609, 0.000877416) {};
\node (v8) at (1.28609, 0.0387157) {};
\node (v9) at (1.30947, -0.0369609) {};
\node (v10) at (1.32392, 0.00980982) {};
\node (v11) at (1.66432, -0.124004) {};
\node (v12) at (1.6644, -0.0619402) {};
\node (v13) at (1.66454, 0.0628176) {};
\node (v14) at (1.66461, 0.124882) {};
\node (v15) at (1.66467, 0.832715) {};
\node (v16) at (1.66474, 1.29787) {};
\node (v17) at (2.04285, 0.) {};
\node (v18) at (2.37188, 1.06519) {};
\node (v19) at (2.70674, 2.21982) {};
\node (v20) at (3.1941, 0.345284) {};
\node (v21) at (3.33003, 1.44411) {};
\draw (v7.center) to (v4.center);
\draw (v7.center) to (v17.center);
\draw (v20.center) to (v18.center);
\draw (v18.center) to (v19.center);
\draw (v18.center) to (v21.center);
\draw (v18.center) to (v17.center);
\draw (v2.center) to (v4.center);
\draw (v4.center) to (v3.center);
\draw (v4.center) to (v1.center);
\draw (v7.center) .. controls (v14.center)  .. (v17.center);
\draw (v7.center) .. controls (v13.center)  .. (v17.center);
\draw (v7.center) .. controls (v12.center)  .. (v17.center);
\draw (v7.center) .. controls (v11.center)  .. (v17.center);
\draw (v18.center) .. controls (v15.center)  .. (v4.center);
\draw (v18.center) .. controls (v16.center)  .. (v4.center);
\filldraw[black] (v20) circle (0.0325604);
\filldraw[black] (v18) circle (0.0325604);
\filldraw[black] (v19) circle (0.0325604);
\filldraw[black] (v21) circle (0.0325604);
\filldraw[black] (v2) circle (0.0325604);
\filldraw[black] (v4) circle (0.0325604);
\filldraw[black] (v3) circle (0.0325604);
\filldraw[black] (v1) circle (0.0325604);
\filldraw[black] (v17) circle (0.0325604);
\filldraw[red] (v8.center) -- (v5.center) -- (v6.center) -- (v9.center) -- (v10.center) -- (v8.center) -- cycle;
\end{tikzpicture}

				      & $\frac{2}{9 \ep ^2}-\frac{40}{27 \ep }$ \\
	\nickel{eee112|3333|eee33||} &
\begin{tikzpicture}[baseline=(v4),scale=\scalefigure]
\node (v1) at (0., 1.44352) {};
\node (v2) at (0.135198, 0.345212) {};
\node (v3) at (0.621651, 2.22002) {};
\node (v4) at (0.946754, 0.479108) {};
\node (v5) at (0.957524, 1.0654) {};
\node (v6) at (1.24825, 0.00980982) {};
\node (v7) at (1.2627, -0.0369609) {};
\node (v8) at (1.28609, 0.000877416) {};
\node (v9) at (1.28609, 0.0387157) {};
\node (v10) at (1.29686, 0.587166) {};
\node (v11) at (1.30947, -0.0369609) {};
\node (v12) at (1.32392, 0.00980982) {};
\node (v13) at (1.66432, -0.124004) {};
\node (v14) at (1.66442, -0.0411672) {};
\node (v15) at (1.66452, 0.0420447) {};
\node (v16) at (1.66461, 0.124882) {};
\node (v17) at (2.03221, 0.5867) {};
\node (v18) at (2.04285, 0.) {};
\node (v19) at (2.37188, 1.06519) {};
\node (v20) at (2.38253, 0.478488) {};
\node (v21) at (2.70674, 2.21982) {};
\node (v22) at (3.1941, 0.345284) {};
\node (v23) at (3.33003, 1.44411) {};
\draw (v22.center) to (v19.center);
\draw (v19.center) to (v21.center);
\draw (v19.center) to (v23.center);
\draw (v19.center) to (v5.center);
\draw (v2.center) to (v5.center);
\draw (v5.center) to (v3.center);
\draw (v5.center) to (v1.center);
\draw (v8.center) .. controls (v4.center)  .. (v5.center);
\draw (v8.center) .. controls (v10.center)  .. (v5.center);
\draw (v8.center) .. controls (v16.center)  .. (v18.center);
\draw (v8.center) .. controls (v15.center)  .. (v18.center);
\draw (v8.center) .. controls (v14.center)  .. (v18.center);
\draw (v8.center) .. controls (v13.center)  .. (v18.center);
\draw (v19.center) .. controls (v20.center)  .. (v18.center);
\draw (v19.center) .. controls (v17.center)  .. (v18.center);
\filldraw[black] (v22) circle (0.0325604);
\filldraw[black] (v19) circle (0.0325604);
\filldraw[black] (v21) circle (0.0325604);
\filldraw[black] (v23) circle (0.0325604);
\filldraw[black] (v2) circle (0.0325604);
\filldraw[black] (v5) circle (0.0325604);
\filldraw[black] (v3) circle (0.0325604);
\filldraw[black] (v1) circle (0.0325604);
\filldraw[black] (v18) circle (0.0325604);
\filldraw[red] (v9.center) -- (v6.center) -- (v7.center) -- (v11.center) -- (v12.center) -- (v9.center) -- cycle;
\end{tikzpicture}

				      & $-\frac{2 \pi ^2}{3 \ep }$ \\
	\nickel{eee123|eee23|3333||} &
\begin{tikzpicture}[baseline=(v4),scale=\scalefigure]
\node (v1) at (0., 1.21099) {};
\node (v2) at (0.442187, 0.000834941) {};
\node (v3) at (0.442917, 2.42077) {};
\node (v4) at (1.09244, 1.21025) {};
\node (v5) at (1.67442, 1.20883) {};
\node (v6) at (1.76883, 1.20871) {};
\node (v7) at (1.77364, 0.787316) {};
\node (v8) at (1.78971, 0.735311) {};
\node (v9) at (1.81572, 0.777384) {};
\node (v10) at (1.81572, 0.819457) {};
\node (v11) at (1.81679, 1.63991) {};
\node (v11p1) at (1.21679, 2.03991) {};
\node (v11p2) at (2.41679, 2.03991) {};
\node (v12) at (1.84172, 0.735311) {};
\node (v13) at (1.85779, 0.787316) {};
\node (v14) at (1.86368, 1.20859) {};
\node (v15) at (1.95809, 1.20847) {};
\node (v16) at (2.54031, 1.21028) {};
\node (v17) at (3.19074, 2.42128) {};
\node (v18) at (3.19108, 0.) {};
\node (v19) at (3.63192, 1.21071) {};
\draw (v9.center) to (v16.center);
\draw (v9.center) to (v4.center);
\draw (v19.center) to (v16.center);
\draw (v16.center) to (v17.center);
\draw (v16.center) to (v18.center);
\draw (v16.center) .. controls (v11p2.center) and (v11p1.center) .. (v4.center);
\draw (v16.center) to (v11.center);
\draw (v2.center) to (v4.center);
\draw (v4.center) to (v3.center);
\draw (v4.center) to (v1.center);
\draw (v4.center) to (v11.center);
\draw (v9.center) .. controls (v5.center)  .. (v11.center);
\draw (v9.center) .. controls (v6.center)  .. (v11.center);
\draw (v9.center) .. controls (v14.center)  .. (v11.center);
\draw (v9.center) .. controls (v15.center)  .. (v11.center);
\filldraw[black] (v19) circle (0.034606);
\filldraw[black] (v16) circle (0.034606);
\filldraw[black] (v17) circle (0.034606);
\filldraw[black] (v18) circle (0.034606);
\filldraw[black] (v2) circle (0.034606);
\filldraw[black] (v4) circle (0.034606);
\filldraw[black] (v3) circle (0.034606);
\filldraw[black] (v1) circle (0.034606);
\filldraw[black] (v11) circle (0.034606);
\filldraw[red] (v10.center) -- (v7.center) -- (v8.center) -- (v12.center) -- (v13.center) -- (v10.center) -- cycle;
\end{tikzpicture}

				      & $\frac{4}{3 \ep ^2}-\frac{16}{3 \ep }$ \\
	\nickel{eee112|ee33|e3333||} &
\begin{tikzpicture}[baseline=(v4),scale=\scalefigure]
\node (v1) at (0., 1.34627) {};
\node (v2) at (0.268225, 1.91229) {};
\node (v3) at (0.503209, 0.) {};
\node (v4) at (0.931306, 0.608105) {};
\node (v5) at (0.956827, 1.40712) {};
\node (v6) at (1.08198, 2.07356) {};
\node (v7) at (1.34252, 0.772554) {};
\node (v8) at (1.34829, 0.680632) {};
\node (v9) at (1.35408, 0.588295) {};
\node (v10) at (1.35985, 0.496373) {};
\node (v11) at (1.5489, 1.10127) {};
\node (v12) at (1.61487, 1.5124) {};
\node (v13) at (1.73975, 0.668215) {};
\node (v14) at (1.75171, 0.629503) {};
\node (v15) at (1.77106, 0.660821) {};
\node (v16) at (1.77106, 0.692139) {};
\node (v17) at (1.79042, 0.629503) {};
\node (v18) at (1.80238, 0.668215) {};
\node (v19) at (1.89926, 1.00536) {};
\node (v20) at (2.07874, 0.862008) {};
\node (v21) at (2.20694, 1.20655) {};
\node (v22) at (2.97739, 1.61126) {};
\node (v23) at (3.11035, 0.952854) {};
\draw (v1.center) to (v5.center);
\draw (v5.center) to (v2.center);
\draw (v5.center) to (v6.center);
\draw (v5.center) to (v4.center);
\draw (v22.center) to (v21.center);
\draw (v21.center) to (v23.center);
\draw (v3.center) to (v4.center);
\draw (v15.center) .. controls (v19.center)  .. (v21.center);
\draw (v15.center) .. controls (v20.center)  .. (v21.center);
\draw (v15.center) .. controls (v10.center)  .. (v4.center);
\draw (v15.center) .. controls (v9.center)  .. (v4.center);
\draw (v15.center) .. controls (v8.center)  .. (v4.center);
\draw (v15.center) .. controls (v7.center)  .. (v4.center);
\draw (v5.center) .. controls (v12.center)  .. (v21.center);
\draw (v5.center) .. controls (v11.center)  .. (v21.center);
\filldraw[black] (v1) circle (0.0310256);
\filldraw[black] (v5) circle (0.0310256);
\filldraw[black] (v2) circle (0.0310256);
\filldraw[black] (v6) circle (0.0310256);
\filldraw[black] (v22) circle (0.0310256);
\filldraw[black] (v21) circle (0.0310256);
\filldraw[black] (v23) circle (0.0310256);
\filldraw[black] (v3) circle (0.0310256);
\filldraw[black] (v4) circle (0.0310256);
\filldraw[red] (v16.center) -- (v13.center) -- (v14.center) -- (v17.center) -- (v18.center) -- (v16.center) -- cycle;
\end{tikzpicture}

				      & $-\frac{2 \pi ^2}{3 \ep }$ \\
	\nickel{eee112|ee23|3333|e|} &
\begin{tikzpicture}[baseline=(v4),scale=\scalefigure]
\node (v1) at (0., 0.90078) {};
\node (v2) at (0.329825, 1.72985) {};
\node (v3) at (0.522585, 0.17061) {};
\node (v4) at (1.01481, 1.02383) {};
\node (v5) at (1.55841, 1.42673) {};
\node (v6) at (1.69144, 1.02541) {};
\node (v7) at (1.92742, 0.78916) {};
\node (v8) at (1.94084, 0.745746) {};
\node (v9) at (1.96254, 0.780868) {};
\node (v10) at (1.96254, 0.815991) {};
\node (v11) at (1.98425, 0.745746) {};
\node (v12) at (1.99767, 0.78916) {};
\node (v13) at (2.23505, 1.42831) {};
\node (v14) at (2.31106, 2.40978) {};
\node (v15) at (2.36434, 0.542481) {};
\node (v16) at (2.38533, 0.637347) {};
\node (v17) at (2.40642, 0.732643) {};
\node (v18) at (2.42741, 0.82751) {};
\node (v19) at (2.82921, 0.589122) {};
\node (v20) at (3.11116, 1.92543) {};
\node (v21) at (3.61468, 0.) {};
\draw (v9.center) to (v4.center);
\draw (v9.center) to (v13.center);
\draw (v3.center) to (v4.center);
\draw (v4.center) to (v2.center);
\draw (v4.center) to (v1.center);
\draw (v20.center) to (v13.center);
\draw (v13.center) to (v14.center);
\draw (v13.center) to (v19.center);
\draw (v21.center) to (v19.center);
\draw (v9.center) .. controls (v18.center)  .. (v19.center);
\draw (v9.center) .. controls (v17.center)  .. (v19.center);
\draw (v9.center) .. controls (v16.center)  .. (v19.center);
\draw (v9.center) .. controls (v15.center)  .. (v19.center);
\draw (v4.center) .. controls (v5.center)  .. (v13.center);
\draw (v4.center) .. controls (v6.center)  .. (v13.center);
\filldraw[black] (v3) circle (0.034491);
\filldraw[black] (v4) circle (0.034491);
\filldraw[black] (v2) circle (0.034491);
\filldraw[black] (v1) circle (0.034491);
\filldraw[black] (v20) circle (0.034491);
\filldraw[black] (v13) circle (0.034491);
\filldraw[black] (v14) circle (0.034491);
\filldraw[black] (v21) circle (0.034491);
\filldraw[black] (v19) circle (0.034491);
\filldraw[red] (v10.center) -- (v7.center) -- (v8.center) -- (v11.center) -- (v12.center) -- (v10.center) -- cycle;
\end{tikzpicture}
				      & $\frac{2}{3 \ep ^2}-\frac{16}{3 \ep }$ \\
 \hline
\end{tabular}
}
\end{center}
\caption{
    The second part of the six-loop logarithmically divergent diagrams in Nickel notation contributing to the \auxdia{6}{6} auxiliary graphs and their $\KRP$. We highlight the operator insertion by a red pentagon.
    These diagrams do not appear in the case of fixed-charge operators \cite{Jack:2020wvs} due to vanishing contribution.
}
\label{tab:6-6_diagrams_2}
\end{table}
\def\scalefigure{0.5}
\begin{table}[H]
	\begin{center}
{\renewcommand{\arraystretch}{1.5}
\begin{tabular}{|l|c|c|}
	\hline
	Nickel notation & Diagram & $(64 \pi^2)^3 \cdot \KRP$ \\
	\hline
\nickel{eee112|e22222||:}&
	\multirow{2}{*}{\begin{tikzpicture}[baseline=(v4),scale=\scalefigure]
\node (v1) at (0., 1.07468) {};
\node (v2) at (0.419117, 0.) {};
\node (v3) at (0.837605, 1.86404) {};
\node (v4) at (1.02231, 0.85351) {};
\node (v5) at (1.61936, 1.05128) {};
\node (v6) at (1.6202, 0.658282) {};
\node (v7) at (1.74178, 0.326944) {};
\node (v8) at (1.89251, 0.669684) {};
\node (v9) at (1.93591, 0.630689) {};
\node (v10) at (2.02313, 0.552304) {};
\node (v11) at (2.06653, 0.513309) {};
\node (v12) at (2.18169, 0.864446) {};
\node (v13) at (2.19527, 0.820482) {};
\node (v14) at (2.21726, 0.85605) {};
\node (v15) at (2.21726, 0.891617) {};
\node (v16) at (2.23924, 0.820482) {};
\node (v17) at (2.25282, 0.864446) {};
\node (v18) at (3.22503, 0.973987) {};
\draw (v18.center) to (v14.center);
\draw (v14.center) to (v7.center);
\draw (v2.center) to (v4.center);
\draw (v4.center) to (v3.center);
\draw (v4.center) to (v1.center);
\draw (v4.center) to (v7.center);
\draw (v14.center) .. controls (v6.center)  .. (v4.center);
\draw (v14.center) .. controls (v5.center)  .. (v4.center);
\draw (v14.center) .. controls (v11.center)  .. (v7.center);
\draw (v14.center) .. controls (v10.center)  .. (v7.center);
\draw (v14.center) .. controls (v9.center)  .. (v7.center);
\draw (v14.center) .. controls (v8.center)  .. (v7.center);
\filldraw[black] (v18) circle (0.0318318);
\filldraw[black] (v2) circle (0.0318318);
\filldraw[black] (v4) circle (0.0318318);
\filldraw[black] (v3) circle (0.0318318);
\filldraw[black] (v1) circle (0.0318318);
\filldraw[black] (v7) circle (0.0318318);
\filldraw[red] (v15.center) -- (v12.center) -- (v13.center) -- (v16.center) -- (v17.center) -- (v15.center) -- cycle;
\end{tikzpicture}
}
				      & \multirow{2}{*}{$\frac{4 \left(q_1+q_2+q_3\right){}^2}{27 \ep }$} \\
		\nickel{1\_2\_3\_0\_0\_0|-1\_0\_0\_0\_0\_0||} & &\\[2mm] 
\nickel{ee1112|e2222|e|:} &
	\multirow{2}{*}{\begin{tikzpicture}[baseline=(v4),scale=\scalefigure]
\node (v1) at (0., 0.62643) {};
\node (v2) at (0.00311552, 1.83421) {};
\node (v3) at (0.810474, 1.22901) {};
\node (v4) at (1.20412, 0.850026) {};
\node (v5) at (1.35223, 1.15766) {};
\node (v6) at (1.59934, 1.22812) {};
\node (v7) at (1.69765, 1.22786) {};
\node (v8) at (1.70096, 0.789272) {};
\node (v9) at (1.71811, 0.733764) {};
\node (v10) at (1.74587, 0.778671) {};
\node (v11) at (1.74587, 0.823578) {};
\node (v12) at (1.74818, 1.6768) {};
\node (v13) at (1.77362, 0.733764) {};
\node (v14) at (1.79077, 0.789272) {};
\node (v15) at (1.7964, 1.22761) {};
\node (v16) at (1.89471, 1.22736) {};
\node (v17) at (2.37196, 0.) {};
\node (v18) at (2.37647, 2.45439) {};
\draw (v17.center) to (v10.center);
\draw (v10.center) to (v3.center);
\draw (v1.center) to (v3.center);
\draw (v3.center) to (v2.center);
\draw (v3.center) to (v12.center);
\draw (v18.center) to (v12.center);
\draw (v10.center) .. controls (v4.center)  .. (v3.center);
\draw (v10.center) .. controls (v5.center)  .. (v3.center);
\draw (v10.center) .. controls (v6.center)  .. (v12.center);
\draw (v10.center) .. controls (v7.center)  .. (v12.center);
\draw (v10.center) .. controls (v15.center)  .. (v12.center);
\draw (v10.center) .. controls (v16.center)  .. (v12.center);
\filldraw[black] (v17) circle (0.0261947);
\filldraw[black] (v1) circle (0.0261947);
\filldraw[black] (v3) circle (0.0261947);
\filldraw[black] (v2) circle (0.0261947);
\filldraw[black] (v18) circle (0.0261947);
\filldraw[black] (v12) circle (0.0261947);
\filldraw[red] (v11.center) -- (v8.center) -- (v9.center) -- (v13.center) -- (v14.center) -- (v11.center) -- cycle;
\end{tikzpicture}
}
					& \multirow{2}{*}{$\frac{q_3^2}{9 \ep ^2}+\frac{4 \left[3 (q_1+q_2)^2-7 q_3^2-6 (q_1+q_2) q_3\right]}{27 \ep }$} \\
	\nickel{1\_2\_0\_0\_0\_0|-1\_0\_0\_0\_0|3|} & & \\[3mm] 
 \hline
\end{tabular}
}
\end{center}
\caption{Six-loop quadratically divergent diagrams in Nickel notation \cite{Batkovich:2014bla} contributing to the \auxdia{7}{3} auxiliary graphs and their $\KRP$. 
	The internal lines are denoted as \nickel{0}, the external leg of the operator as \nickel{-1} and an external leg with $q_i$ momentum as $i=\nickel{1,2,3}$.
} \label{tab:7-3_diagrams}
\end{table}
\begin{table}[H]
	\begin{center}
{\renewcommand{\arraystretch}{1.5}
\begin{tabular}{|l|c|c|}
	\hline
	Nickel notation & Diagram & $(64 \pi^2)^3 \cdot \KRP$ \\
	\hline
    \nickel{ee1122|e2222||:} &
    \multirow{2}{*}{\begin{tikzpicture}[baseline=(v3),scale=0.6]
\node (v1) at (0., 0.787969) {};
\node (v2) at (0.978238, 0.611878) {};
\node (v3) at (0.993171, 0.563555) {};
\node (v4) at (1.01733, 0.602649) {};
\node (v5) at (1.01733, 0.641743) {};
\node (v6) at (1.04149, 0.563555) {};
\node (v7) at (1.05643, 0.611878) {};
\node (v8) at (1.1673, 0.219409) {};
\node (v9) at (1.23327, 0.273937) {};
\node (v10) at (1.29954, 0.328712) {};
\node (v11) at (1.3655, 0.38324) {};
\node (v12) at (1.51548, 0.) {};
\node (v13) at (1.59376, 0.409396) {};
\node (v14) at (1.59594, 0.789267) {};
\node (v15) at (1.74592, 0.406028) {};
\node (v16) at (1.94193, 0.189987) {};
\node (v17) at (2.17237, 0.596015) {};
\node (v18) at (2.77051, 1.45058) {};
\node (v19) at (3.10162, 0.0991225) {};
\draw (v1.center) to (v4.center);
\draw (v18.center) to (v17.center);
\draw (v17.center) to (v19.center);
\draw (v4.center) .. controls (v14.center)  .. (v17.center);
\draw (v4.center) .. controls (v13.center)  .. (v17.center);
\draw (v4.center) .. controls (v11.center)  .. (v12.center);
\draw (v4.center) .. controls (v10.center)  .. (v12.center);
\draw (v4.center) .. controls (v9.center)  .. (v12.center);
\draw (v4.center) .. controls (v8.center)  .. (v12.center);
\draw (v17.center) .. controls (v16.center)  .. (v12.center);
\draw (v17.center) .. controls (v15.center)  .. (v12.center);
\filldraw[black] (v1) circle (0.0309638);
\filldraw[black] (v18) circle (0.0309638);
\filldraw[black] (v17) circle (0.0309638);
\filldraw[black] (v19) circle (0.0309638);
\filldraw[black] (v12) circle (0.0309638);
\filldraw[red] (v5.center) -- (v2.center) -- (v3.center) -- (v6.center) -- (v7.center) -- (v5.center) -- cycle;
\end{tikzpicture}}

							      & \multirow{2}{*}{$0$} \\
    \nickel{1\_2\_0\_0\_0\_0|-1\_0\_0\_0\_0||} & & \\[2mm]
    \nickel{e11122|e222|e|:} &
    \multirow{2}{*}{
	\begin{tikzpicture}[baseline=(v3),scale=0.6]
\node (v1) at (0., 2.39764) {};
\node (v2) at (0.878215, 1.87385) {};
\node (v3) at (0.97698, 1.36913) {};
\node (v4) at (1.25729, 1.52628) {};
\node (v5) at (1.30722, 1.03308) {};
\node (v6) at (1.32587, 0.972726) {};
\node (v7) at (1.34373, 0.3) {};
\node (v8) at (1.35605, 1.02155) {};
\node (v9) at (1.35605, 1.07038) {};
\node (v10) at (1.36479, 1.70646) {};
\node (v11) at (1.36916, 2.02795) {};
\node (v12) at (1.38623, 0.972726) {};
\node (v13) at (1.40488, 1.03308) {};
\node (v14) at (1.46792, 1.52322) {};
\node (v15) at (1.74386, 1.35889) {};
\node (v16) at (1.85573, 1.86056) {};
\node (v17) at (2.7496, 2.35814) {};
\draw (v7.center) to (v8.center);
\draw (v8.center) to (v16.center);
\draw (v8.center) to (v2.center);
\draw (v17.center) to (v16.center);
\draw (v1.center) to (v2.center);
\draw (v8.center) .. controls (v14.center)  .. (v16.center);
\draw (v8.center) .. controls (v15.center)  .. (v16.center);
\draw (v8.center) .. controls (v3.center)  .. (v2.center);
\draw (v8.center) .. controls (v4.center)  .. (v2.center);
\draw (v16.center) .. controls (v10.center)  .. (v2.center);
\draw (v16.center) .. controls (v11.center)  .. (v2.center);
\filldraw[black] (v7) circle (0.0284165);
\filldraw[black] (v17) circle (0.0284165);
\filldraw[black] (v16) circle (0.0284165);
\filldraw[black] (v1) circle (0.0284165);
\filldraw[black] (v2) circle (0.0284165);
\filldraw[red] (v9.center) -- (v5.center) -- (v6.center) -- (v12.center) -- (v13.center) -- (v9.center) -- cycle;
\end{tikzpicture}
}

				    & \multirow{2}{*}{$\frac{q_1^2+q_2^2}{9 \ep ^2}-\frac{4 \left(7 q_1^2+12 q_2 q_1+7 q_2^2\right)}{27 \ep }$} \\
		\nickel{1\_0\_0\_0\_0\_0|-1\_0\_0\_0|2|} && \\[3mm]
 \hline
\end{tabular}
}
\end{center}
\caption{
Six-loop quadratically divergent diagrams in Nickel notation \cite{Batkovich:2014bla} contributing to the \auxdia{6}{2} auxiliary graphs and their $\KRP$. 
The internal lines are denoted as \nickel{0}, the external leg of the operator as \nickel{-1} and an external leg with $q_i$ momentum as $i=\nickel{1,2}$.
}\label{tab:6-2_diagrams}
\end{table}


\begin{thebibliography}{10}
\expandafter\ifx\csname url\endcsname\relax
  \def\url#1{\texttt{#1}}\fi
\expandafter\ifx\csname urlprefix\endcsname\relax\def\urlprefix{URL }\fi
\expandafter\ifx\csname href\endcsname\relax
  \def\href#1#2{#2} \def\path#1{#1}\fi

\bibitem{Antipin:2025ekk}
O.~Antipin, J.~Bersini, F.~Sannino, {Exact results for scaling dimensions of
  neutral operators in scalar conformal field theories}, Phys. Rev. D 111~(4)
  (2025) L041701.
\newblock \href {http://arxiv.org/abs/2408.01414} {\path{arXiv:2408.01414}},
  \href {https://doi.org/10.1103/PhysRevD.111.L041701}
  {\path{doi:10.1103/PhysRevD.111.L041701}}.

\bibitem{Isidori:2023pyp}
G.~Isidori, F.~Wilsch, D.~Wyler, {The standard model effective field theory at
  work}, Rev. Mod. Phys. 96~(1) (2024) 015006.
\newblock \href {http://arxiv.org/abs/2303.16922} {\path{arXiv:2303.16922}},
  \href {https://doi.org/10.1103/RevModPhys.96.015006}
  {\path{doi:10.1103/RevModPhys.96.015006}}.

\bibitem{Henriksson:2025vyi}
J.~Henriksson, S.~R. Kousvos, J.~Roosmale~Nepveu, {EFT meets CFT: Multiloop
  renormalization of higher-dimensional operators in general $\phi^4$ theories}
  (11 2025).
\newblock \href {http://arxiv.org/abs/2511.16740} {\path{arXiv:2511.16740}}.

\bibitem{Pelissetto:2000ek}
A.~Pelissetto, E.~Vicari, {Critical phenomena and renormalization group
  theory}, Phys. Rept. 368 (2002) 549--727.
\newblock \href {http://arxiv.org/abs/cond-mat/0012164}
  {\path{arXiv:cond-mat/0012164}}, \href
  {https://doi.org/10.1016/S0370-1573(02)00219-3}
  {\path{doi:10.1016/S0370-1573(02)00219-3}}.

\bibitem{Vasiliev:2006}
A.~N. Vasili'ev, The Field Theoretic Renormalization Group in Critical Behavior
  Theory and Stochastic Dynamics, Chapman and Hall/CRC, London, 2004,
  originally published in Russia in 1998 by St. Petersburg Institute of Nuclear
  Physics Press; translated by Patricia A. de Forcrand-Millard. (see 1.14,
  1.16, 4.20--4.23).

\bibitem{Wilson:1971dc}
K.~G. Wilson, M.~E. Fisher, {Critical exponents in 3.99 dimensions}, Phys. Rev.
  Lett. 28 (1972) 240--243.
\newblock \href {https://doi.org/10.1103/PhysRevLett.28.240}
  {\path{doi:10.1103/PhysRevLett.28.240}}.

\bibitem{Alvarez-Gaume:2016vff}
L.~Alvarez-Gaume, O.~Loukas, D.~Orlando, S.~Reffert, {Compensating strong
  coupling with large charge}, JHEP 04 (2017) 059.
\newblock \href {http://arxiv.org/abs/1610.04495} {\path{arXiv:1610.04495}},
  \href {https://doi.org/10.1007/JHEP04(2017)059}
  {\path{doi:10.1007/JHEP04(2017)059}}.

\bibitem{Gaume:2020bmp}
L.~{\'A}. Gaum{\'e}, D.~Orlando, S.~Reffert, {Selected topics in the large
  quantum number expansion}, Phys. Rept. 933 (2021) 1--66.
\newblock \href {http://arxiv.org/abs/2008.03308} {\path{arXiv:2008.03308}},
  \href {https://doi.org/10.1016/j.physrep.2021.08.001}
  {\path{doi:10.1016/j.physrep.2021.08.001}}.

\bibitem{Antipin:2020abu}
O.~Antipin, J.~Bersini, F.~Sannino, Z.-W. Wang, C.~Zhang, {Charging the $O(N)$
  model}, Phys. Rev. D 102~(4) (2020) 045011.
\newblock \href {http://arxiv.org/abs/2003.13121} {\path{arXiv:2003.13121}},
  \href {https://doi.org/10.1103/PhysRevD.102.045011}
  {\path{doi:10.1103/PhysRevD.102.045011}}.

\bibitem{Badel:2019khk}
G.~Badel, G.~Cuomo, A.~Monin, R.~Rattazzi, {Feynman diagrams and the large
  charge expansion in $3-\varepsilon$ dimensions}, Phys. Lett. B 802 (2020)
  135202.
\newblock \href {http://arxiv.org/abs/1911.08505} {\path{arXiv:1911.08505}},
  \href {https://doi.org/10.1016/j.physletb.2020.135202}
  {\path{doi:10.1016/j.physletb.2020.135202}}.

\bibitem{Badel:2019oxl}
G.~Badel, G.~Cuomo, A.~Monin, R.~Rattazzi, {The Epsilon Expansion Meets
  Semiclassics}, JHEP 11 (2019) 110.
\newblock \href {http://arxiv.org/abs/1909.01269} {\path{arXiv:1909.01269}},
  \href {https://doi.org/10.1007/JHEP11(2019)110}
  {\path{doi:10.1007/JHEP11(2019)110}}.

\bibitem{Jack:2020wvs}
I.~Jack, D.~R.~T. Jones, {Anomalous dimensions for $\phi^n$ in scale invariant
  $d=3$ theory}, Phys. Rev. D 102~(8) (2020) 085012.
\newblock \href {http://arxiv.org/abs/2007.07190} {\path{arXiv:2007.07190}},
  \href {https://doi.org/10.1103/PhysRevD.102.085012}
  {\path{doi:10.1103/PhysRevD.102.085012}}.

\bibitem{Jin:2022nqq}
Q.~Jin, Y.~Li, {Five-loop anomalous dimensions of {\ensuremath{\phi}}$^{Q}$
  operators in a scalar theory with $O(N)$ symmetry}, JHEP 10 (2022) 084.
\newblock \href {http://arxiv.org/abs/2205.02535} {\path{arXiv:2205.02535}},
  \href {https://doi.org/10.1007/JHEP10(2022)084}
  {\path{doi:10.1007/JHEP10(2022)084}}.

\bibitem{Bednyakov:2022guj}
A.~Bednyakov, A.~Pikelner, {Six-loop anomalous dimension of the
  {\ensuremath{\phi}}$^Q$ operator in the $O(N)$ symmetric model}, Phys. Rev. D
  106~(7) (2022) 076015.
\newblock \href {http://arxiv.org/abs/2208.04612} {\path{arXiv:2208.04612}},
  \href {https://doi.org/10.1103/PhysRevD.106.076015}
  {\path{doi:10.1103/PhysRevD.106.076015}}.

\bibitem{Huang:2024hsn}
R.~Huang, Q.~Jin, Y.~Li, {From operator product expansion to anomalous
  dimensions}, JHEP 06 (2025) 135.
\newblock \href {http://arxiv.org/abs/2410.03283} {\path{arXiv:2410.03283}},
  \href {https://doi.org/10.1007/JHEP06(2025)135}
  {\path{doi:10.1007/JHEP06(2025)135}}.

\bibitem{Antipin:2025ilv}
O.~Antipin, J.~Bersini, J.~Hafjall, G.~Muco, F.~Sannino, {Exact Results for the
  Spectrum of the Ising Conformal Field Theory} (11 2025).
\newblock \href {http://arxiv.org/abs/2511.08276} {\path{arXiv:2511.08276}}.

\bibitem{b:PTandCP9}
C.~Domb, J.~L. Lebowitz (Eds.), PHASE TRANSITIONS AND CRITICAL PHENOMENA. VOL.
  9, 1985, (1. Theory of Tricritical Points, Authors: I. D. Lawrie and S.
  Sarbach).

\bibitem{Riedel:1972_1}
E.~K. Riedel, Scaling approach to tricritical phase transitions, Phys. Rev.
  Lett. 28 (1972) 675--678.
\newblock \href {https://doi.org/10.1103/PhysRevLett.28.675}
  {\path{doi:10.1103/PhysRevLett.28.675}}.

\bibitem{Riedel:1972_2}
E.~K. Riedel, F.~J. Wegner, Tricritical exponents and scaling fields, Phys.
  Rev. Lett. 29 (1972) 349--352.
\newblock \href {https://doi.org/10.1103/PhysRevLett.29.349}
  {\path{doi:10.1103/PhysRevLett.29.349}}.

\bibitem{a:STEPHEN1973}
M.~Stephen, J.~McCauley, Feynman graph expansion for tricritical exponents,
  Physics Letters A 44~(2) (1973) 89--90.
\newblock \href {https://doi.org/10.1016/0375-9601(73)90799-8}
  {\path{doi:10.1016/0375-9601(73)90799-8}}.

\bibitem{LewisAdams:1978}
A.~L. Lewis, F.~W. Adams, Tricritical behavior in two dimensions. ii. universal
  quantities from the $\ensuremath{\epsilon}$ expansion, Phys. Rev. B 18 (1978)
  5099--5111.
\newblock \href {https://doi.org/10.1103/PhysRevB.18.5099}
  {\path{doi:10.1103/PhysRevB.18.5099}}.

\bibitem{Hager:1999}
J.~Hager, L.~Sch\"afer, \ensuremath{\Theta}-point behavior of diluted polymer
  solutions: Can one observe the universal logarithmic corrections predicted by
  field theory?, Phys. Rev. E 60 (1999) 2071--2085.
\newblock \href {https://doi.org/10.1103/PhysRevE.60.2071}
  {\path{doi:10.1103/PhysRevE.60.2071}}.

\bibitem{Hager:2002}
J.~S. Hager, {Six-loop renormalization group functions of O(n)-symmetric
  $\phi^6$-theory and epsilon-expansions of tricritical exponents up to
  $\epsilon^3$}, J. Phys. A 35 (2002) 2703--2711.
\newblock \href {https://doi.org/10.1088/0305-4470/35/12/301}
  {\path{doi:10.1088/0305-4470/35/12/301}}.

\bibitem{Adzhemyan:2026}
L.~Adzhemyan, M.~Kompaniets, A.~Trenogin, {Renormalization group analysis of
  tricritical behavior of the $O(n)$-symmetric $\varphi^4 + \varphi^6$ theory
  up to six loops (in preparation)} (2026).

\bibitem{Henriksson:2025kws}
J.~Henriksson, {The tricritical Ising CFT and conformal bootstrap}, JHEP 08
  (2025) 031.
\newblock \href {http://arxiv.org/abs/2501.18711} {\path{arXiv:2501.18711}},
  \href {https://doi.org/10.1007/JHEP08(2025)031}
  {\path{doi:10.1007/JHEP08(2025)031}}.

\bibitem{a:Moueddene_2024_d2}
L.~Moueddene, N.~G~Fytas, Y.~Holovatch, R.~Kenna, B.~Berche, Critical and
  tricritical singularities from small-scale monte carlo simulations: the
  blume–capel model in two dimensions, Journal of Statistical Mechanics:
  Theory and Experiment 2024~(2) (2024) 023206.
\newblock \href {http://arxiv.org/abs/2401.02720} {\path{arXiv:2401.02720}},
  \href {https://doi.org/10.1088/1742-5468/ad1d60}
  {\path{doi:10.1088/1742-5468/ad1d60}}.

\bibitem{a:Moueddene_2024_d3}
L.~Moueddene, N.~G. Fytas, B.~Berche, {Critical and tricritical behavior of the
  $d=3$ Blume-Capel model: Results from small-scale Monte Carlo simulations},
  Phys. Rev. E 110~(6) (2024) 064144.
\newblock \href {http://arxiv.org/abs/2410.01710} {\path{arXiv:2410.01710}},
  \href {https://doi.org/10.1103/PhysRevE.110.064144}
  {\path{doi:10.1103/PhysRevE.110.064144}}.

\bibitem{tHooft:1972tcz}
G.~'t~Hooft, M.~J.~G. Veltman, {Regularization and Renormalization of Gauge
  Fields}, Nucl. Phys. B 44 (1972) 189--213.
\newblock \href {https://doi.org/10.1016/0550-3213(72)90279-9}
  {\path{doi:10.1016/0550-3213(72)90279-9}}.

\bibitem{Bardeen:1978yd}
W.~A. Bardeen, A.~J. Buras, D.~W. Duke, T.~Muta, {Deep Inelastic Scattering
  Beyond the Leading Order in Asymptotically Free Gauge Theories}, Phys. Rev. D
  18 (1978) 3998.
\newblock \href {https://doi.org/10.1103/PhysRevD.18.3998}
  {\path{doi:10.1103/PhysRevD.18.3998}}.

\bibitem{Antipin:2025rsr}
O.~Antipin, J.~Bersini, J.~Hafjall, G.~Muco and F.~Sannino,
{`Semiclassical Canovaccio for Composite Operators}
\newblock \href {http://arxiv.org/abs/2512.23539}
{\path{arXiv:1512.23539 [hep-th]}}.

\bibitem{Cao:2021cdt}
W.~Cao, F.~Herzog, T.~Melia, J.~R. Nepveu, {Renormalization and
  non-renormalization of scalar EFTs at higher orders}, JHEP 09 (2021) 014.
\newblock \href {http://arxiv.org/abs/2105.12742} {\path{arXiv:2105.12742}},
  \href {https://doi.org/10.1007/JHEP09(2021)014}
  {\path{doi:10.1007/JHEP09(2021)014}}.

\bibitem{RoosmaleNepveu:2024zlz}
J.~Roosmale~Nepveu, {Renormalization and the Double Copy of Effective Field
  Theories}, Ph.D. thesis, Humboldt-Universit{\"a}t zu Berlin, Humboldt U.,
  Berlin (2024).
\newblock \href {https://doi.org/10.18452/29107} {\path{doi:10.18452/29107}}.

\bibitem{Bern:2019wie}
Z.~Bern, J.~Parra-Martinez, E.~Sawyer, {Nonrenormalization and Operator Mixing
  via On-Shell Methods}, Phys. Rev. Lett. 124~(5) (2020) 051601.
\newblock \href {http://arxiv.org/abs/1910.05831} {\path{arXiv:1910.05831}},
  \href {https://doi.org/10.1103/PhysRevLett.124.051601}
  {\path{doi:10.1103/PhysRevLett.124.051601}}.

\bibitem{Nogueira:1991ex}
P.~Nogueira, {Automatic Feynman Graph Generation}, J. Comput. Phys. 105 (1993)
  279--289.
\newblock \href {https://doi.org/10.1006/jcph.1993.1074}
  {\path{doi:10.1006/jcph.1993.1074}}.

\bibitem{Batkovich:2014bla}
D.~Batkovich, Y.~Kirienko, M.~Kompaniets, S.~Novikov, {GraphState -- a tool for
  graph identification and labelling} (9 2014).
\newblock \href {http://arxiv.org/abs/1409.8227} {\path{arXiv:1409.8227}}.

\bibitem{Vermaseren:2000nd}
J.~A.~M. Vermaseren, {New features of FORM} (10 2000).
\newblock \href {http://arxiv.org/abs/math-ph/0010025}
  {\path{arXiv:math-ph/0010025}}.

\bibitem{Tentyukov:2007mu}
M.~Tentyukov, J.~A.~M. Vermaseren, {The Multithreaded version of FORM}, Comput.
  Phys. Commun. 181 (2010) 1419--1427.
\newblock \href {http://arxiv.org/abs/hep-ph/0702279}
  {\path{arXiv:hep-ph/0702279}}, \href
  {https://doi.org/10.1016/j.cpc.2010.04.009}
  {\path{doi:10.1016/j.cpc.2010.04.009}}.

\bibitem{Kuipers:2012rf}
J.~Kuipers, T.~Ueda, J.~A.~M. Vermaseren, J.~Vollinga, {FORM version 4.0},
  Comput. Phys. Commun. 184 (2013) 1453--1467.
\newblock \href {http://arxiv.org/abs/1203.6543} {\path{arXiv:1203.6543}},
  \href {https://doi.org/10.1016/j.cpc.2012.12.028}
  {\path{doi:10.1016/j.cpc.2012.12.028}}.

\bibitem{Ruijl:2017dtg}
B.~Ruijl, T.~Ueda, J.~Vermaseren, {FORM version 4.2} (7 2017).
\newblock \href {http://arxiv.org/abs/1707.06453} {\path{arXiv:1707.06453}}.

\bibitem{Vladimirov:1979zm}
A.~A. Vladimirov, {Method for Computing Renormalization Group Functions in
  Dimensional Renormalization Scheme}, Theor. Math. Phys. 43 (1980) 417.
\newblock \href {https://doi.org/10.1007/BF01018394}
  {\path{doi:10.1007/BF01018394}}.

\bibitem{Basu:2015gpa}
P.~Basu, C.~Krishnan, {$\epsilon$-expansions near three dimensions from
  conformal field theory}, JHEP 11 (2015) 040.
\newblock \href {http://arxiv.org/abs/1506.06616} {\path{arXiv:1506.06616}},
  \href {https://doi.org/10.1007/JHEP11(2015)040}
  {\path{doi:10.1007/JHEP11(2015)040}}.

\bibitem{O_Dwyer_2008}
J.~O’Dwyer, H.~Osborn, Epsilon expansion for multicritical fixed points and
  exact renormalisation group equations, Annals of Physics 323~(8) (2008)
  1859–1898.
\newblock \href {https://doi.org/10.1016/j.aop.2007.10.005}
  {\path{doi:10.1016/j.aop.2007.10.005}}.

\bibitem{Demidov:2018czx}
S.~V. Demidov, B.~R. Farkhtdinov, {Numerical study of multiparticle scattering
  in $\lambda\phi^4$ theory}, JHEP 11 (2018) 068.
\newblock \href {http://arxiv.org/abs/1806.10996} {\path{arXiv:1806.10996}},
  \href {https://doi.org/10.1007/JHEP11(2018)068}
  {\path{doi:10.1007/JHEP11(2018)068}}.

\bibitem{GFunctions}
K.~Chetyrkin, A.~Kataev, F.~Tkachov, New approach to evaluation of multiloop
  feynman integrals: The gegenbauer polynomial $x$-space technique, Nuclear
  Physics B 174~(2) (1980) 345--377.
\newblock \href {https://doi.org/https://doi.org/10.1016/0550-3213(80)90289-8}
  {\path{doi:https://doi.org/10.1016/0550-3213(80)90289-8}}.

\bibitem{Herzog:2017bjx}
F.~Herzog, B.~Ruijl, {The R$^{*}$-operation for Feynman graphs with generic
  numerators}, JHEP 05 (2017) 037.
\newblock \href {http://arxiv.org/abs/1703.03776} {\path{arXiv:1703.03776}},
  \href {https://doi.org/10.1007/JHEP05(2017)037}
  {\path{doi:10.1007/JHEP05(2017)037}}.

\bibitem{Chetyrkin:1982nn}
K.~G. Chetyrkin, F.~V. Tkachov, {INFRARED R OPERATION AND ULTRAVIOLET
  COUNTERTERMS IN THE MS SCHEME}, Phys. Lett. B 114 (1982) 340--344.
\newblock \href {https://doi.org/10.1016/0370-2693(82)90358-6}
  {\path{doi:10.1016/0370-2693(82)90358-6}}.

\bibitem{Chetyrkin:1984xa}
K.~G. Chetyrkin, V.~A. Smirnov, {R* OPERATION CORRECTED}, Phys. Lett. B 144
  (1984) 419--424.
\newblock \href {https://doi.org/10.1016/0370-2693(84)91291-7}
  {\path{doi:10.1016/0370-2693(84)91291-7}}.

\end{thebibliography}

\end{document}